\documentclass[
 reprint,
 amsmath,
 amssymb,
 aps,
 prx,
 superscriptaddress,
 longbibliography,
 floatfix
]{revtex4-2}
\usepackage[T1]{fontenc}
\usepackage[utf8]{inputenc}
\usepackage{graphicx}
\usepackage{dcolumn}
\usepackage{siunitx}
\DeclareSIUnit{\atom}{atom}
\DeclareSIUnit{\angstrom}{\text{\AA}}
\DeclareSIUnit{\mebibyte}{MiB}
\usepackage{xcolor}
\usepackage{microtype}
\usepackage{xr-hyper}
\usepackage[hidelinks]{hyperref}

\definecolor{maceblue}{HTML}{2F6690}
\definecolor{fnored}{HTML}{C44E52}
\definecolor{lesgreen}{HTML}{4C956C}

\begin{document}

\title{When do machine-learning potentials need a configuration-dependent kernel? Learning long-range interactions with Fourier neural operators}
\author{Wei Chen}
\email{wei.chen@uclouvain.be}
\affiliation{Institute of Condensed Matter and Nanosciences (IMCN), UCLouvain, Louvain-la-Neuve, Belgium}
\author{Gian-Marco Rignanese}
\email{gian-marco.rignanese@uclouvain.be}
\affiliation{Institute of Condensed Matter and Nanosciences (IMCN), UCLouvain, Louvain-la-Neuve, Belgium}
\affiliation{WEL Research Institute, Wavre, Belgium}

\date{\today}

\begin{abstract}
Long-range interactions in machine-learning interatomic potentials are commonly
treated by coupling learned sources through either a prescribed electrostatic
kernel or a learned but configuration-independent one.
Whether such a kernel suffices depends on the underlying screened interaction.
We address this question by using Fourier neural operators (FNOs) to learn
how latent source fields, built from the descriptors of a local backbone
(here MACE), interact across the simulation cell.
For a synthetic ionic system with tunable screening,
a prescribed Coulomb kernel underestimates the long-wavelength
charge fluctuations by \ensuremath{43\,\%} under strong screening, whereas the FNO
reproduces them within \ensuremath{3\,\%} and recovers
\ensuremath{95\text{--}99\,\%}
of the variance of the residual left by the local backbone.
A linear FNO, which learns a configuration-independent kernel,
performs equally well in the homogeneous case.
Under inhomogeneous screening, rearranging the remote medium
modulates the coupling between distant ions,
a dependence that no configuration-independent kernel can represent.
Among the models tested, only the nonlinear FNO responds to this modulation,
and its force errors are
\ensuremath{27\text{--}41\,\%}
lower than those of configuration-independent models.
In first-principles benchmarks of liquid water, lithium lanthanum zirconate, and Au$_2$/MgO(001),
MACE--\ensuremath{\mathrm{FNO}} is comparable to a prescribed-kernel model for the bulk systems
and more accurate for the interface.
It further captures the reversal of the Au$_2$ wetting preference
induced by remote Al dopants,
with deviations below \ensuremath{5\,\mathrm{meV}} per cell from density functional theory.
These results indicate that a configuration-dependent kernel
becomes necessary once the screening medium varies,
even though the energy remains quadratic in the charges for a fixed medium.
Nonlinear FNOs, which learn such a kernel from energies and forces alone,
offer a flexible route to environment-dependent long-range interactions.
\end{abstract}
\maketitle
\raggedbottom

\section{Introduction}
\label{sec:intro}

Machine-learning interatomic potentials (MLIPs) have made accurate atomistic
simulations accessible at length and time scales beyond first-principles
electronic-structure calculations. A large class of neural-network MLIPs
constructs the energy from local atomic environments.
In message-passing models such as MACE~\cite{batatia2022mace},
information is exchanged within a finite receptive field determined by the radial cutoff
and the number of interaction layers.
The spatial information available to each atom-centered representation
is restricted to this finite receptive field.

This finite range poses a strong limitation when local MLIPs are applied to systems
where long-range interactions become important, such as
polar liquids, ionic solids, interfaces, and adsorbate
systems~\cite{ko2021fourth,unke2021spookynet, cheng2025les}.
Their use can also be problematic when nonlocal correlation arises, e.g., through
a remote charge state or collective polarization beyond the receptive field of the model.

\begin{table*}[!t]
\caption{Comparison of long-range MLIPs that couple learned charges,
multipoles, or latent source features through global operators.
Charge-equilibration models, which determine the sources self-consistently,
are not listed.
Configuration-dependent propagation indicates whether
the effective propagation between two sources can change with the
rest of the configuration through induction or nonlinear updates
for given sources in a fixed cell.
The long-range energy is classified as quadratic or nonquadratic in the source
amplitudes at fixed positions and response coefficients.
Interleaved models, unlike additive ones, feed the propagated features back
into message passing.
Training targets are energies ($E$), forces ($\mathbf F$), stresses
($\boldsymbol\sigma$), total-charge penalties ($Q$), and Born effective charges
(BEC), with parentheses marking targets used only for some systems.
The last two rows correspond to the present work.}
\label{tab:long-range-methods}
\small
\setlength{\tabcolsep}{2.5pt}
\renewcommand{\arraystretch}{1.25}

\begin{ruledtabular}
\begin{tabular}{lllllll}
Method & Sources & Global propagation & \shortstack[l]{Configuration-\\dependent\\propagation} & \shortstack[l]{Energy in\\sources} & \shortstack[l]{Coupling to\\local model} & \shortstack[l]{Training\\targets}\\
\colrule
\parbox[t]{0.14\textwidth}{\raggedright LES~\cite{cheng2025les,king2025charges}} & \parbox[t]{0.13\textwidth}{\raggedright multichannel latent charges} & \parbox[t]{0.19\textwidth}{\raggedright prescribed Ewald} & no & quadratic & additive & $E$, $\mathbf F$ ($\boldsymbol\sigma$)\\[3pt]
\parbox[t]{0.14\textwidth}{\raggedright Extended LES~\cite{park2026multipoles}} & \parbox[t]{0.13\textwidth}{\raggedright latent multipoles} & \parbox[t]{0.19\textwidth}{\raggedright prescribed Ewald} & induction\footnote{Induced charges and dipoles respond linearly, without self-consistency, to the potential and field of the permanent multipoles with locally predicted hardness and polarizability.} & quadratic & additive & $E$, $\mathbf F$ ($\boldsymbol\sigma$)\\[3pt]
\parbox[t]{0.14\textwidth}{\raggedright qNEP~\cite{fan2026qnep}} & \parbox[t]{0.13\textwidth}{\raggedright learned atomic charges} & \parbox[t]{0.19\textwidth}{\raggedright prescribed Ewald/PPPM} & no & quadratic & additive & $E$, $\mathbf F$, $\boldsymbol\sigma$, $Q$\footnote{Charge neutrality is additionally enforced exactly before the electrostatic energy is evaluated.} (BEC)\\[8pt]
\parbox[t]{0.14\textwidth}{\raggedright RSNN (RSP)~\cite{guo2026reciprocal}} & \parbox[t]{0.13\textwidth}{\raggedright latent charges} & \parbox[t]{0.19\textwidth}{\raggedright learned kernel of $|\mathbf k|$} & no & quadratic & additive & $E$, $\mathbf F$, $\boldsymbol\sigma$, $Q$\\[3pt]
\parbox[t]{0.14\textwidth}{\raggedright SOG-Net~\cite{ji2025sognet}} & \parbox[t]{0.13\textwidth}{\raggedright multichannel latent charges} & \parbox[t]{0.19\textwidth}{\raggedright learned sum of Gaussians per channel} & no & quadratic & additive & $\mathbf F$ ($E$)\\[8pt]
\parbox[t]{0.14\textwidth}{\raggedright Ewald MP~\cite{kosmala2023ewaldmp}} & \parbox[t]{0.13\textwidth}{\raggedright atomic embeddings} & \parbox[t]{0.19\textwidth}{\raggedright learned filters} & yes\footnote{The long-range message is linear in the embeddings, whereas the atomic updates within each block are nonlinear.} & nonquadratic & interleaved\footnote{In the DimeNet++~\cite{gasteiger2020dimenetpp} integration, the long-range embeddings form a separate stream that is combined in the output blocks.} & $E$ ($\mathbf F$)\\[3pt]
\parbox[t]{0.14\textwidth}{\raggedright Neural P$^3$M~\cite{wang2024neuralp3m}} & \parbox[t]{0.13\textwidth}{\raggedright mesh features} & \parbox[t]{0.19\textwidth}{\raggedright learned nonlinear Fourier layer per block} & yes & nonquadratic & interleaved & $E$ ($\mathbf F$)\\[3pt]
\colrule
\parbox[t]{0.14\textwidth}{\raggedright MACE--\ensuremath{\mathrm{FNO}}, linear} & \parbox[t]{0.13\textwidth}{\raggedright latent source fields} & \parbox[t]{0.19\textwidth}{\raggedright learned spectral kernel with channel mixing} & no & quadratic & additive & $E$, $\mathbf F$\\[3pt]
\parbox[t]{0.14\textwidth}{\raggedright MACE--\ensuremath{\mathrm{FNO}}, nonlinear} & \parbox[t]{0.13\textwidth}{\raggedright latent source fields} & \parbox[t]{0.19\textwidth}{\raggedright stacked nonlinear Fourier layers} & yes & nonquadratic & additive & $E$, $\mathbf F$\\
\end{tabular}
\end{ruledtabular}
\end{table*}

Long-range-aware MLIPs address this limitation through several complementary
strategies. Environment-dependent atomic charges can be learned from reference
charges or molecular moments and used in explicit electrostatic
energies~\cite{unke2019physnet,sifain2018charge,gong2025bamboo}. Global charge
redistribution can instead be obtained through charge equilibration schemes
~\cite{ko2021fourth,shaidu2024variational,batatia2026macepolar}.
Electronic response can also be represented through
learned maximally localized Wannier centers, as in DPLR~\cite{zeng2022dplr}, or
through a self-consistently updated long-range field, as in
SCFNN~\cite{gao2022scfnn}.
Approaches that do not assign explicit atomic charges
include long-distance equivariant descriptors~\cite{grisafi2019longrange,
huguenin2023nonbonded},
Ewald-based long-range message passing~\cite{kosmala2023ewaldmp,rumiantsev2026lorem},
density-based long-range descriptors~\cite{faller2024density}, and attention-based global
relay nodes~\cite{caruso2025range}.

A recent long-range-aware MLIP family learns environment-dependent sources
from local descriptors and propagates them through a prescribed electrostatic kernel.
Within this family, latent Ewald summation (LES) maps local invariant features
to latent charge-like variables and evaluates their global coupling through a prescribed Coulomb or Ewald
kernel~\cite{cheng2025les}. The framework has been used to augment several
local MLIPs and to recover polarization-related response from energy--force
training~\cite{king2025charges,kim2025universal,zhong2025electrical}. Its
source--kernel factorization is compact and interpretable when the missing
interaction is well described as electrostatics.
Park et al.\ extend this framework to latent atomic dipoles and quadrupoles,
and induced charges and dipoles describe nonlocal charge
redistribution and polarization through a non-self-consistent
linear-response treatment~\cite{park2026multipoles}.
The qNEP framework applies the same broad source--kernel principle within the
neuroevolution potential (NEP) formalism~\cite{fan2021nep}. It predicts
environment-dependent atomic charges and couples them through Ewald
summation or particle--particle particle--mesh electrostatics~\cite{fan2026qnep}.
LES and qNEP demonstrate that local chemical descriptors can supply
the sources of an explicit long-range operator.

While a prescribed electrostatic kernel provides a strong physical prior,
it confines global propagation to a chosen functional form.
In a first-principles description, the bare interaction is Coulombic,
and environment-dependent charges can partly represent the screening that
emerges from electronic polarization~\cite{grasselli2026electrostatics}.
The effective interaction between learned sources, however, need not
take the Coulomb form between point charges.
This is the case, e.g., for metallic Thomas--Fermi screening,
charge transfer at interfaces, and implicit-solvent models.
Learned propagation kernels offer an alternative.
The reciprocal-space neural network (RSNN) couples latent atomic charge
features through an isotropic reciprocal-space kernel represented by a
fully connected network, which describes mixed Coulomb and Lennard-Jones
interactions within a single learned form~\cite{guo2026reciprocal}.
SOG-Net combines latent sources with Fourier-space
kernels represented by sums of Gaussians for which the amplitudes and widths are
learned to incorporate the long-range effects~\cite{ji2025sognet}.
The long-range energy in both approaches is quadratic in the latent sources.
A learned Fourier-space kernel also underlies Neural P$^3$M, which uses nonlinear Fourier
neural operator (FNO) blocks to account for global interactions~\cite{wang2024neuralp3m}.
Table~\ref{tab:long-range-methods} summarizes these approaches
and how they propagate the local sources through global operators.

Building on the principle of learned global propagation,
here we formulate a nonlocal energy correction to a complete local backbone
using the FNO~\cite{li2021fno,kovachki2023neuraloperator} to propagate
latent source fields derived from local atomic descriptors
through spectral mixing and pointwise transformations.
Unlike Neural P$^3$M,
the propagated fields do not feed back into the atomic message-passing layers,
and the learned FNO supplies an additive correction to the local backbone.
Compared with SOG-Net, which propagates each source channel independently
through a sum of Gaussians, the FNO mixes the source channels
at each wavevector and learns its spectral coupling for every wavelength without
an assumed radial form.
Stacked Fourier layers with pointwise nonlinearities further give rise to
an energy that is nonquadratic in the sources, so that the effective propagation
between the sources of two atoms is configuration dependent.
For slabs, the operator couples sources at different heights without
imposing periodicity along the surface normal.
We first construct a synthetic ionic system with a Yukawa interaction
to establish what the FNO learns from the residual interactions
missed by the local backbone.
We then introduce spatially inhomogeneous screening to assess the role
of configuration-dependent propagation when the effective interaction depends on the surrounding medium.
Finally, we evaluate energy and force accuracy for liquid water,
Li$_7$La$_3$Zr$_2$O$_{12}$ (LLZO), and an Au$_2$ dimer on MgO(001).
The latter also tests whether the learned correction captures the
effect of remote dopants on the wetting preference of the adsorbate.
Our results show that a linear FNO with a single configuration-independent
propagation kernel already suffices to describe nonlocal interactions
under homogeneous screening, without prior knowledge of their functional form.
The nonlinear FNO, the effective propagation of which can adapt to
the configuration, remains accurate for screening lengths between
\num{2} and \ensuremath{20\,\text{\AA}} and from weakly to strongly inhomogeneous screening.

\section{Methods}
\label{sec:methods}

\subsection{MACE--\ensuremath{\mathrm{FNO}} architecture}

The central idea is to augment a local MLIP with a nonlocal energy
correction learned through an FNO.
Here we use MACE as a representative backbone
to predict the local energy $E_\mathrm{MACE}$.
The construction can be coupled to other local MLIPs
as long as they provide differentiable and rotationally invariant
atomic descriptors.
The residual energy term is recast into a correction term
$\Delta E_\mathrm{FNO}$ from the learned field response.
For a given atomic configuration, the total energy can be written as
a sum of the two contributions \begin{equation}
 E=E_{\mathrm{MACE}}+\Delta E_{\mathrm{FNO}}.
 \label{eq:decomposition}
\end{equation}

\begin{figure*}[tp]
\centering
\includegraphics[width=\textwidth]{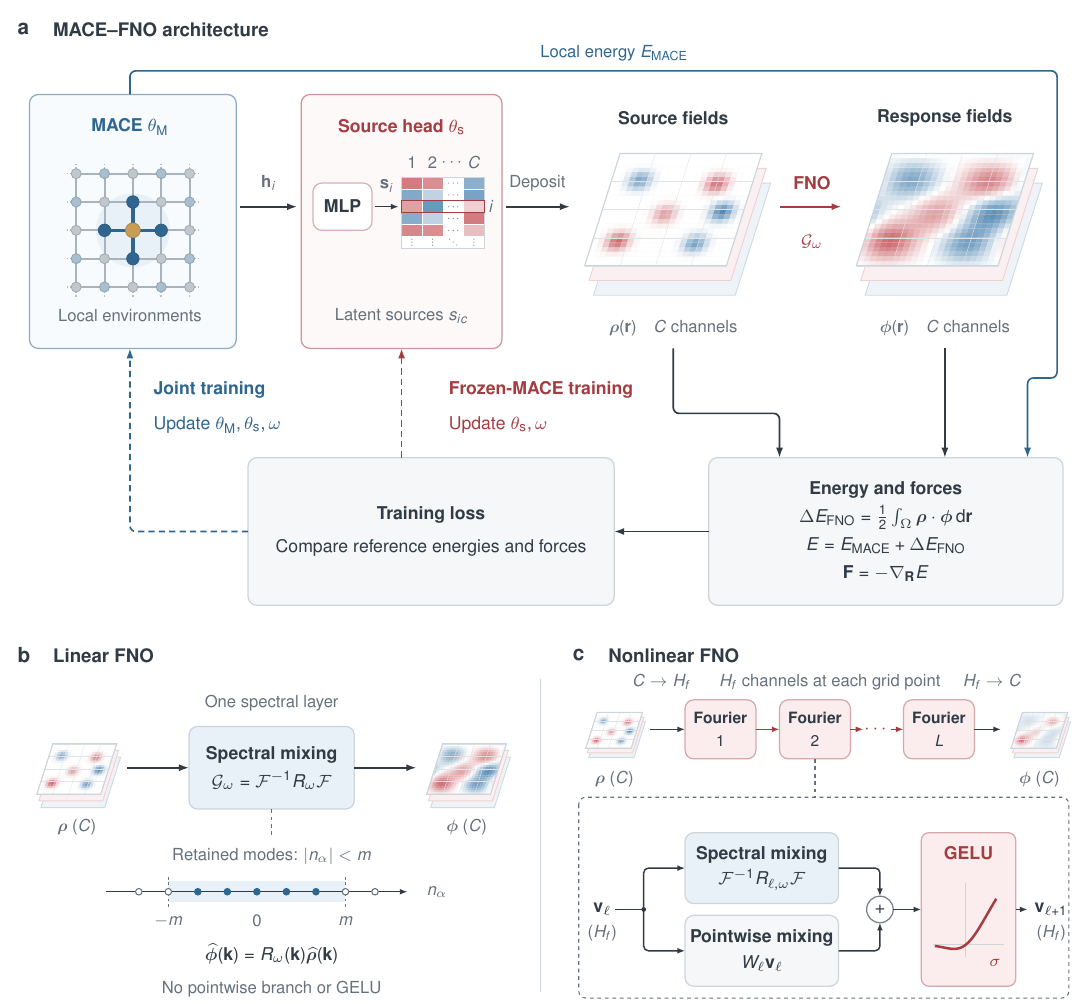}
\caption{Schematics of the MACE--\ensuremath{\mathrm{FNO}} architecture and the two operator variants.
(a) Local MACE descriptors $\mathbf h_i$ are mapped to latent atomic
sources $s_{ic}$ and deposited on a mesh. The FNO transforms the source
fields into response fields, and their contraction gives the energy correction.
Solid arrows show forward propagation, and dashed red and blue arrows
indicate frozen-MACE and joint training, respectively.
(b) Linear FNO applies a single spectral operation to the source channels.
No pointwise branch or nonlinear activation is used.
The inset illustrates the retained modes along one Fourier direction.
(c) Nonlinear FNO maps the sources to hidden channels, applies successive
Fourier layers, and projects back to the response channels.
The dashed box shows a single Fourier layer, in which the spectral and
pointwise branch outputs are summed before applying the GELU activation.}
\label{fig:fno-workflow}
\end{figure*}

Figure~\ref{fig:fno-workflow}(a) illustrates our MACE--\ensuremath{\mathrm{FNO}} architecture.
Following the particle--mesh method, we deposit atomic sources on a
regular grid and evaluate global coupling through Fourier-space
operations~\cite{darden1993pme,essmann1995pme}.
For atom $i$, MACE provides a rotationally invariant descriptor
$\mathbf h_i$ that encodes its local chemical environment.
A trainable source head maps this descriptor to $C$ scalar source
amplitudes $s_{ic}$, one per channel $c=1,\ldots,C$.
These amplitudes are then distributed on a mesh using smooth
cubic B-spline weights centered on the atomic position.
Summing these contributions produces a source field $\rho_c(\mathbf r)$
at a mesh position $\mathbf r$.
Motivated by overall electrostatic neutrality, we constrain each latent
source field to have a zero mesh integral.

The FNO transforms the combined source field
$\boldsymbol\rho=(\rho_1,\ldots,\rho_C)$ into a response field $\boldsymbol\phi$
through a learned operator $\mathcal G_\omega$
\begin{equation}
 \boldsymbol\phi(\mathbf r)
 =\mathcal G_\omega[\boldsymbol\rho;\mathbf H](\mathbf r),
 \label{eq:learned-global-map}
\end{equation}
where $\mathbf H$ is the cell matrix, and $\omega$ denotes the learnable
operator parameters.
The response field $\boldsymbol\phi$ contains $C$ channels, matching
those of the source fields.
We build $\mathcal G_\omega$ by passing the output fields of one Fourier
layer to the next.
In layer $\ell$, the input $\mathbf v_\ell(\mathbf r)$ is a real-space vector
containing $H_f$ feature values at mesh position $\mathbf r$.
In particular, the first layer receives $H_f$ hidden fields obtained by a learned,
pointwise mapping of the source channels $\boldsymbol \rho$.
Within a single Fourier layer, spectral mixing
proceeds sequentially through a Fourier transform, learned channel mixing, and an inverse transform
\begin{equation}
 \begin{aligned}
 \widehat{\mathbf v}_\ell(\mathbf k)
 &=\mathcal F_{\mathbf r\to\mathbf k}
   [\mathbf v_\ell(\mathbf r)],\\
 \widehat{\mathbf u}_\ell(\mathbf k)
 &=R_{\ell,\omega}(\mathbf k;\mathbf H)\,
   \widehat{\mathbf v}_\ell(\mathbf k),\\
 \mathbf u_\ell(\mathbf r)
 &=\mathcal F^{-1}_{\mathbf k\to\mathbf r}
   [\widehat{\mathbf u}_\ell(\mathbf k)],
 \end{aligned}
 \label{eq:main-spectral-mixing}
\end{equation}
where the learned matrix $R_{\ell,\omega}$ combines the hidden channels
at a reciprocal vector $\mathbf k$, and $\mathbf u_\ell(\mathbf r)$ is
the resulting real-space spectral response.
This procedure is illustrated schematically in Fig.~\ref{fig:fno-workflow}(c).
In real space, this operation corresponds to a convolution throughout the periodic unit cell.
The response in one region can therefore depend on fields beyond the MACE receptive field.
This provides the mechanism for learning nonlocal interactions using the FNO.

To complement the nonlocal spectral response, we further consider a pointwise contribution
$W_\ell\mathbf v_\ell(\mathbf r)$, as shown in Fig.~\ref{fig:fno-workflow}(c).
$W_\ell$ is a learned $H_f\times H_f$ matrix that is shared across all mesh points
within a specific Fourier layer $\ell$.
With the standard Fourier-layer update, the input to layer $\ell+1$ is
\begin{equation}
 \mathbf v_{\ell+1}(\mathbf r)
 =\sigma\!\left[
 \mathbf u_\ell(\mathbf r)+W_\ell\mathbf v_\ell(\mathbf r)
 \right],
 \label{eq:main-layer-update}
\end{equation}
where $\sigma$ applies the Gaussian error linear unit (GELU) activation
componentwise to the combined contributions.

For two-dimensional systems such as Au$_2$/MgO(001), we apply the Fourier
transform only to the periodic in-plane coordinates and jointly mix all
hidden channels and height points along the surface normal.
Writing $z_j$ for the $j$th height point, the spectral response becomes
\begin{equation}
 \begin{aligned}
 \widehat u_{\ell,a}(\mathbf k_\parallel,z_j)
 ={}&\sum_{b=1}^{H_f}\sum_{j'=1}^{N_z}
 R_{\ell,\omega}^{ab}(|\mathbf k_\parallel|;z_j,z_{j'})\\
 &\qquad\times\widehat v_{\ell,b}(\mathbf k_\parallel,z_{j'}),
 \end{aligned}
 \label{eq:main-slab-spectral-mixing}
\end{equation}
where $a$ and $b$ label hidden channels, and $N_z$ is the number of height points.
We refer to this formulation as full $z$-mixing.
The dependence of $R_{\ell,\omega}$ on both heights and wavelength allows the coupling between
different slab heights to vary with the lateral length scale.
An inverse in-plane transform gives $\mathbf u_\ell$ for
Eq.~\eqref{eq:main-layer-update}, with $W_\ell$ mixing channels at each mesh point.
No Fourier transform or periodic wrapping is applied along $z$.
Alternatively, we separate channel and height mixing so that
\begin{equation}
 \begin{aligned}
 \widehat u_{\ell,a}(\mathbf k_\parallel,z_j)
 ={}&\sum_{b=1}^{H_f}\widetilde R_{\ell,\omega}^{ab}(|\mathbf k_\parallel|)
 \widehat v_{\ell,b}(\mathbf k_\parallel,z_j)\\
 &+\sum_{j'=1}^{N_z}Z_{\ell,a}(j,j')
 \widehat v_{\ell,a}(\mathbf k_\parallel,z_{j'}),
 \end{aligned}
 \label{eq:main-slab-global-mixing}
\end{equation}
where $\widetilde R_{\ell,\omega}$ mixes channels at each height and is
shared across all heights,
and the learned matrices $Z_{\ell,a}$ mix heights within each channel and are
independent of the in-plane wavevector.
Within one Fourier layer, the channel-wise mixing
of Eq.~\eqref{eq:main-slab-global-mixing} couples different heights only within
the same channel, while a height in one channel couples to a different height
in another channel only through successive Fourier layers.
Appendix~\ref{app:slab-full-height} analyzes the computational
complexity of the two different mixing schemes,
and Appendix~\ref{app:slab-mixing-comparison} compares the numerical accuracy
and GPU inference costs.

Once the final response field $\boldsymbol\phi$ is produced,
the energy correction can be obtained by contracting the source and response
fields over the mesh domain $\Omega$
\begin{equation}
 \Delta E_{\mathrm{FNO}}
 =\frac12\int_\Omega
 \boldsymbol\rho(\mathbf r)\cdot\boldsymbol\phi(\mathbf r)
 \,\mathrm d\mathbf r,
 \label{eq:main-energy-readout}
\end{equation}
where the dot product sums over the $C$ channels.
We emphasize that the source ($\boldsymbol\rho$) and response fields ($\boldsymbol\phi$)
are latent variables and need not represent physical charge densities or electrostatic potentials.
Rather than prescribing electrostatics explicitly as in LES and qNEP,
our approach learns both the sources and the global response operator.
The total forces follow from
\begin{equation}
 \mathbf F_i=-\frac{\partial}{\partial\mathbf R_i}
 \left(E_{\mathrm{MACE}}+\Delta E_{\mathrm{FNO}}\right),
 \label{eq:main-force-readout}
\end{equation}
where $\mathbf R_i$ is the position of atom $i$.

The FNO hitherto defines a nonlinear mapping from the source to response fields
due to the nonlinear activation between the Fourier layers.
By replacing the stacked Fourier layers and their input and output
mappings with a single spectral operation directly on the $C$ source
channels [cf.\ Fig.~\ref{fig:fno-workflow}(b)],
we can construct a \emph{linear} FNO, denoted \ensuremath{\mathrm{FNO}^{(\mathrm{l})}},
for which the response satisfies
$\widehat{\boldsymbol\phi}(\mathbf k)
=R_\omega(\mathbf k;\mathbf H)\widehat{\boldsymbol\rho}(\mathbf k)$,
where $R_\omega$ is a learned $C\times C$ matrix.
The linear \ensuremath{\mathrm{FNO}^{(\mathrm{l})}} therefore omits the pointwise branch
and nonlinear activation of the field operator.
A linear operator with more than one Fourier layer would not be more expressive,
since stacked layers without activation compose into
a single wavevector-dependent channel-mixing matrix,
apart from a local term that the backbone can represent.
The difference in accuracy between the linear and nonlinear FNO
therefore cannot be attributed to the depth of the operator.
As the source head remains nonlinear while the contraction follows
Eq.~\eqref{eq:main-energy-readout},
the energy correction obtained with a linear FNO is quadratic only in the \emph{source fields}.

The MACE--\ensuremath{\mathrm{FNO}} optimization spans the parameter space defined by
$\theta_{\mathrm M}$, $\theta_{\mathrm s}$, and $\omega$, denoting the
parameters of MACE, the source head, and $\mathcal G_\omega$, respectively.
Within this space, we consider two training strategies initialized from
a pretrained MACE model and fitted to reference energies and forces.
In frozen-MACE training, we keep $\theta_{\mathrm M}$ fixed and update
$\theta_{\mathrm s}$ and $\omega$ through backpropagation.
This confines optimization to the source head and FNO, isolating
the learned correction relative to an unchanged local baseline.
In joint MACE--\ensuremath{\mathrm{FNO}} training, we update all three parameter
sets together by backpropagating the same energy--force loss through the
full coupled model.
Here, the MACE parameter gradients receive contributions from both its local
energy prediction and the descriptor-dependent FNO correction.
Through the latter path, the FNO correction can reshape the local descriptors
as the local representation and global operator adapt together.
These alternatives correspond to the two return routes in
Fig.~\ref{fig:fno-workflow}(a).
In the tables, the superscript $(0)$ in
\mbox{\ensuremath{\mathrm{MACE}_{1}^{(0)}}--\ensuremath{\mathrm{FNO}}} and \mbox{\ensuremath{\mathrm{MACE}_{2}^{(0)}}--\ensuremath{\mathrm{FNO}}} denotes a frozen one- or
two-interaction MACE backbone. FNO-augmented entries without this
superscript denote joint training.

The source construction, complete layer equations, and derivative
expressions are given in Appendix~\ref{app:technical-formulation}.
The geometry-specific operators and symmetry treatment are detailed in
Appendix~\ref{sec:geometries}.

\subsection{Spectral-response diagnostic}
\label{sec:spectral-diagnostic-method}

Although $\mathcal G_\omega$ is not tied to a prescribed kernel,
the learned response can still be characterized by the changes in the FNO energy
upon Fourier perturbations to the latent source fields.
For a nonzero reciprocal vector $\mathbf k$, we construct a cosine probe $f_{\mathbf k}$
at that wavevector~\cite{baroni2001dfpt,dornheim2021harmonics}, and add it to the deposited fields
\begin{equation}
 \boldsymbol\rho_{\mathbf a,\mathbf k}(\mathbf r)
 =\boldsymbol\rho(\mathbf r)+f_{\mathbf k}(\mathbf r)\mathbf a,
 \label{eq:spectral-field-probe}
\end{equation}
where $\mathbf a$ contains the perturbation amplitudes of the $C$ latent source channels.
The probe has a zero mesh integral and is normalized to an RMS amplitude of one,
preserving the zero mesh integral of the source fields.
The learned response then follows from the channel-space energy curvature
\begin{equation}
 H_{cc'}(\mathbf k)
 =\frac{1}{\mathcal V_{\mathrm{eff}}}
 \left.
 \frac{\partial^2\Delta E_{\mathrm{FNO}}
 [\boldsymbol\rho_{\mathbf a,\mathbf k}]}
 {\partial a_c\,\partial a_{c'}}
 \right|_{\mathbf a=0},
 \label{eq:spectral-channel-curvature}
\end{equation}
where $c$ and $c'$ label source channels, and $\mathcal V_{\mathrm{eff}}=V$
is the cell volume.
For a nonlinear FNO, this describes the energy curvature around the supplied
fields and may vary between configurations.

Diagonalizing $H(\mathbf k)$ gives eigenvalues $\lambda_j$ and corresponding
eigenvectors that specify normalized combinations of source channels.
Each eigenvalue is the quadratic energy cost per unit volume for a
perturbation along its channel combination.
We follow the largest positive eigenvalue $\lambda_+(\mathbf k)$,
the strongest positive curvature at wavelength $2\pi/|\mathbf k|$,
and discard wavevectors at which no eigenvalue is positive.
The radial dependence of this response can be summarized by the
approximate power law
\begin{equation}
 \lambda_+(k)\simeq Ak^{-p},
 \label{eq:spectral-power-law}
\end{equation}
where $k=|\mathbf k|$, $A$ is a fitted prefactor,
and $p$ is the spectral exponent over the chosen wavevector range.
It is instructive to compare the spectral exponent with the wavevector
dependence of a reference interaction, e.g., $p_{\mathrm{ref}}=2$
for a pure three-dimensional Coulomb kernel.

We apply the spectral analysis to the synthetic benchmark of Sec.~\ref{sec:synthetic-results},
for which the reference kernel is known.
For the first-principles benchmarks, no reference kernel is available,
and the fitted exponent of the latent response cannot be checked
against the interaction it represents.

The fit targets the lowest nonzero wavevectors where the accessible
wavelengths are among the longest.
We express wavevector magnitudes as
$\widetilde k=kL_{\mathrm{eff}}/(2\pi)$ with $L_{\mathrm{eff}}=V^{1/3}$.
Within a shell $s$ of equal wavevector magnitude $k_s$,
we average the positive response to obtain $\bar\lambda_s$.
An unweighted linear regression of $\ln\bar\lambda_s$ against $\ln k_s$
yields $p$ as the negative slope.
The upper cutoff $\widetilde k_{\max}=\num{1.8}$ admits the three
lowest shells at $\widetilde k=1,\sqrt{2},\sqrt{3}$.
We average the structure-level exponents within a checkpoint before
calculating means and sample standard deviations across training seeds.
Note that these three-point fits are pertinent to the accessible low-$k$ response
and they do not necessarily establish the asymptotic $k\to0$ limit.
Appendix~\ref{app:spectral-response-diagnostic} gives the finite-difference
procedure and the detailed sampling and regression settings.

\subsection{Implementation and training protocols}
\label{sec:implementation}

\begin{table*}[t]
\caption{Principal FNO hyperparameters for the four benchmark systems.
The synthetic settings apply to all three screening parameters $\kappa$.
$N$ denotes the number of mesh points per periodic direction, and $L$ the
number of Fourier layers. Bulk meshes contain $N^3$ points, while the slab
mesh has $N\times N\times24$ points, with the last dimension resolving height.}
\label{tab:computational-settings}
\begin{ruledtabular}
\begin{tabular}{lcccc}
Parameter & Synthetic ionic system & Liquid water & LLZO & Au$_2$/MgO(001) \\
\hline
Mesh points $N$ & 44 & 24 & 32 & 24 \\
Mode parameter $m$ & 11 & 8 & 8 & 8 (in-plane) \\
Source channels $C$ & 4 & 16 & 16 & 16 \\
Fourier layers $L$ & 2 & 2 & 4 & 4 \\
\end{tabular}
\end{ruledtabular}
\end{table*}

We implement MACE--\ensuremath{\mathrm{FNO}} in \textsc{PyTorch}~\cite{paszke2019pytorch}.
Following the equivariance treatment in EqGINO~\cite{kim2026eqgino},
we impose isotropic spectral coupling
by sharing channel-mixing matrices among modes with the same physical wavelength
(Appendix~\ref{sec:geometries}).
Our MACE--\ensuremath{\mathrm{FNO}} code interfaces with \textsc{ase}~\cite{larsen2017ase} for energy--force inference
through a calculator interface.

The mode parameter $m$ includes integer indices
$n_\alpha=-(m-1),\ldots,m-1$ along the Fourier-transformed directions.
For orthorhombic cells, these indices give
$k_\alpha=2\pi n_\alpha/L_\alpha$, where $L_\alpha$ is the cell length along
direction $\alpha$.
The source-channel count $C$ specifies the number of deposited fields.
The source head has \num{64} hidden units in all models.
Appendix~\ref{app:hyperparameter-convergence} logs the sensitivity to the
mesh resolution, mode setting, and source-channel count.
Table~\ref{tab:computational-settings} summarizes the principal FNO
hyperparameters used for the synthetic and physical benchmarks.

For a given dataset and interaction depth, the frozen and joint
runs start from the same MACE checkpoint and share the data
split, mesh, and loss normalization.
We perform independent FNO training runs per backbone with
five random seeds.
We report the mean and sample standard deviation of test root-mean-square
errors (RMSEs) obtained
by evaluating the five trained models separately.
These statistics describe training variability for a given starting MACE model.

The source head and FNO are optimized with \texttt{Adam}~\cite{kingma2015adam}.
We evaluate validation errors in batches of eight structures and select
checkpoints by the combined normalized energy and force error.
The step budgets, batch sizes, loss normalization, learning-rate schedule, and selection
rules are given in Appendix~\ref{app:training-selection}.
Double-precision arithmetic is used throughout training and inference.

\section{Homogeneous screening beyond a prescribed kernel}
\label{sec:synthetic-results}

Drawing on the homogeneous continuum description of screening in many bulk materials,
we first construct a synthetic ionic system with spatially uniform screening.
The system comprises disordered, charge-neutral NaCl-like
configurations interacting through a short-range Lennard--Jones (LJ) potential and a
Gaussian-smeared Yukawa potential with a tunable inverse screening length $\kappa$.
Screening is made homogeneous by construction since $\kappa$
is spatially constant and independent of the ionic configuration.

The short-range interaction of the synthetic systems is described
by a smoothly truncated LJ interaction
\begin{equation}
u_{\mathrm{LJ}}(r)=4\varepsilon\left[
\left(\frac{\sigma_{\mathrm{LJ}}}{r}\right)^{12}
-\left(\frac{\sigma_{\mathrm{LJ}}}{r}\right)^{6}\right]S(r),
\label{eq:synthetic-lj}
\end{equation}
where $\varepsilon=\ensuremath{0.20\,\mathrm{eV}}$ defines the well depth,
and $\sigma_{\mathrm{LJ}}=\ensuremath{3.0\,\text{\AA}}$ is the characteristic LJ distance
that places the potential minimum at $2^{1/6}\sigma_{\mathrm{LJ}}=\ensuremath{3.37\,\text{\AA}}$.
The switching function is a quintic smoothstep that truncates the LJ interactions smoothly to zero beyond
a cutoff $r_{\mathrm{cut}}$
\begin{equation}
S=1-10y^3+15y^4-6y^5,
\end{equation}
where $y=(r-3.6)/(r_{\mathrm{cut}}-3.6)$, and $r_{\mathrm{cut}}=\ensuremath{4.5\,\text{\AA}}$ following the
MACE backbone.
The switching function guarantees that the energy and its first two derivatives vanish continuously
at the cutoff.
As a result, the short-range interaction is confined by construction
to the \ensuremath{\mathrm{MACE}_{1}} receptive field without introducing a discontinuity.

The Gaussian-smeared Yukawa interaction is specified by the reciprocal-space kernel
\begin{equation}
 V_Y(k)\,g(k)=\frac{4\pi}{k^2+\kappa^2}\,e^{-\sigma^2k^2},
\label{eq:synthetic-kernel}
\end{equation}
where $\kappa$ refers to the inverse screening length, and $\sigma=\ensuremath{0.6\,\text{\AA}}$
is a smearing parameter that regularizes the interaction at contact.
The pointwise logarithmic slope of Eq.~\eqref{eq:synthetic-kernel}
has the analytical form
\begin{equation}
\begin{aligned}
p_{Y}(k)
&=-\frac{\mathrm d\ln[V(k)g(k)]}{\mathrm d\ln k}\\
&=\frac{2k^2}{k^2+\kappa^2}+2\sigma^2 k^2.
\end{aligned}
\label{eq:p_Vg}
\end{equation}
This exponent provides an analytical reference, the wavevector dependence
of which can be varied through $\kappa$.
Without appreciable Gaussian damping,
it crosses approximately from zero when $k\ll\kappa$
to two when $k\gg\kappa$.
Here we use $\kappa=\ensuremath{0.05\,\text{\AA}^{-1}\text{, }0.2\,\text{\AA}^{-1}\text{, and }0.5\,\text{\AA}^{-1}}$, which correspond to
screening lengths $\kappa^{-1}$ of \ensuremath{20\,\text{\AA}\text{, }5\,\text{\AA}\text{, and }2\,\text{\AA}} that straddle the backbone cutoff.
This variation tests whether a learned operator can follow the change in
screening without a prescribed kernel.
LES keeps its Coulomb propagation kernel fixed, although its learned charges
and local energy contribution can adapt to the different training targets.

To keep the nonlocal energy-variation scale constant across $\kappa$,
we multiply the nonlocal energies and forces by a dimensionless factor
$a_\kappa$, leaving the LJ contribution unchanged.
The factors are \numlist{1.8512;2.0483;2.9920} for
$\kappa=\ensuremath{0.05\,\text{\AA}^{-1}\text{, }0.2\,\text{\AA}^{-1}\text{, and }0.5\,\text{\AA}^{-1}}$, respectively.
They fix the standard deviation of the nonlocal energy per atom across
the generated configurations at \ensuremath{5\,\mathrm{meV}\,\mathrm{atom}^{-1}}.
This rescaling changes the interaction strength but not its wavevector
dependence or logarithmic slope.
With this normalization, we calculate the periodic Yukawa energy as
\begin{equation}
\begin{aligned}
E_{\mathrm{Y}}
&=\frac{a_\kappa C_{\mathrm{Coul}}}{2V}
\sum_{\substack{\mathbf k\\0<|\mathbf k|\le k_{\mathrm{cut}}}}
V_Y(k)g(k)|\rho_{\mathrm{ref}}(\mathbf k)|^2\\
&\quad-a_\kappa E_{\mathrm{self}}.
\end{aligned}
\label{eq:synthetic-yukawa-energy}
\end{equation}
Here $\rho_{\mathrm{ref}}(\mathbf k)=\sum_i q_i e^{-i\mathbf k\cdot\mathbf R_i}$
is the Fourier amplitude of the reference charges $q_i$ at positions $\mathbf R_i$,
and $C_{\mathrm{Coul}}$ is the Coulomb unit-conversion factor.
The sum includes both $\mathbf k$ and $-\mathbf k$ up to
$k_{\mathrm{cut}}=\ensuremath{3.5\,\text{\AA}^{-1}}$.
The term $E_{\mathrm{self}}$ removes the unscaled self-interaction energies
of the Gaussian-smeared charges.

The total reference energy of the synthetic ionic system is
\begin{equation}
E_{\mathrm{ref}}=E_{\mathrm{LJ}}+ E_{\mathrm{Y}},
\end{equation}
where $E_{\mathrm{LJ}}=\sum_{i<j}u_{\mathrm{LJ}}(r_{ij})$.
The Yukawa contribution follows Eq.~\eqref{eq:synthetic-yukawa-energy},
which already accounts for the scaling through $a_\kappa$.
The residual energy $E_\mathrm{res}={E_\mathrm{ref}}-E_\mathrm{MACE}$
forms the target of the FNO correction, which includes both
missing nonlocal contributions and remaining errors of the local backbone.

We model \num{1146} atoms with charges $\pm\num{0.4}\,e$ in a periodic
cubic cell with a side length of $L_a=\ensuremath{32\,\text{\AA}}$.
This cell size gives a smallest nonzero wavevector magnitude of
$k_{\min}=2\pi/L_a=\ensuremath{0.196\,\text{\AA}^{-1}}$.
The corresponding wavelength is about seven times the descriptor cutoff
of the backbone.
From random initial positions, we generate \num{40} independent Langevin
molecular-dynamics (MD) trajectories at \ensuremath{400\,\mathrm{K}} driven by the unscaled
Yukawa interaction with $\kappa=\ensuremath{0.05\,\text{\AA}^{-1}}$ and $a_\kappa=1$,
and sample \num{30} configurations per trajectory after a \ensuremath{3\,\mathrm{ps}} burn-in.
These trajectories are subsequently assigned to training, validation, and
testing, resulting in \num{660}, \num{240}, and \num{300} configurations, respectively.
No trajectory is shared between subsets.
The trajectories are then relabeled with the scaled energies and forces evaluated at each $\kappa$.
The atomic configurations and the splits are thus identical across the three datasets.
We note that the datasets at $\kappa=\ensuremath{0.2\,\text{\AA}^{-1}\text{ and }0.5\,\text{\AA}^{-1}}$ carry the
suppressed long-wavelength charge fluctuations of the nearly Coulombic fluid
as the configurations are sampled at $\kappa=\ensuremath{0.05\,\text{\AA}^{-1}}$.
This is adequate for a static benchmark,
whereas the MD simulations of Sec.~\ref{sec:S_zz} sample each interaction
in its own ensemble.

Throughout this benchmark, we use a simple one-interaction \ensuremath{\mathrm{MACE}_{1}} model
with a cutoff of \ensuremath{4.5\,\text{\AA}} as the local backbone.
Its hidden representation contains \num{128} scalar ($0e$) and
\num{128} vector ($1o$) channels.
We adopt the FNO settings in Table~\ref{tab:computational-settings}.
The Fourier layers have a hidden-channel width $H_f$ of \num{32}.
Spectral mixing uses a single group and couples all hidden channels.
The static energy--force benchmark is evaluated on a single mesh
without interlacing.

To first order, the agreement of the FNO correction and the residual energy
can be evaluated through the residual variance
$1-\mathrm{Var}[E_\mathrm{res}-\Delta E_\mathrm{FNO}]/\mathrm{Var}[E_\mathrm{res}]$.
We calculate the residual variance across the \num{300} test configurations
using the five FNO seeds on the same frozen backbone.
The FNO correction explains \ensuremath{99.3\,\%},
\ensuremath{98.2\,\%}, and
\ensuremath{95.1\,\%} of the residual variance for
$\kappa=\ensuremath{0.05\,\text{\AA}^{-1}\text{, }0.2\,\text{\AA}^{-1}\text{, and }0.5\,\text{\AA}^{-1}}$, respectively,
thereby confirming that the FNO closely follows the residual energy as a whole.
Nevertheless, such an agreement does not address the wavelengths at which
the FNO recovers the residual.
Knowing the reference charges of the synthetic system,
we now proceed to project the residual and FNO energies onto reciprocal-space
shell features.

\subsection{Shell-resolved residual recovery}
\label{sec:shell_residual}
\begin{figure*}[t]
\includegraphics[width=\textwidth]{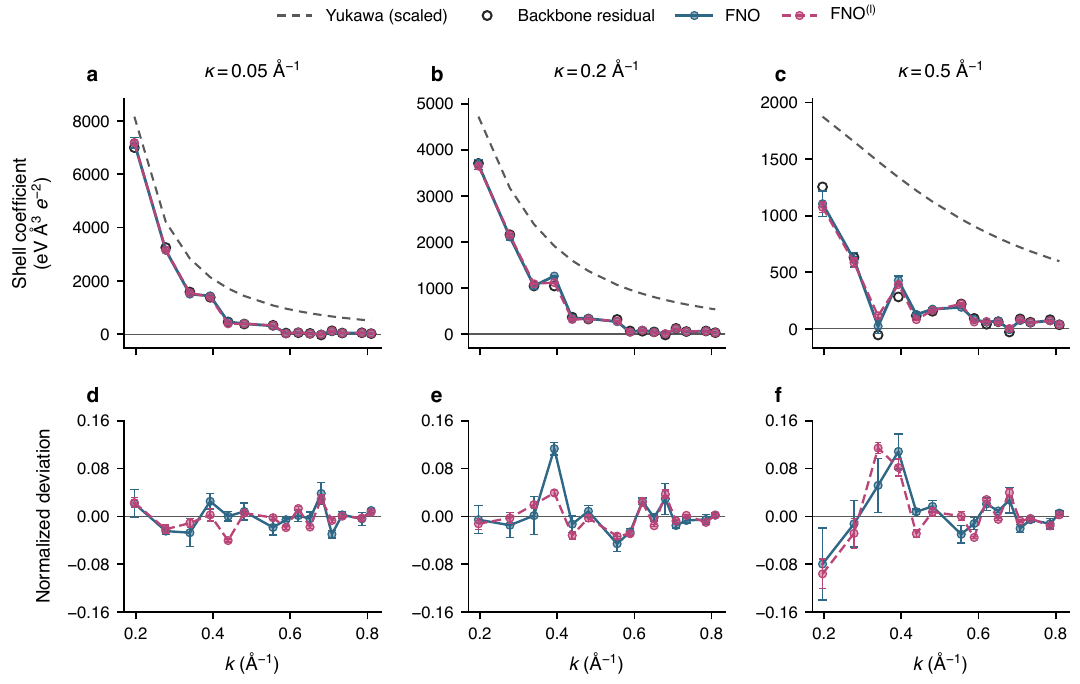}
\caption{Shell-resolved recovery of the frozen-MACE residual in the synthetic ionic system.
Columns correspond to $\kappa=\ensuremath{0.05\,\text{\AA}^{-1}\text{, }0.2\,\text{\AA}^{-1}\text{, and }0.5\,\text{\AA}^{-1}}$.
(a--c) Raw coefficients. The gray dashed line gives the reference Yukawa coefficient
$c_{\mathrm Y,s}=a_\kappa C_{\mathrm{Coul}}V(k_s)$.
Black circles are coefficients fitted to $E_{\mathrm{ref}}-E_{\mathrm{MACE}}$
for the MACE backbone (seed \num{17}),
whereas filled blue and purple circles are the mean coefficients for FNO and
\ensuremath{\mathrm{FNO}^{(\mathrm{l})}}, respectively, over five seeds on this frozen backbone.
Error bars denote one sample standard deviation for both variants.
(d--f) Deviations from the residual, $(c_{A,s}-c_{\mathrm{res},s})/c_{\mathrm Y,s}$,
for the two FNO variants, with error bars normalized by the same reference coefficient.
The horizontal zero lines mark exact agreement with the residual.
The coefficients are shown for the $\num{15}$ lowest shells.
Lines are guides to the eye.
}
\label{fig:synthetic-shell-recovery}
\end{figure*}

For configuration $j$, we construct the shell features from the known
reference charges,
\begin{equation}
\begin{aligned}
X_{js}
&=\frac{1}{V}\sum_{\mathbf k\in\mathcal S_s}
g(k)|\rho_{\mathrm{ref},j}(\mathbf k)|^2,\\
\rho_{\mathrm{ref},j}(\mathbf k)
&=\sum_i q_i e^{-i\mathbf k\cdot\mathbf R_i^{(j)}},
\end{aligned}
\label{eq:synthetic-shell-features}
\end{equation}
where $V$ is the cell volume,
$q_i$ is the reference charge of atom $i$ at position $\mathbf R_i^{(j)}$,
and $g(k)=\exp(-\sigma^2k^2)$ accounts for the Gaussian smearing.
The shell $\mathcal S_s$ groups nonzero wavevectors of equal magnitude $k_s$.
Only one member of a $\pm\mathbf k$ pair enters the sum, giving
the prefactor $1/V$ instead of $1/(2V)$.
Thus $X_{js}$ measures the weighted charge-density modulation at $k_s$
with the reference charges.

We express the Yukawa reference, the frozen-backbone residual, and the FNO corrections
in a common shell feature representation
\begin{equation}
y_{A,j}\simeq b_A+\sum_{s=1}^{S}c_{A,s}X_{js}+t_A T_j,
\label{eq:synthetic-shell-projection}
\end{equation}
where $A\in\{\mathrm Y,\mathrm{res},\mathrm{FNO}\}$ identifies the energy being analyzed.
We resolve the lowest $S=\num{15}$ shells individually and
include the higher-$k$ reference contribution in
$T_j=\sum_{s>S}c_{\mathrm Y,s}X_{js}$ up to the cutoff
$k_{\mathrm{cut}}=\ensuremath{3.5\,\text{\AA}^{-1}}$.
For the Yukawa reference, the coefficient corresponds essentially to the scaled Yukawa kernel
$c_{\mathrm Y,s}=a_\kappa C_{\mathrm{Coul}}V(k_s)$, and thus no fitting is required.
For the residual and FNO energies, we determine $b_A$, $c_{A,s}$, and $t_A$ simultaneously
by ordinary least squares over the same \num{300} test configurations.
Note that the choice of energy reference does not affect the fitted shell coefficients
as the energy offset is described by $b_A$.
Within this shell representation, $c_{\mathrm{res},s}$ measures
the interaction strength remaining after subtracting the MACE backbone at $k_s$.
The corresponding coefficient $c_{\mathrm{FNO},s}$ measures the strength
supplied by the FNO correction.

Figure~\ref{fig:synthetic-shell-recovery} shows that the residual coefficients are
substantial at small wavevectors and generally decrease towards high $k$.
To examine the extent to which the reference Yukawa coefficient can be captured by
the MACE backbone alone, we introduce $\alpha_s=1-c_{\mathrm{res},s}/c_{\mathrm Y,s}$.
At the lowest shell $k_1=\ensuremath{0.196\,3\,\text{\AA}^{-1}}$,
the local backbone only recovers $\alpha_1=\ensuremath{14.2\,\%}$
of the reference for $\kappa=\ensuremath{0.05\,\text{\AA}^{-1}}$,
leaving the interactions largely unaccounted for at this small wavevector.
On the other hand, the backbone absorbs \ensuremath{93.8\,\%} of the reference
for the high-$k$ tail above the \num{15} lowest shells.
Figure~\ref{fig:synthetic-shell-recovery} hence provides direct evidence of
a pronounced wavevector dependence of the local MACE backbone, i.e.,
the responses at short wavelengths are better described than those at longer wavelengths.
This is further supported by the $\kappa$ dependence of the residual,
which shows that the local backbone is more adequate as the screening strengthens
($\alpha_1=\ensuremath{21.5\,\%\text{ and }33.1\,\%}$
for $\kappa=\ensuremath{0.2\,\text{\AA}^{-1}\text{ and }0.5\,\text{\AA}^{-1}}$, respectively).

The FNO coefficients closely follow the overall magnitude and wavevector dependence
of the residual across the resolved shells, irrespective of whether the field operator
is linear or nonlinear [cf.\ Fig.~\ref{fig:synthetic-shell-recovery}(d--f)].
By reproducing the departure of the residual from the Yukawa reference,
both FNO variants complement the uneven local description at the varying wavelengths
without explicitly prescribing the Yukawa kernel or supervising the shell coefficients.
Over the three lowest shells, the nonlinear FNO gives relative RMS deviations
from the residual coefficients of
\ensuremath{2.8\,\%\text{, }1.2\,\%\text{, and }12.0\,\%}
for $\kappa=\ensuremath{0.05\,\text{\AA}^{-1}\text{, }0.2\,\text{\AA}^{-1}\text{, and }0.5\,\text{\AA}^{-1}}$, respectively.
The corresponding deviations for \ensuremath{\mathrm{FNO}^{(\mathrm{l})}} are
\ensuremath{2.7\,\%\text{, }1.7\,\%\text{, and }17.9\,\%}.
The largest deviation occurs at $\kappa=\ensuremath{0.5\,\text{\AA}^{-1}}$,
where the residual exhibits a pronounced non-monotonic shell dependence
as shown in Fig.~\ref{fig:synthetic-shell-recovery}(c).

\subsection{Spectral response of latent fields}

\begin{table}[t]
\centering
\setlength{\tabcolsep}{3pt}
\caption{Reference and latent-field spectral exponents and log--log fit
scores $R^2$ for the synthetic benchmark, fitted over the three shells
at $\widetilde k=1,\sqrt{2},\sqrt{3}$.
Learned $p$ and $R^2$ are averaged over eight validation configurations per
seed, then reported as means and sample standard deviations over five seeds
of \ensuremath{\mathrm{FNO}^{(\mathrm{l})}} and FNO on a frozen seed-17 backbone at a given $\kappa$.
The diagnostics are evaluated at the validation-selected checkpoints.}
\label{tab:synthetic-spectral}
\begin{ruledtabular}
\begin{tabular}{D{.}{.}{1.2}rrlrr}
\multicolumn{1}{c}{$\kappa$ (\unit{\per\angstrom})} & $p_{\mathrm{ref}}$ & $R^2_{\mathrm{ref}}$ & & $p$ & $R^2$\\
\colrule
0.05 & \num{1.97} & \num{1.000} & \ensuremath{\mathrm{FNO}^{(\mathrm{l})}} & \num{2.10 +- 0.07} & $\num{1.000}\pm\num{0.000}$\\
 & & & \ensuremath{\mathrm{FNO}} & \num{2.07 +- 0.05} & $\num{0.999}\pm\num{0.001}$\\
0.2 & \num{1.28} & \num{0.997} & \ensuremath{\mathrm{FNO}^{(\mathrm{l})}} & \num{1.46 +- 0.06} & $\num{0.996}\pm\num{0.003}$\\
 & & & \ensuremath{\mathrm{FNO}} & \num{1.42 +- 0.04} & $\num{0.994}\pm\num{0.001}$\\
0.5 & \num{0.47} & \num{0.986} & \ensuremath{\mathrm{FNO}^{(\mathrm{l})}} & \num{0.60 +- 0.12} & $\num{0.965}\pm\num{0.021}$\\
 & & & \ensuremath{\mathrm{FNO}} & \num{0.66 +- 0.04} & $\num{0.961}\pm\num{0.009}$\\
\end{tabular}
\end{ruledtabular}
\end{table}

We now probe the latent-field energy curvature
using the spectral diagnostic described in Sec.~\ref{sec:spectral-diagnostic-method}.
The fit is performed over the three lowest shells, the wavevector magnitudes
of which span \ensuremath{0.196\text{--}0.340\,\text{\AA}^{-1}}.
Table~\ref{tab:synthetic-spectral} compares the resulting exponent $p$ with
$p_{\mathrm{ref}}$, the reference exponent obtained as the negative slope
of a log--log fit of the analytical smeared Yukawa kernel in
Eq.~\eqref{eq:synthetic-kernel} over the same shells.
The screening-induced modification in the wavevector dependence of the kernel
is clearly visible through $p_\mathrm{ref}$.
Over the sampled low-$k$ range, $p_\mathrm{ref}$ is close to the unscreened Coulomb
value of two at $\kappa=\ensuremath{0.05\,\text{\AA}^{-1}}$
and falls below \num{0.5} for $\kappa=\ensuremath{0.5\,\text{\AA}^{-1}}$.
The trend is in agreement with the pointwise analytical exponent given in Eq.~\eqref{eq:p_Vg}
although $p_\mathrm{ref}$ here represents a fitted slope over a finite window.

The spectral exponents largely agree with $p_{\mathrm{ref}}$ for all three $\kappa$ values,
albeit being systematically larger by 0.1--0.2.
Exact equality is admittedly not warranted by the present architecture.
Indeed, Fig.~\ref{fig:synthetic-shell-recovery} shows that the local MACE backbone
already describes a share of the reference Yukawa interaction even at the lowest shells,
thus modifying the learned target.
More fundamentally, the perturbation in the spectral diagnostic applies directly
to the deposited latent fields.
It does not propagate through the source head, which also contributes to the full residual
energy response by changing the latent fields as the atomic configuration varies.
To put it into perspective, we extract $p_\mathrm{res}$ from the $k$-dependence of the
residual coefficients $c_\mathrm{res,s}$ shown in Fig.~\ref{fig:synthetic-shell-recovery}.
We perform a log--log fit of $g(k_s)c_\mathrm{res,s}$ over the three lowest shells,
including the Gaussian factor $g(k_s)$ for consistency with $p_\mathrm{ref}$.
This leads to $p_\mathrm{res}=\num{2.70}$ and \num{2.28} at $\kappa=\num{0.05}$
and \ensuremath{0.2\,\text{\AA}^{-1}}, respectively.
Both values exceed $p_{\mathrm{ref}}$ as the backbone absorbs an increasing share of the
reference from the first to the third shell.
The learned spectral $p$ lies between $p_{\mathrm{ref}}$ and $p_{\mathrm{res}}$.
The log--log fit at $\kappa=\ensuremath{0.5\,\text{\AA}^{-1}}$ is precluded by the negative
third-shell residual coefficient.
In contrast to the spectral exponent $p$, which characterizes the FNO energy response
to latent-field perturbations, $p_\mathrm{res}$ describes the wavevector
dependence of the residual expressed through physical-charge shell features.
The two analyses provide complementary descriptions,
and accurate residual recovery does not require the two to coincide.
For the present benchmark, the spectral diagnostic provides a numerically more
robust characterization across the three screening strengths.

Despite a positive offset, the spectral exponent $p$ decreases
systematically as screening strengthens in both linear and nonlinear FNOs,
closely following the trend in $p_\mathrm{ref}$ as shown in Table~\ref{tab:synthetic-spectral}.
The smaller $p$ is consistent with the flattening of the analytical
Yukawa kernel over the sampled long-wavelength window at larger $\kappa$.
As both FNO variants yield comparable exponents, we conclude that a linear
operator alone is able to track the screening dependency in this
homogeneous-screening benchmark.

\subsection{Energy--force accuracy}
\begin{figure*}[t]
\includegraphics[width=\textwidth]{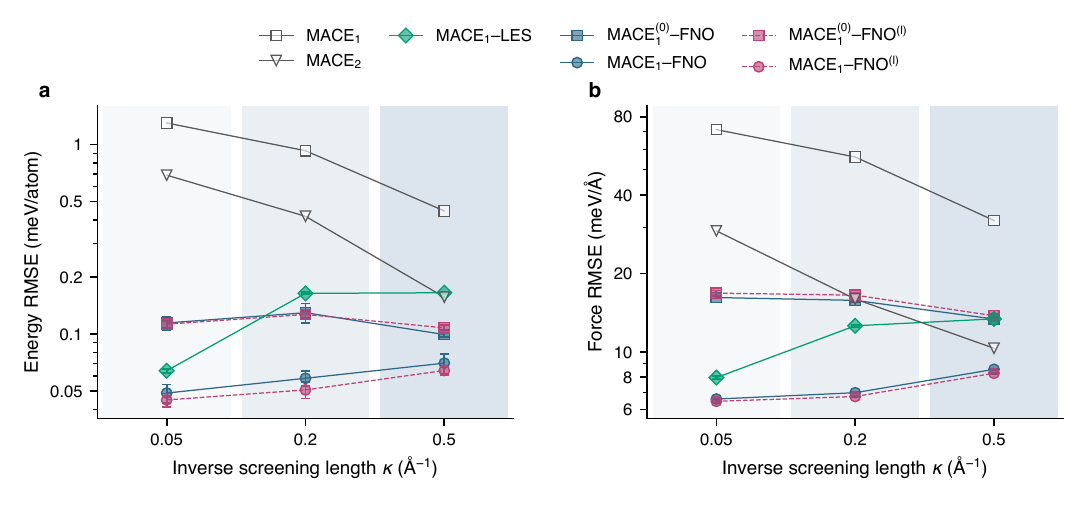}
\caption{Accuracy of energy and forces for the synthetic benchmark.
Energy and force RMSEs are shown in panels (a) and (b), respectively,
at three inverse screening lengths.
Both RMSE axes are logarithmic.
Darker backgrounds indicate stronger screening as $\kappa$ increases.
Error bars denote sample standard deviations over five seeds,
whereas \ensuremath{\mathrm{MACE}_{1}} and \ensuremath{\mathrm{MACE}_{2}} are single-seed baselines.
Lines are guides to the eye.}
\label{fig:synthetic-results}
\end{figure*}

\begin{table*}[t]
\caption{Energy and force RMSEs on the synthetic benchmark for the three
inverse screening lengths $\kappa$.
Spreads are sample standard deviations over five seeds,
whereas \ensuremath{\mathrm{MACE}_{1}} and \ensuremath{\mathrm{MACE}_{2}} are single-seed baselines.
Bold means denote the lowest RMSE or a mean $\pm$ standard-deviation interval
overlapping that of the lowest-RMSE model.}
\label{tab:synthetic-results}
\small
\setlength{\tabcolsep}{4pt}

\begin{ruledtabular}
\begin{tabular}{lrr@{\hspace{1.8em}}rr@{\hspace{1.8em}}rr}
 & \multicolumn{2}{c}{$\kappa=\ensuremath{0.05\,\text{\AA}^{-1}}$}
 & \multicolumn{2}{c}{$\kappa=\ensuremath{0.2\,\text{\AA}^{-1}}$}
 & \multicolumn{2}{c}{$\kappa=\ensuremath{0.5\,\text{\AA}^{-1}}$}\\
\cline{2-3}\cline{4-5}\cline{6-7}
\noalign{\vskip 3pt}
Model & \shortstack{Energy\\(\unit{\milli\electronvolt\per\atom})} & \shortstack{Force\\(\unit{\milli\electronvolt\per\angstrom})}
 & \shortstack{Energy\\(\unit{\milli\electronvolt\per\atom})} & \shortstack{Force\\(\unit{\milli\electronvolt\per\angstrom})}
 & \shortstack{Energy\\(\unit{\milli\electronvolt\per\atom})} & \shortstack{Force\\(\unit{\milli\electronvolt\per\angstrom})}\\
\colrule
\ensuremath{\mathrm{MACE}_{1}} & \num{1.30} & \num{71.37} & \num{0.93} & \num{56.03} & \num{0.45} & \num{31.95}\\
\ensuremath{\mathrm{MACE}_{2}} & \num{0.69}\footnotemark[1] & \num{29.16}\footnotemark[1] & \num{0.42} & \num{15.95} & \num{0.16} & \num{10.32}\\[0.6em]
\ensuremath{\mathrm{MACE}_{1}}--\ensuremath{\mathrm{LES}} &   $\num{0.06}\pm\num{0.00}$ & \num{7.96 +- 0.11} &   $\num{0.16}\pm\num{0.00}$ & \num{12.59 +- 0.10} &   $\num{0.17}\pm\num{0.00}$ & \num{13.38 +- 0.09}\\[0.6em]
\mbox{\ensuremath{\mathrm{MACE}_{1}^{(0)}}--\ensuremath{\mathrm{FNO}^{(\mathrm{l})}}} &   $\num{0.11}\pm\num{0.00}$ & \num{16.81 +- 0.40} &   $\num{0.13}\pm\num{0.00}$ & \num{16.51 +- 0.26} &   $\num{0.11}\pm\num{0.00}$ & \num{13.76 +- 0.20}\\
\mbox{\ensuremath{\mathrm{MACE}_{1}^{(0)}}--\ensuremath{\mathrm{FNO}}} &   $\num{0.11}\pm\num{0.01}$ & \num{16.16 +- 0.15} &   $\num{0.13}\pm\num{0.02}$ & \num{15.77 +- 0.23} &   $\num{0.10}\pm\num{0.00}$ & \num{13.36 +- 0.16}\\
\ensuremath{\mathrm{MACE}_{1}}--\ensuremath{\mathrm{FNO}^{(\mathrm{l})}} &   $\text{\bfseries\num[mode=text,detect-weight=true]{0.04}}  \pm\num{0.00}$ &   $\text{\bfseries\num[mode=text,detect-weight=true]{6.45}}  \pm\num{0.06}$ &   $\text{\bfseries\num[mode=text,detect-weight=true]{0.05}}  \pm\num{0.01}$ &   $\text{\bfseries\num[mode=text,detect-weight=true]{6.74}}  \pm\num{0.03}$ &   $\text{\bfseries\num[mode=text,detect-weight=true]{0.06}}  \pm\num{0.00}$ &   $\text{\bfseries\num[mode=text,detect-weight=true]{8.28}}  \pm\num{0.04}$\\
\ensuremath{\mathrm{MACE}_{1}}--\ensuremath{\mathrm{FNO}} &   $\text{\bfseries\num[mode=text,detect-weight=true]{0.05}}  \pm\num{0.01}$ & \num{6.59 +- 0.01} &   $\text{\bfseries\num[mode=text,detect-weight=true]{0.06}}  \pm\num{0.01}$ & \num{6.97 +- 0.03} &   $\text{\bfseries\num[mode=text,detect-weight=true]{0.07}}  \pm\num{0.01}$ & \num{8.56 +- 0.04}\\
\end{tabular}
\end{ruledtabular}
\footnotetext[1]{\ensuremath{\mathrm{MACE}_{2}} at $\kappa=\ensuremath{0.05\,\text{\AA}^{-1}}$ uses the
stage-one checkpoint, which yields lower validation energy and force RMSEs
than the checkpoint during stage two.
Other local MACE and LES results use stage-two checkpoints.}
\end{table*}

Figure~\ref{fig:synthetic-results} shows the energy and force RMSEs
obtained with the one-interaction \ensuremath{\mathrm{MACE}_{1}} backbone, the frozen \mbox{\ensuremath{\mathrm{MACE}_{1}^{(0)}}--\ensuremath{\mathrm{FNO}}},
and the joint \ensuremath{\mathrm{MACE}_{1}}--\ensuremath{\mathrm{FNO}}, at the three inverse screening lengths.
The figure and Table~\ref{tab:synthetic-results} also include the corresponding linear-FNO
models with the same seed set, training budgets, and checkpoint-selection rule.
We further include \ensuremath{\mathrm{MACE}_{1}}--\ensuremath{\mathrm{LES}} and a two-interaction \ensuremath{\mathrm{MACE}_{2}} backbone
for comparison.
The local MACE backbones use a nonresidual first interaction block
(\texttt{RealAgnostic\allowbreak InteractionBlock}), which we also adopt for
MACE--\ensuremath{\mathrm{LES}}.
The local MACE backbones and \ensuremath{\mathrm{MACE}_{1}}--\ensuremath{\mathrm{LES}} are trained in two stages,
with an increased relative weight of the energy error during the second stage.
MACE--\ensuremath{\mathrm{LES}} models use the default values for the Gaussian
smoothing length ($\sigma=\ensuremath{1.0\,\text{\AA}}$) and the reciprocal resolution
parameter ($dl=\ensuremath{2.0\,\text{\AA}}$).
In accord with the FNO models, we train the LES models independently with five random seeds,
which sample variability in both the local backbone and the latent-charge prediction.
\ensuremath{\mathrm{MACE}_{2}} uses the same cutoff and hidden representation as \ensuremath{\mathrm{MACE}_{1}}.
The numerical RMSE values are reported in Table~\ref{tab:synthetic-results}.

With the frozen \ensuremath{\mathrm{MACE}_{1}} backbone, nonlinear FNO lowers the energy RMSE by
\ensuremath{78\text{--}91\,\%}
and the force RMSE by
\ensuremath{58\text{--}77\,\%}
relative to the local baseline.
The energy and force accuracy is largely maintained with linear FNO.
By allowing the backbone to adapt in joint optimization, both FNO variants achieve
lower energy and force RMSEs than \ensuremath{\mathrm{MACE}_{1}}--\ensuremath{\mathrm{LES}} and \ensuremath{\mathrm{MACE}_{2}}
across all three screening strengths.

While a second interaction layer in \ensuremath{\mathrm{MACE}_{2}} extends the receptive radius
from \num{4.5} to \ensuremath{9\,\text{\AA}},
it remains less accurate than jointly trained \ensuremath{\mathrm{MACE}_{1}}--\ensuremath{\mathrm{FNO}} across all three
screening lengths.
Likewise, \ensuremath{\mathrm{MACE}_{1}}--\ensuremath{\mathrm{LES}} outperforms \ensuremath{\mathrm{MACE}_{2}} at $\kappa=\ensuremath{0.05\,\text{\AA}^{-1}\text{ and }0.2\,\text{\AA}^{-1}}$
although its force RMSE is higher at $\kappa=\ensuremath{0.5\,\text{\AA}^{-1}}$.
A global correction thus provides a clear advantage over a deeper local backbone,
particularly in the weakly screened regime with pronounced long-range interactions.
Compared to \ensuremath{\mathrm{MACE}_{2}}, LES exhibits a contrasting trend with screening strength.
It approaches the accuracy of the jointly trained FNO models at $\kappa=\ensuremath{0.05\,\text{\AA}^{-1}}$,
and becomes less reliable as the reference interaction departs further from the unscreened
Coulomb form.

The fact that all the FNO models maintain high accuracy across varying screening strengths
clearly supports the flexibility of FNO augmentation in learning the propagation kernel.
In line with the residual recovery analysis and spectral diagnostics,
we find that a linear FNO suffices for the present system,
and introducing nonlinearity into the field operator provides no systematic improvement in
energy--force accuracy.

\subsection{Charge--charge correlations from molecular dynamics}
\label{sec:S_zz}

\begin{figure*}[!t]
\centering
\includegraphics[width=\textwidth]{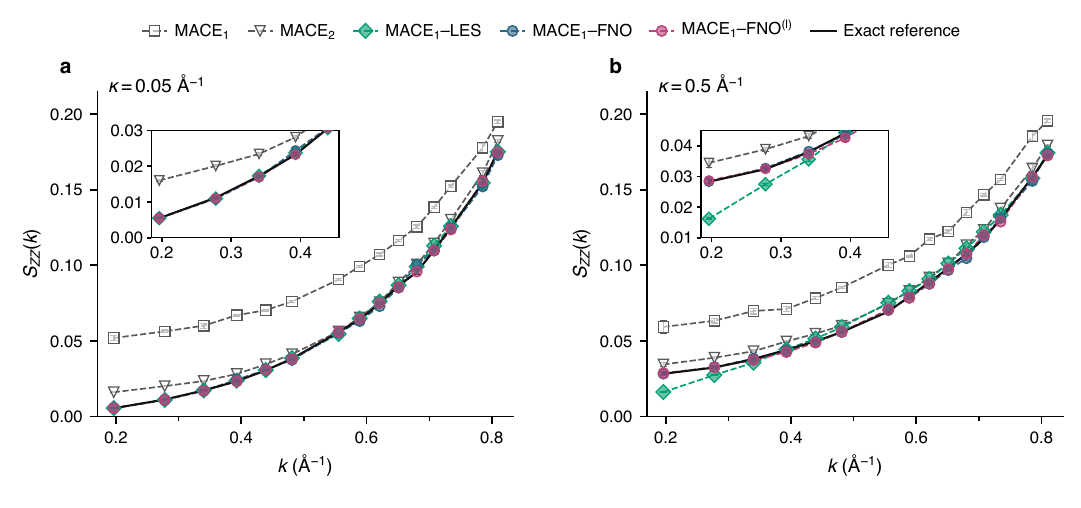}
\caption{Charge--charge structure factors of the synthetic ionic fluid at
\ensuremath{2\,800\,\mathrm{K}} in a cubic cell ($L_a=\ensuremath{32\,\text{\AA}}$).
(a) $\kappa=\ensuremath{0.05\,\text{\AA}^{-1}}$;
(b) $\kappa=\ensuremath{0.5\,\text{\AA}^{-1}}$.
Insets show a zoomed-in view of the lowest shells.
The exact reference denotes MD with the analytical interaction potential.
Error bars indicate one standard error of the mean
for MLIPs trained with a single seed (ten \ensuremath{100\,\mathrm{ps}} blocks of a \ensuremath{1\,\mathrm{ns}} trajectory)
and the reference (ten \ensuremath{100\,\mathrm{ps}} replicas),
and are mostly smaller than the symbols.}
\label{fig:md-structure}
\end{figure*}

To understand how the learned nonlocal response carries over to the
equilibrium structure, we consider the static charge--charge structure
factor $S_{ZZ}(k)$~\cite{copestake1982chargeordering,hansen2013simpleliquids}
in an ionic fluid.
For nonzero reciprocal wavevectors, we calculate
\begin{equation}
S_{ZZ}(k)=
\frac{\left\langle\left|\sum_i q_i
e^{-i\mathbf k\cdot\mathbf R_i(t)}\right|^2\right\rangle_{t,|\mathbf k|=k}}
{\sum_i q_i^2},
\label{eq:md-charge-structure-factor}
\end{equation}
where $q_i$ is the fixed reference charge of ion $i$ and $\mathbf R_i(t)$
is its position at time $t$.
The brackets refer to an average over production MD frames and over wavevectors
of equal magnitude.
The charge--charge structure factor therefore measures the charge-density
fluctuations at a given wavelength.
With a pure long-range Coulomb interaction, $S_{ZZ}$ vanishes as $k^2$ for
$k\to0$, whereas it stays finite at long wavelength
when the system is subject to a screened interaction~\cite{grasselli2026electrostatics}.
The long-wavelength behavior of $S_{ZZ}$ therefore sheds light on how well a model
reproduces the collective charge correlations associated with long-range interactions.

To this end, we perform MD simulations in the canonical ensemble (NVT) at \ensuremath{2\,800\,\mathrm{K}}
using the same cubic cell with $L_a=\ensuremath{32\,\text{\AA}}$.
At this temperature, the synthetic ionic system is a diffusive fluid as evidenced by
self-diffusion coefficients of about \ensuremath{4.2\times 10^{-5}\,\mathrm{cm}^{2}\,\mathrm{s}^{-1}}
for both ion species using the reference potential.
The self-diffusion coefficients are largely invariant to the screening parameter $\kappa$.
We use a Langevin thermostat and a time step of \ensuremath{2\,\mathrm{fs}}.
For MD simulations with the synthetic ionic fluid, the FNO energy is averaged
over eight mesh origins to reduce the spurious net force from the mesh.
The models are nonetheless trained on a single mesh.
For the MLIPs, $S_{ZZ}$ is sampled over a trajectory of \ensuremath{1\,\mathrm{ns}} (with an interval
of \ensuremath{0.5\,\mathrm{ps}}) after an equilibration of \ensuremath{2\,\mathrm{ps}}.
We estimate the standard error from ten consecutive \ensuremath{100\,\mathrm{ps}} segments of the trajectory.
The reference MD using the analytical potential is sampled by ten independent
\ensuremath{100\,\mathrm{ps}} replicas, with its standard errors taken over these replicas.
None of the \ensuremath{\mathrm{MACE}_{1}}--\ensuremath{\mathrm{FNO}} trajectories shows an appreciable change in potential energy,
which is within about \ensuremath{1\,\mathrm{meV}\,\mathrm{atom}^{-1}} between the first and second halves
of the \ensuremath{1\,\mathrm{ns}} runs.
The mean temperature lies within \ensuremath{10\,\mathrm{K}} of the target.
To verify energy conservation, we additionally run \ensuremath{100\,\mathrm{ps}} NVE trajectories
with a time step of \ensuremath{2\,\mathrm{fs}} from the equilibrated states.
The total energy of \ensuremath{\mathrm{MACE}_{1}}--\ensuremath{\mathrm{FNO}} drifts by at most \ensuremath{0.04\,\mathrm{meV}\,\mathrm{atom}^{-1}}
over \ensuremath{100\,\mathrm{ps}}, whereas \ensuremath{\mathrm{MACE}_{1}} shows no measurable drift (below \ensuremath{0.001\,\mathrm{meV}\,\mathrm{atom}^{-1}}).
This drift is comparable to the total-energy fluctuation of both models
(about \ensuremath{0.04\,\mathrm{meV}\,\mathrm{atom}^{-1}}) and below \ensuremath{1\,\%} of the kinetic-energy fluctuation.
Distinct from the static benchmark checkpoints,
all MLIPs for MD simulations are trained with a single seed on a separate dataset.
Specifically, we generate \num{40} Langevin trajectories at \ensuremath{2\,800\,\mathrm{K}},
and sample every \ensuremath{1\,\mathrm{ps}} after a \ensuremath{10\,\mathrm{ps}} burn-in.
These configurations are then split into \num{660} training, \num{240} validation, and \num{300}
test configurations.

Figure~\ref{fig:md-structure} shows $S_{ZZ}(k)$ on the \num{15} lowest shells
obtained with the short-range and long-range MLIPs
in comparison to the analytical reference potential.
The reference MD shows the expected dependence on screening:
$S_{ZZ}=\num{0.0054}$ at the lowest shell for $\kappa=\ensuremath{0.05\,\text{\AA}^{-1}}$, and
it increases about fivefold to \num{0.028} for $\kappa=\ensuremath{0.5\,\text{\AA}^{-1}}$
as the shorter-ranged Yukawa interaction penalizes long-wavelength charge fluctuations less.
Lacking the long-range interaction, the local \ensuremath{\mathrm{MACE}_{1}} strongly overestimates
$S_{ZZ}$ by factors of \num{9.5} and \num{2.1} at the lowest shell for $\kappa=\num{0.05}$
and \ensuremath{0.5\,\text{\AA}^{-1}}, respectively, compared to the reference.
The excess decays only gradually with $k$.
At $k=\ensuremath{0.81\,\text{\AA}^{-1}}$, or a wavelength of about \ensuremath{8\,\text{\AA}},
$S_{ZZ}$ is still \ensuremath{11\text{--}13\,\%} above the reference.
The deeper local \ensuremath{\mathrm{MACE}_{2}} extends, but does not complete, the
long-range response: $S_{ZZ}$ at the lowest shell is still
overestimated by a factor of \num{2.9} for $\kappa=\ensuremath{0.05\,\text{\AA}^{-1}}$
and \num{1.2} for $\kappa=\ensuremath{0.5\,\text{\AA}^{-1}}$.

Thanks to the prescribed Coulomb kernel, the long-range \ensuremath{\mathrm{MACE}_{1}}--\ensuremath{\mathrm{LES}}
reproduces the reference within \ensuremath{4\,\%} on all
shells for $\kappa=\ensuremath{0.05\,\text{\AA}^{-1}}$.
For $\kappa=\ensuremath{0.5\,\text{\AA}^{-1}}$, however, the long-wavelength charge
fluctuations are suppressed too strongly,
leading to an underestimated $S_{ZZ}$ at the lowest shell
reaching only \ensuremath{57\,\%} of the reference value.
By contrast, linear and nonlinear \ensuremath{\mathrm{MACE}_{1}}--\ensuremath{\mathrm{FNO}} follow the reference
within \ensuremath{2\,\%\text{ and }3\,\%}, respectively, on the three lowest shells
at both screening strengths.
Since $S_{ZZ}$ is not a training target,
such an agreement indicates that the learned FNO infers the reference
interaction successfully from energy and force labels alone.
In particular, whereas LES adapts only its charges and local energy
to a fixed Coulomb kernel, the FNO also learns the propagation itself.

\section{Inhomogeneous screening and nonlinear response}
\label{sec:heterogeneous-results}

With a spatially uniform $\kappa$ across atomic configurations,
the Yukawa interaction can be represented by a configuration-independent
\emph{linear} propagation kernel.
We show this by writing the screened Poisson equation
\begin{equation}
\label{eq:poisson}
(-\nabla^2 + \kappa^2)\varphi(\mathbf r) = 4\pi \rho_\sigma(\mathbf r),
\end{equation}
where $\rho_\sigma$ is the Gaussian-smeared physical charge density,
and $a_\kappa C_\mathrm{Coul}$ is omitted for simplicity.
The reference potential satisfying Eq.~\eqref{eq:poisson}
can be expressed in terms of the propagation kernel $G_\kappa$
\begin{equation}
\varphi(\mathbf r) = \int_\Omega G_{\kappa}(\mathbf r - \mathbf r') \rho_\sigma(\mathbf r')\,\mathrm d\mathbf r',
\end{equation}
where $\Omega$ denotes the periodic simulation cell.
As $G_{\kappa}$ depends only on the separation, it is diagonal in Fourier space
\begin{equation}
\widehat{G}_{\kappa}(\mathbf k) = \frac{4\pi}{k^2 + \kappa^2}.
\end{equation}
At fixed screening, the potential thus depends \emph{linearly} on $\rho_\sigma$,
which further leads to the quadratic energy dependence
$\sum_{\mathbf k}\widehat{G}_\kappa(\mathbf k)|\widehat{\rho}_\sigma(\mathbf k)|^2$.
This explains the success of linear \ensuremath{\mathrm{FNO}^{(\mathrm{l})}} for the homogeneous benchmark
in Sec.~\ref{sec:synthetic-results},
and we expect an equally good description of the homogeneous reference
by other sufficiently expressive, learned but configuration-independent
kernels, such as those of SOG-Net~\cite{ji2025sognet} and
RSNN~\cite{guo2026reciprocal}.

In many systems, screening is neither uniform nor independent of the atomic configuration.
In electrolytes near charged surfaces
or dielectric interfaces~\cite{buyukdagli2016beyond,aluru2023fluids},
spatial variations in ion concentration can produce a screening response
that depends on the screening environment.
Importantly, rearranging the screening medium can change the effective
interactions between two charges, even when their local environments remain intact.
Capturing this effect requires a configuration-dependent propagation that takes into
account the screening medium beyond the local environment of both charges.

Here, we extend the synthetic system of Sec.~\ref{sec:synthetic-results}
by incorporating a charge-neutral screening species $S$ into the NaCl-like ionic compound.
Inspired by Debye--H\"{u}ckel theory~\cite{debye1923electrolytes}, in which the squared inverse screening length
is proportional to ionic strength, we prescribe a screening law
in which the local concentration of $S$ determines the inverse screening length,
\begin{equation}
\kappa^2(\mathbf r)=\kappa_0^2+\kappa_1^2\,c_S(\mathbf r),\qquad
c_S(\mathbf r)=\frac{1}{\bar n_S}\sum_{m\in S}w_\ell(\mathbf r-\mathbf R_m),
\label{eq:het-kappa}
\end{equation}
where $c_S(\mathbf r)$ is the smoothed concentration relative to its cell average,
$w_\ell$ is a normalized Gaussian of width $\ell=\ensuremath{3\,\text{\AA}}$ summed
over periodic images, and the number density $\bar n_S=N_S/V$ of $S$ ensures
$\langle c_S(\mathbf r) \rangle=1$ over the cell.
We choose $\kappa_0=\ensuremath{0.05\,\text{\AA}^{-1}}$ to keep the $S$-depleted regions
Coulombic, and $\kappa_1=\ensuremath{0.295\,\text{\AA}^{-1}}$ so that \ensuremath{95\,\%}
of the volume of the training configurations has $\kappa\le\ensuremath{0.5\,\text{\AA}^{-1}}$.
The local screening thus spans the range covered separately by the three
homogeneous benchmarks of Sec.~\ref{sec:synthetic-results}.
All the other parameters, such as the unit cell, the particle density,
the ionic charges and their Gaussian smearing,
and the species-independent LJ interaction, follow exactly the ones used in
Sec.~\ref{sec:synthetic-results}.

With $A=-\nabla^2+\kappa^2(\mathbf r)$, the screened electrostatic energy reads
\begin{equation}
E_{\mathrm{scr}}
=\frac{C_{\mathrm{Coul}}}{2}\int_\Omega\rho_\sigma(\mathbf r)\,\varphi(\mathbf r)\,\mathrm d\mathbf r
=2\pi C_{\mathrm{Coul}}\left\langle\rho_\sigma,A^{-1}\rho_\sigma\right\rangle.
\label{eq:het-energy}
\end{equation}
The reference energy is then
$E_{\mathrm{ref}}=E_{\mathrm{LJ}}+a_{\mathrm h}\,(E_{\mathrm{scr}}-E_{\mathrm{self}}^{(0)})$.
The constant $E_{\mathrm{self}}^{(0)}$ refers to the self-energy of the smeared
ionic charges in an $S$-free medium ($\kappa=\kappa_0$).
Subtracting this constant ensures that isolated atoms keep zero energy, consistent with the
isolated-atom reference energies used in training
(as for $E_{\mathrm{self}}$ in Eq.~\eqref{eq:synthetic-yukawa-energy}).
The medium-dependent part of the self-energy stays in $E_{\mathrm{scr}}$,
lowering the energy of an ion as the screening medium accumulates around it.
As in Sec.~\ref{sec:synthetic-results}, the factor $a_{\mathrm h}=\num{2.039}$
fixes the standard deviation of the nonlocal energy per atom across the training
configurations at \ensuremath{5\,\mathrm{meV}\,\mathrm{atom}^{-1}}.

For a fixed arrangement of $S$, $\varphi$ remains linear in $\rho_\sigma$, and
$E_{\mathrm{scr}}$ is quadratic in the charges, as for the homogeneous Yukawa
interaction.
In reciprocal space, however, the screened Poisson equation becomes
\begin{equation}
(k^2+\kappa_0^2)\,\widehat\varphi(\mathbf k)
+\kappa_1^2\sum_{\mathbf q}\widehat c_S(\mathbf k-\mathbf q)\,\widehat\varphi(\mathbf q)
=4\pi\widehat\rho_\sigma(\mathbf k),
\label{eq:het-poisson-k}
\end{equation}
in which the Fourier components $\widehat c_S$ of the concentration field couple
different wavevectors.
Unlike $\widehat G_\kappa(\mathbf k)$ above, no diagonal kernel can thus
describe the response, and the operator $A^{-1}$ changes whenever the screening
medium rearranges.
The inhomogeneity hence enters through the propagation of the response, not
through a nonlinear dependence on the charges.
Note that this construction is phenomenological (the $S$ particles prescribe the
screening without carrying explicit charge), and it is not meant as a
thermodynamically consistent coarse-graining of a specific electrolyte.

We solve Eq.~\eqref{eq:het-poisson-k} numerically in the plane-wave basis of
Eq.~\eqref{eq:synthetic-yukawa-energy} ($|\mathbf k|\le\ensuremath{3.5\,\text{\AA}^{-1}}$),
with an iterative solver converged to near machine precision,
and obtain the forces on both the ions and the $S$ particles from the converged potential.
For a uniform medium, the solution reproduces the Yukawa energies and forces of
Sec.~\ref{sec:synthetic-results} at $\kappa^2=\kappa_0^2+\kappa_1^2$,
and finite differences confirm the forces in strongly inhomogeneous media
(cf.\ Appendix~\ref{app:heterogeneous-solver}).

To generate configurations with inhomogeneous screening,
we start with \num{80} frames obtained from the \num{40} MD trajectories of the pure ionic systems
equilibrated at \ensuremath{2\,800\,\mathrm{K}} and $\kappa=\ensuremath{0.05\,\text{\AA}^{-1}}$
as described in Sec.~\ref{sec:S_zz}.
Within every frame, we randomly relabel \num{286} of the \num{1146} ions
as the charge-neutral species $S$.
We then repeatedly propose transferring an $S$ label to a non-$S$ site,
both chosen at random.
Acceptance of the Monte Carlo (MC) move is subject to the Metropolis criterion based
on the energy change in an external potential acting exclusively on $S$.
Here we use a randomized long-wavelength Fourier potential~\cite{sammuller2023neural}
\begin{equation}
V_{\mathrm{ext}}(\mathbf r)=V_0\,k_{\mathrm B}T\sum_{\alpha=x,y,z}a_\alpha\cos\!\left(\frac{2\pi r_\alpha}{L_a}+\theta_\alpha\right),
\label{eq:het-bias}
\end{equation}
where $r_\alpha$ denotes the Cartesian coordinate along axis $\alpha$, and $T=\ensuremath{2\,800\,\mathrm{K}}$.
The Gaussian random weights $a_\alpha$ are normalized so that the sinusoidal sum has unit spatial variance,
and the phases $\theta_\alpha$ are uniform on $[0,2\pi)$.
The amplitude $V_0$, drawn uniformly between zero and four for every trajectory,
controls the segregation of $S$ and thus the screening contrast.
Upon acceptance, the original $S$ site recovers its template ionic species to maintain
the initial charge ordering as much as possible.
As this procedure generally leaves an excess of one ion species over the other,
we restore charge neutrality by randomly relabeling ions of the more abundant species
after all \num{60} MC sweeps are performed.

For each of the \num{80} relabeled configurations,
we run Langevin MD at \ensuremath{2\,800\,\mathrm{K}}
and sample \num{15} frames every \ensuremath{0.5\,\mathrm{ps}} after a \ensuremath{6\,\mathrm{ps}} burn-in.
During MD, all particles interact through the LJ potential.
The ions additionally experience the homogeneous Yukawa interaction of
Eq.~\eqref{eq:synthetic-yukawa-energy} at $\kappa=\ensuremath{0.05\,\text{\AA}^{-1}}$.
The external potential $V_\mathrm{ext}$ acts only on $S$ to maintain its spatial pattern.
For labeling, we recompute the reference energies and forces using the LJ interaction
and the inhomogeneous screening model.
Splitting by whole trajectories as in Sec.~\ref{sec:synthetic-results},
we obtain \numlist{660;240;300} configurations for training, validation, and test, respectively.
Over all configurations, the 5th and 95th percentiles of the spatial
distribution of $\kappa$ are \ensuremath{0.066\,\text{\AA}^{-1}\text{ and }0.498\,\text{\AA}^{-1}}, respectively.
We quantify the inhomogeneity of a configuration by the spatial standard
deviation of $\kappa(\mathbf r)$ over the cell, which, averaged over the frames
of a trajectory, ranges from \num{0.039} to \ensuremath{0.167\,\text{\AA}^{-1}} across
the \num{80} trajectories.
This dataset thus spans weakly to strongly inhomogeneous screening environments.

\subsection{Energy--force accuracy and screening inhomogeneity}
\label{sec:heterogeneous-accuracy}

\begin{table}[t]
\centering
\caption{Energy and force RMSEs on the \num{300} test configurations of the
inhomogeneous-screening benchmark.
Spreads are sample standard deviations over five seeds, whereas \ensuremath{\mathrm{MACE}_{1}}
and \ensuremath{\mathrm{MACE}_{2}} are single-seed baselines.
The local CACE backbone uses a cutoff of \ensuremath{4.5\,\text{\AA}} without
message passing.
Bold values mark the lowest RMSE.}
\label{tab:inhomogeneous-accuracy}
\small
\setlength{\tabcolsep}{4pt}

\begin{ruledtabular}
\begin{tabular}{lrr}
\noalign{\vskip 3pt}
Model & \shortstack{Energy\\(\unit{\milli\electronvolt\per\atom})} & \shortstack{Force\\(\unit{\milli\electronvolt\per\angstrom})}\\
\colrule
\ensuremath{\mathrm{MACE}_{1}} & \num{1.49} & \num{51.14}\\
\ensuremath{\mathrm{MACE}_{2}} & \num{0.50} & \num{19.58}\\
CACE &   $\num{2.24}\pm\num{0.36}$ &   $\num{51.28}\pm\num{0.75}$\\[0.6em]
\ensuremath{\mathrm{MACE}_{1}}--\ensuremath{\mathrm{LES}} &   $\num{1.07}\pm\num{0.04}$ &   $\num{15.76}\pm\num{0.11}$\\
\ensuremath{\mathrm{CACE}}--\ensuremath{\mathrm{SOG}} &   $\num{0.18}\pm\num{0.01}$ &   $\num{12.95}\pm\num{0.83}$\\[0.6em]
\mbox{\ensuremath{\mathrm{MACE}_{1}^{(0)}}--\ensuremath{\mathrm{FNO}^{(\mathrm{l})}}} &   $\num{0.20}\pm\num{0.01}$ &   $\num{18.19}\pm\num{0.30}$\\
\mbox{\ensuremath{\mathrm{MACE}_{1}^{(0)}}--\ensuremath{\mathrm{FNO}}} &   $\num{0.14}\pm\num{0.01}$ &   $\num{14.87}\pm\num{0.29}$\\
\ensuremath{\mathrm{MACE}_{1}}--\ensuremath{\mathrm{FNO}^{(\mathrm{l})}} &   $\num{0.15}\pm\num{0.00}$ &   $\num{12.73}\pm\num{0.07}$\\
\ensuremath{\mathrm{MACE}_{1}}--\ensuremath{\mathrm{FNO}} &   $\text{\bfseries\num[mode=text,detect-weight=true]{0.08}}  \pm\num{0.00}$ &   $\text{\bfseries\num[mode=text,detect-weight=true]{9.25}}  \pm\num{0.03}$\\
\end{tabular}
\end{ruledtabular}
\end{table}

Unlike for homogeneous screening (Sec.~\ref{sec:synthetic-results}),
an accurate description of inhomogeneous screening is only attainable
with the nonlinear FNO, as shown by the energy and force RMSEs in Table~\ref{tab:inhomogeneous-accuracy}.
Joint \ensuremath{\mathrm{MACE}_{1}}--\ensuremath{\mathrm{FNO}} achieves the lowest energy and force RMSEs,
\ensuremath{0.08\,\mathrm{meV}\,\mathrm{atom}^{-1}}
and \ensuremath{9.25\,\mathrm{meV}\,\text{\AA}^{-1}},
which are
\ensuremath{49\,\%}
and \ensuremath{27\,\%}
below those of joint \ensuremath{\mathrm{MACE}_{1}}--\ensuremath{\mathrm{FNO}^{(\mathrm{l})}}, respectively.
The nonlinear FNO is also more accurate with a frozen backbone.
As both FNO variants share the same source-head architecture, the consistent
gain in accuracy can be attributed to the nonlinear field propagation
for inhomogeneous screening.

As introduced in Sec.~\ref{sec:intro}, SOG-Net~\cite{ji2025sognet}
propagates latent sources linearly through a learned sum of Gaussians,
as the linear FNO does through its learned spectral kernel.
Here we assess CACE--SOG, a SOG-Net implementation with the
Cartesian atomic cluster expansion (CACE) local backbone~\cite{cheng2024cartesian},
on the inhomogeneous-screening benchmark.
To match the receptive radius of \ensuremath{\mathrm{MACE}_{1}}, we use a cutoff of
\ensuremath{4.5\,\text{\AA}} for the CACE backbone with no message passing.
We use four latent source channels, each propagated through a shared
kernel parametrized as a sum of twelve trainable Gaussian terms.
The energy--force weights are \num{1}:\num{100} for the first \num{45} epochs,
and increase to \num{100}:\num{100} thereafter up to \num{1000} epochs.
The local CACE baseline follows the same two-stage schedule.
Table~\ref{tab:inhomogeneous-accuracy} shows
that the force RMSE obtained with CACE--SOG agrees with that of joint \ensuremath{\mathrm{MACE}_{1}}--\ensuremath{\mathrm{FNO}^{(\mathrm{l})}} within the seed
spread, while the energy RMSE is larger to some extent.
Given that the improvement over the local CACE baseline is comparable to that of \ensuremath{\mathrm{MACE}_{1}}--\ensuremath{\mathrm{FNO}^{(\mathrm{l})}} over \ensuremath{\mathrm{MACE}_{1}},
we conclude that a configuration-independent propagation saturates at this level
irrespective of the kernel form and the local backbone.

\begin{figure*}[!t]
\centering
\includegraphics[width=\textwidth]{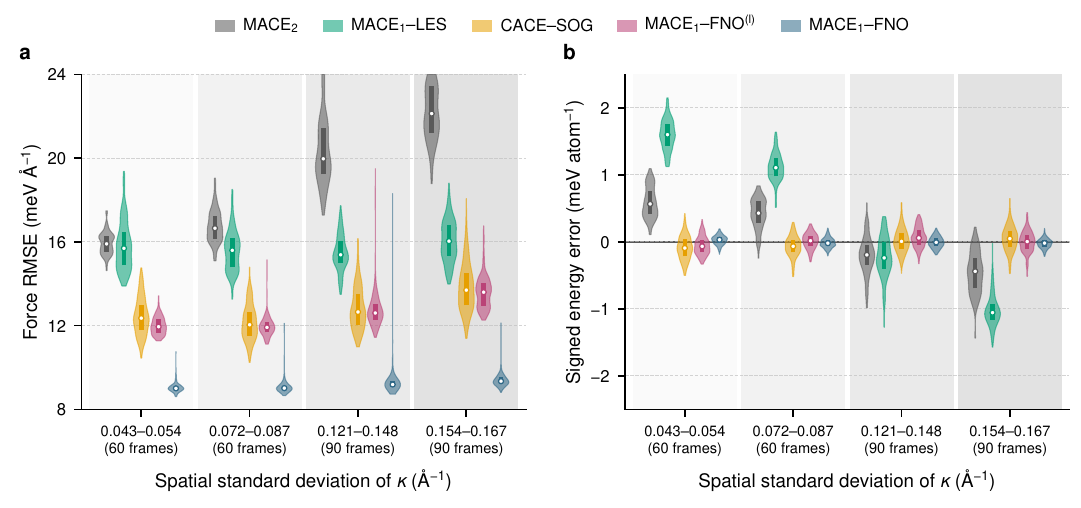}
\caption{Distributions of (a) per-configuration force RMSEs and
(b) signed energy errors per atom (model minus reference) in four windows of
the spatial standard deviation of $\kappa(\mathbf r)$.
The solid horizontal line in (b) marks zero error.
Darker background shading indicates greater screening inhomogeneity.
The equally spaced groups are labeled by their observed screening-spread
ranges.
Grouping is based on the per-configuration screening spread.
Violins pool configurations and training seeds, with widths normalized independently.
White dots and thick bars indicate medians and interquartile
ranges.
All methods have five seeds except the single-seed \ensuremath{\mathrm{MACE}_{2}} baseline.
Only jointly trained FNO models are shown.
}
\label{fig:force-rmse-vs-cs}
\end{figure*}

To examine how these differences depend on the degree of inhomogeneity,
Fig.~\ref{fig:force-rmse-vs-cs}(a) groups the per-configuration force RMSEs
into four windows of the spatial standard deviation of $\kappa(\mathbf r)$.
The error of joint \ensuremath{\mathrm{MACE}_{1}}--\ensuremath{\mathrm{FNO}} depends weakly on the inhomogeneity,
and its median rises by only \ensuremath{4\,\%}
(from \num{9.00} to \ensuremath{9.35\,\mathrm{meV}\,\text{\AA}^{-1}}) between the
least and the most inhomogeneous windows.
Over the same range, the medians rise by \ensuremath{14\,\%} for
\ensuremath{\mathrm{MACE}_{1}}--\ensuremath{\mathrm{FNO}^{(\mathrm{l})}} and by \ensuremath{11\,\%} for CACE--SOG (mostly beyond
the second window), and by \ensuremath{39\,\%} for the local \ensuremath{\mathrm{MACE}_{2}},
whereas those of \ensuremath{\mathrm{MACE}_{1}}--\ensuremath{\mathrm{LES}} stay between \num{15.4} and
\ensuremath{16.0\,\mathrm{meV}\,\text{\AA}^{-1}} without a systematic trend.
Hence, the gap between linear and nonlinear FNO widens from
\num{2.9} to \ensuremath{4.3\,\mathrm{meV}\,\text{\AA}^{-1}}, and
the two learned fixed kernels are more accurate than LES.
The rising \ensuremath{\mathrm{MACE}_{2}} error under higher screening inhomogeneity
in turn suggests that, as the screening
medium $S$ segregates, a larger fraction of the ions resides
in weakly screened regions, in which the forces depend on
charges and on the medium beyond the receptive field of the local model.

The energy RMSE shows no consistent monotonic trend with screening
inhomogeneity when evaluated over the same four windows.
The signed energy errors in Fig.~\ref{fig:force-rmse-vs-cs}(b), however,
reveal systematic shifts that the RMSE obscures.
While the models with a learned kernel (FNO and CACE--SOG)
maintain a mean signed energy error below \ensuremath{0.09\,\mathrm{meV}\,\mathrm{atom}^{-1}} in magnitude,
\ensuremath{\mathrm{MACE}_{1}}--\ensuremath{\mathrm{LES}} records a signed error of \ensuremath{1.60\,\mathrm{meV}\,\mathrm{atom}^{-1}}
in the least inhomogeneous window, gradually shifting to \ensuremath{-1.02\,\mathrm{meV}\,\mathrm{atom}^{-1}}
in the most inhomogeneous one.
In fact, this offset accounts for most of the LES energy RMSE in
Table~\ref{tab:inhomogeneous-accuracy}, which exceeds even that of the local \ensuremath{\mathrm{MACE}_{2}}.
The same trend is observed with the two local MACE models.
Since the errors are defined as model minus reference, this shift indicates
that LES and the local models underestimate the extent to which the reference
energy rises as the screening becomes more inhomogeneous.
Indeed, segregation at a higher degree of inhomogeneity reduces the energetic
stabilization provided by the screening medium.
We see this via Eq.~\eqref{eq:het-energy-variation},
which reduces to
$\delta E_{\mathrm{scr}}=-(C_{\mathrm{Coul}}\kappa_1^2/8\pi)
\int_\Omega\varphi^2\,\delta c_S\,\mathrm d\mathbf r$
when the ionic positions are held fixed.
A local change of the $S$ concentration $c_S$ modifies the energy in
proportion to $\varphi^2$, the squared screened potential
of the ions.
Therefore, moving $S$ away from the ions, where $\varphi^2$ is largest,
raises the energy.
\ensuremath{\mathrm{MACE}_{1}}--\ensuremath{\mathrm{LES}} and the two local MACE models capture this medium-dependent
energy increase only partly, missing about \ensuremath{7\text{--}19\,\%} of its
variation across the test configurations, whereas all learned-kernel models
reproduce it to within \ensuremath{1\,\%}.

Thus, the energy and force errors reflect different aspects of the
inhomogeneous screening.
The energy offset follows the overall screening stabilization of the ions,
which the learned kernels appear to capture well,
whereas the increasing force errors with inhomogeneity point to a
missing dependence of the interionic coupling on the screening medium.
We now proceed to probe this dependence directly in Sec.~\ref{sec:remote-screening-probe}.

\subsection{Remote screening medium and interionic coupling}
\label{sec:remote-screening-probe}

\begin{figure*}[t]
\centering
\includegraphics[width=\textwidth]{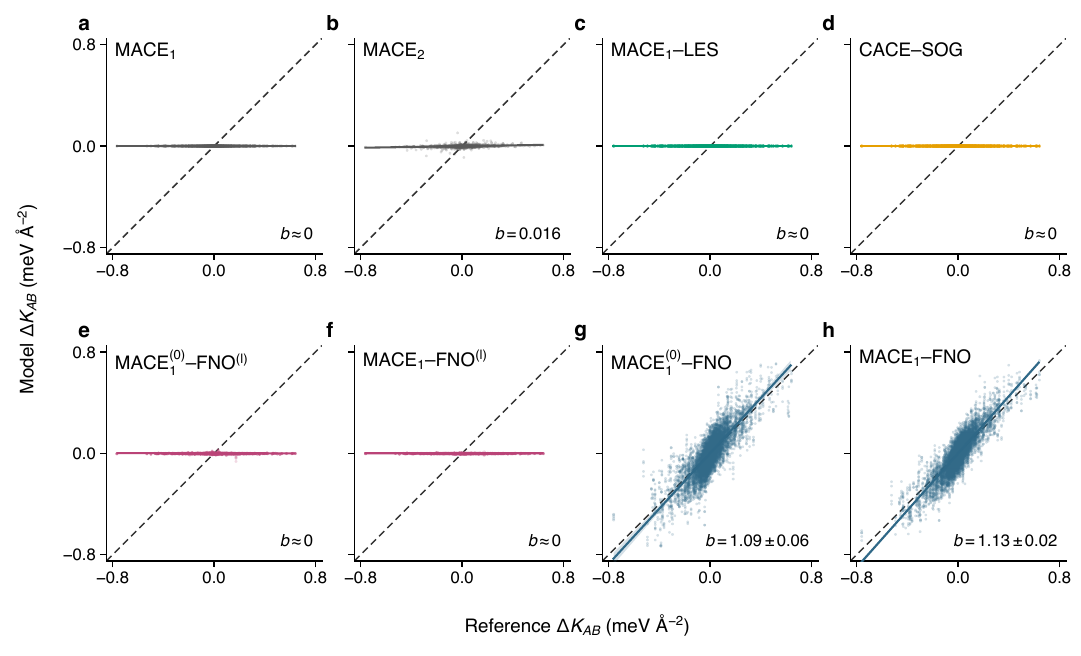}
\caption{Response of the interionic coupling to the remote screening medium.
Elements of the model response $\Delta\mathbf K_{AB}^{\mathrm{model}}$
are plotted against those of the reference $\Delta\mathbf K_{AB}^{\mathrm{ref}}$
on equal axes shared by all panels.
Each panel pools the six environments of all \num{40} probe sets.
All five model seeds are shown except for the single-seed \ensuremath{\mathrm{MACE}_{1}} and
\ensuremath{\mathrm{MACE}_{2}} baselines.
Solid lines show linear fits, and dashed lines indicate $b=1$.
The reported uncertainties in $b$ denote one sample standard
deviation across seeds.
}
\label{fig:remote-probe}
\end{figure*}

In a system with inhomogeneous screening, changes in the screening medium
can modify the coupling between two ions even when they occur far from both
ions.
To probe this dependence directly, we measure how the force on an ion $A$
responds to a small displacement of an ion $B$ when only the medium far from
both ions is rearranged.
The interionic response is the $3\times3$ interatomic force-constant matrix
$\mathbf K_{AB}$, the elements of which are
\begin{equation}
\begin{aligned}
K_{AB}^{\alpha\beta}
&=\frac{\partial^2 E}{\partial R_{A\alpha}\partial R_{B\beta}}
=-\frac{\partial F_{A\alpha}}{\partial R_{B\beta}}\\
&\simeq-\frac{
F_{A\alpha}(\mathbf R_B+\delta\mathbf e_\beta)
-F_{A\alpha}(\mathbf R_B-\delta\mathbf e_\beta)
}{2\delta},
\end{aligned}
\label{eq:remote-cross-force-constant}
\end{equation}
where $R_{A\alpha}$ and $F_{A\alpha}$ are the position and force components of
ion $A$, $\alpha,\beta\in\{x,y,z\}$, and $\mathbf e_\beta$ is a Cartesian unit
vector.
We use $\delta=\ensuremath{0.05\,\text{\AA}}$ for the displacement
of $B$ in the central finite differences.

From one configuration of each of the \num{20} test trajectories,
we select two ion pairs with minimum-image separations between
\num{10} and \ensuremath{12\,\text{\AA}}.
We freeze all species within \ensuremath{10\,\text{\AA}}
of either ion of the probed pair.
Outside these regions, we resample the species by Metropolis swaps at
\ensuremath{2\,800\,\mathrm{K}} using the potential of the MD for over \num{1600} sweeps.
Since the LJ interaction is invariant under the swaps,
the acceptance depends only on the homogeneous Yukawa interaction among the ions
and the external potential acting on $S$.
For each probe set, we generate five remote environments in this way:
four of them are biased by an independently drawn field
$V_{\mathrm{ext}}$ with $V_0=4$ (cf.\ Eq.~\eqref{eq:het-bias})
towards segregation of $S$ into different long-wavelength patterns
away from the probed ions,
whereas the fifth is unbiased, in favor of a uniform $S$
distribution.
Together with the base configuration, each probe set
thus provides six remote environments that differ only
beyond \ensuremath{10\,\text{\AA}} from the probed ions.
The swaps preserve the composition and charge neutrality.
The resampled configurations are representative of the test set,
and their reference energies relative to those of the base configurations
are within the standard deviation of \ensuremath{8.4\,\mathrm{meV}\,\mathrm{atom}^{-1}}
along the corresponding trajectories.

We measure the medium-induced response by $\Delta\mathbf K_{AB}$,
the deviation of $\mathbf K_{AB}$ from its mean over the six environments within a probe set
\begin{equation}
\Delta\mathbf K_{AB}^{X,(v)}
=\mathbf K_{AB}^{X,(v)}
-\left\langle\mathbf K_{AB}^{X,(v')}\right\rangle_{v'\in\mathcal V},
\label{eq:remote-centered-response}
\end{equation}
where $v$ labels the six remote environments in $\mathcal V$,
and $X$ refers to either an MLIP model ($X=\mathrm{model}$)
or the numerical reference ($X=\mathrm{ref}$).
Note that pooling $\Delta\mathbf K_{AB}$ over a probe set is equivalent to comparing
all \num{15} pairwise differences of $\mathbf K_{AB}$ between its six environments.
In the reference, the rearrangements of the remote medium change $\mathbf K_{AB}$ by
\ensuremath{0.11\,\mathrm{meV}\,\text{\AA}^{-2}} (RMS of
$\Delta\mathbf K_{AB}$), about \ensuremath{8\,\%} of $\mathbf K_{AB}$ itself
(RMS of \ensuremath{1.42\,\mathrm{meV}\,\text{\AA}^{-2}}).
This remote modulation stems entirely from the screened interaction
as the LJ interaction between the probed ions is the same in all environments.
As a null control, we generate two additional configurations per probe set
in which only remote ions are exchanged at a fixed medium distribution.
Here the screened Poisson equation depends only on the medium and
remains fixed during ionic displacements.
The quadratic screened energy thus gives a contribution to $\mathbf K_{AB}$
that shows no dependence on the remote charges.
Indeed, the reference shows no response to this null control,
and any change of the model $\mathbf K_{AB}$ would be a spurious response
to the remote charges.

We compare $\Delta\mathbf K_{AB}^{\mathrm{model}}$ with $\Delta\mathbf K_{AB}^{\mathrm{ref}}$
in Fig.~\ref{fig:remote-probe}, pooling all matrix elements of the six environments
in the \num{40} probe sets.
For each model, we report the slope $b$ of a linear fit as a measure of
how strongly a model follows the reference modulation.
Only the nonlinear MACE--\ensuremath{\mathrm{FNO}} attains a slope close to unity (cf.\ Fig.~\ref{fig:remote-probe}(g,h)).
The local models and all the other long-range variants considered here show no appreciable response,
albeit for different reasons.
With a receptive field cutoff $r_\mathrm{cut}=\ensuremath{4.5\,\text{\AA}}$ in \ensuremath{\mathrm{MACE}_{1}},
the matrix elements of $\mathbf K_{AB}$ are effectively zero
since no atomic energy can depend on both ions when they are by construction more than $2r_\mathrm{cut}$ apart.
The deeper local \ensuremath{\mathrm{MACE}_{2}} couples atoms with $4r_{\mathrm{cut}}=\ensuremath{18\,\text{\AA}}$,
and thus yields a nonzero $\mathbf K_{AB}$ for the probed pairs with an RMS of \ensuremath{1.13\,\mathrm{meV}\,\text{\AA}^{-2}}.
The small slope $b=\num{0.016}$ nevertheless shows that \ensuremath{\mathrm{MACE}_{2}} is unable to capture the reference modulation.
Indeed, a response to the remote medium requires atomic-energy contributions centered on other atoms
to depend jointly on the probed pair and the rearranged medium.
This is a more restrictive requirement than coupling the two ions alone,
and the maximum coupling range of $\ensuremath{18\,\text{\AA}}$ does not ensure that shared local
environments capture the medium dependence of this coupling.

\ensuremath{\mathrm{MACE}_{1}}--\ensuremath{\mathrm{LES}}, CACE--SOG, and the linear \ensuremath{\mathrm{MACE}_{1}}--\ensuremath{\mathrm{FNO}} couple the probed ions
through a long-range kernel, as manifested by the $\mathbf K_{AB}$ RMS of
\ensuremath{1.4\text{--}1.6\,\mathrm{meV}\,\text{\AA}^{-2}},
comparable to that of the reference.
Although $S$ atoms contribute to the latent source field,
the independence of these kernels from the screening medium prevents remote $S$ rearrangements
from modifying the propagation operator coupling the probed ions.
Hence, these fixed-kernel models capture an effective interionic coupling
but miss its modulation by the remote screening medium, leaving $\mathbf K_{AB}$ constant
across the six environments.

In the linear FNO, remote $S$ sources only add to the field.
In the nonlinear FNO, they also modify the effective propagation between
the probed ions, so $\mathbf K_{AB}$ responds to the remote-medium rearrangements.
With joint training, the nonlinear FNO attains a seed-averaged slope of $b=\num{1.13}\pm\num{0.02}$,
and the slope within each probe set is positive in all \num{40} sets for every seed.
A comparable slope is obtained with the frozen \ensuremath{\mathrm{MACE}_{1}} backbone ($b=\num{1.09}\pm\num{0.06}$).
We further quantify the element-wise agreement by the coefficient of determination of the predictions,
$R^2=1-\sum(\Delta K^{\mathrm{model}}-\Delta K^{\mathrm{ref}})^2/\sum(\Delta K^{\mathrm{ref}})^2$,
which vanishes for a model without response and reaches unity only when every matrix element is reproduced.
The joint nonlinear FNO achieves $R^2=\num{0.60}\pm\num{0.02}$, higher than
that of the frozen one ($R^2=\num{0.42}\pm\num{0.08}$).

The scatter in \ensuremath{\mathrm{MACE}_{1}}--\ensuremath{\mathrm{FNO}} is in part a spurious response to remote ionic
charges at a fixed distribution of the screening medium.
In the null control, the nonlinear FNO changes $\mathbf K_{AB}$ by an RMS of
\ensuremath{0.03\text{--}0.05\,\mathrm{meV}\,\text{\AA}^{-2}},
about half that of the scatter.
This spurious sensitivity presumably arises because the sources of remote ions can
modify the effective propagation in the nonlinear FNO as those of $S$ do,
and the architecture does not enforce the invariance of the reference
to remote ionic rearrangement.
For $\mathbf K_{AB}$ itself, the joint nonlinear FNO gives the lowest RMSE
(\ensuremath{0.29\,\mathrm{meV}\,\text{\AA}^{-2}}, compared with
\num{0.41} for the joint linear FNO, \num{0.43} for CACE--SOG, and
\ensuremath{0.64\,\mathrm{meV}\,\text{\AA}^{-2}} for LES).
Note that none of the models is trained on force constants,
and the probe provides an independent test of the learned coupling.
It supports the interpretation of the force errors in
Sec.~\ref{sec:heterogeneous-accuracy}: among the models
studied here, only the nonlinear FNO captures a substantial part of the reference
modulation of the interionic coupling by the remote screening medium.

\section{First-principles benchmarks}
\label{sec:physical-results}
\label{sec:first-principles-benchmark}

Alongside the synthetic benchmark, we consider liquid water, lithium
lanthanum zirconate (LLZO), and Au$_2$/MgO(001) with remote Al dopants.
These benchmarks are collected from existing datasets using first-principles references.

Liquid water represents a prototypical system in which local hydrogen bonding combines with
long-range electrostatic coupling between polar water molecules~\cite{gao2022scfnn}.
This coupling influences collective polarization and dielectric screening,
which can be misrepresented by short-range MLIPs~\cite{cheng2025les}.
We examine whether our MACE--\ensuremath{\mathrm{FNO}} can improve energy and
force predictions beyond a compact local MACE description.
The liquid-water dataset released with the qNEP study~\cite{fan2026qnep}
contains reference energies and forces calculated at the meta-GGA SCAN
level~\cite{sun2015scan}. The structures contain \num{128} water molecules
(\num{384} atoms) in an aligned cubic cell.
Removing exact duplicate geometries from the released \num{1388}-entry
training pool results in \num{523} distinct configurations to be used in this study.
A cell-length-stratified split with seed \num{17} assigns \num{471} to training and \num{52} to validation.
From the released \num{500}-entry validation file, we select the \num{134}
unique geometries absent from the source training pool as our test set.
All local MACE, MACE--\ensuremath{\mathrm{FNO}}, and MACE--\ensuremath{\mathrm{LES}} models
share this geometry-disjoint split.

LLZO is a garnet-type solid electrolyte with mobile lithium
ions in an oxide framework.
In LLZO, lithium ions interact electrostatically with other lithium ions and the extended La-Zr-O
network, giving rise to long-range interactions that go beyond their
immediate coordination shell.
For LLZO, we take the \num{1978} reference structures originally generated
by Yan and Zhu~\cite{yan2024llzo} and redistributed by Fan et al.\ with
the qNEP study~\cite{fan2026qnep}.
The reference calculations employ the PBEsol functional~\cite{perdew2008pbesol}.
The structures contain \num{192} atoms per periodic cell.
The unit cell sizes and shapes vary across the dataset.
As the release provides no predefined training and test
partition, we assign consecutive source blocks of \num{20} structures using
seed \num{17}, obtaining \num{1578} training, \num{200} validation, and
\num{200} test structures.
Since the release does not identify the trajectories from which the structures
are drawn, test blocks may share a trajectory with training blocks.
The local baselines, frozen and joint MACE--\ensuremath{\mathrm{FNO}}, and MACE--\ensuremath{\mathrm{LES}} models
share this data partition.

The Au$_2$ dimer adsorbed on MgO(001) extends our benchmarks
from bulk systems to a two-dimensional surface system.
Depending on its axis orientation relative to the MgO surface,
the dimer can adopt either a wetting (parallel) or a non-wetting (upright) configuration.
Theoretical studies have shown that electron-donor doping can stabilize flat Au
clusters on MgO through charge transfer~\cite{mammen2011morphology,mammen2018wetting}.
Interestingly, the geometry preference of the dimer can be modified
by subsurface Al dopants located more than \ensuremath{10\,\text{\AA}}
away~\cite{ko2021fourth}.
Such a long-range correlation provides a direct test of
sensitivity to remote chemical changes.
We use \num{5000} configurations with DFT energies
and forces generated by Ko et al.~\cite{ko2021fourth} using the PBE
functional~\cite{perdew1996pbe}.
We adopt the split of \num{4500} training and \num{500} test structures
released with the LES study of Kim et al.~\cite{kim2025universal}.
For FNO and LES training, we reserve \num{225} training structures for internal
validation, leaving \num{4275} for fitting.

For liquid water and LLZO, we train the local MACE baselines from
scratch with a \ensuremath{4.5\,\text{\AA}} cutoff and \num{128} scalar hidden features.
For Au$_2$/MgO(001), we adopt the one-interaction local MACE checkpoint
released with the LES study~\cite{kim2025universal} and train a two-interaction
baseline. Both use a \ensuremath{5.5\,\text{\AA}} cutoff and \num{128} scalar and
\num{128} vector channels per interaction.

In addition to the one-interaction backbone, we
augment the two-interaction backbone with FNO and LES to assess
whether global corrections provide further accuracy gains beyond those
achieved by extending the local receptive field.
The two-interaction FNO extension follows the same fitting protocol as the
one-interaction runs, including the loss scales, learning-rate schedule,
and checkpoint-selection rule.
The two-interaction source head receives \num{256} invariant descriptors from the two
interactions instead of \num{128}.
The source-head hidden width and all FNO settings are unchanged.

We adopt the FNO settings in Table~\ref{tab:computational-settings}.
For all three systems, we use a hidden-channel width $H_f=\num{16}$
for the Fourier layers and mix all hidden channels through a single spectral
channel group ($G=1$).
For Au$_2$/MgO(001), we report the results obtained with full $z$-mixing over \num{24}
height points within the spectral matrix [Eq.~\eqref{eq:main-slab-spectral-mixing}].
The channel-wise mixing results are presented in Appendix~\ref{app:slab-mixing-comparison}.

To reduce spurious energy and force changes under rigid translation
relative to the mesh, we average the FNO energy corrections over four
in-plane mesh origins for Au$_2$/MgO(001).
These correspond to all combinations of zero and half-grid shifts along
the two periodic directions.
This interlacing is applied during both training and inference.
Liquid water and LLZO are evaluated on a single mesh, since interlacing
changes their energy and force RMSEs only marginally
(Appendix~\ref{app:interlacing-comparison}).

\begin{table*}[t!]
\centering
\caption{RMSEs of the energy and forces for liquid water,
LLZO, and Au$_2$/MgO(001).
FNO and LES entries are five-seed means with sample standard deviations,
whereas \ensuremath{\mathrm{MACE}_{1}} and \ensuremath{\mathrm{MACE}_{2}} are single-seed baselines.
For Au$_2$/MgO(001), all FNO entries use dense full $z$-mixing.
Bold lettering marks the lowest RMSE values within their error margins,
separately for each backbone depth.}
\label{tab:materials-results}
\small
\setlength{\tabcolsep}{4pt}

\begin{ruledtabular}
\begin{tabular}{lrr@{\hspace{1.8em}}rr@{\hspace{1.8em}}rr}
 & \multicolumn{2}{c}{Liquid water} & \multicolumn{2}{c}{LLZO} & \multicolumn{2}{c}{Au$_2$/MgO(001)}\\
\cline{2-3}\cline{4-5}\cline{6-7}
\noalign{\vskip 3pt}
Model & \shortstack{Energy\\(\unit{\milli\electronvolt\per\atom})} & \shortstack{Forces\\(\unit{\milli\electronvolt\per\angstrom})}
 & \shortstack{Energy\\(\unit{\milli\electronvolt\per\atom})} & \shortstack{Forces\\(\unit{\milli\electronvolt\per\angstrom})}
 & \shortstack{Energy\\(\unit{\milli\electronvolt\per\atom})} & \shortstack{Forces\\(\unit{\milli\electronvolt\per\angstrom})}\\
\colrule
\ensuremath{\mathrm{MACE}_{1}} & \num{1.58} & \num{106.35} & \num{0.58} & \num{70.50} & \num{2.31} & \num{56.99}\\
\ensuremath{\mathrm{MACE}_{1}}--\ensuremath{\mathrm{LES}} & \num{0.90 +- 0.01} &   $\text{\bfseries\num[mode=text,detect-weight=true]{71.01}}  \pm\num{0.12}$ &   $\text{\bfseries\num[mode=text,detect-weight=true]{0.24}}  \pm\num{0.02}$ &   $\text{\bfseries\num[mode=text,detect-weight=true]{25.29}}  \pm\num{0.19}$ &   $\num{0.11}  \pm\num{0.01}$ &   $\num{6.42}  \pm\num{0.04}$\\[0.3em]
\mbox{\ensuremath{\mathrm{MACE}_{1}^{(0)}}--\ensuremath{\mathrm{FNO}^{(\mathrm{l})}}} &   $\num{0.79}  \pm\num{0.02}$ &   $\num{84.45}  \pm\num{0.23}$ &   $\num{0.39}  \pm\num{0.01}$ &   $\num{44.80}  \pm\num{0.26}$ &   $\num{0.07}  \pm\num{0.00}$ &   $\num{7.61}  \pm\num{0.21}$\\
\mbox{\ensuremath{\mathrm{MACE}_{1}^{(0)}}--\ensuremath{\mathrm{FNO}}} & \num{0.77 +- 0.04} & \num{81.79 +- 1.06} & \num{0.35 +- 0.00} & \num{40.29 +- 0.29} &   $\text{\bfseries\num[mode=text,detect-weight=true]{0.07}}  \pm\num{0.00}$ &   $\text{\bfseries\num[mode=text,detect-weight=true]{5.70}}  \pm\num{0.45}$\\
\ensuremath{\mathrm{MACE}_{1}}--\ensuremath{\mathrm{FNO}^{(\mathrm{l})}} &   $\text{\bfseries\num[mode=text,detect-weight=true]{0.72}}  \pm\num{0.02}$ &   $\num{73.87}  \pm\num{0.97}$ &   $\num{0.27}  \pm\num{0.01}$ &   $\num{31.21}  \pm\num{0.89}$ &   $\num{0.07}  \pm\num{0.00}$ &   $\num{6.72}  \pm\num{0.24}$\\
\ensuremath{\mathrm{MACE}_{1}}--\ensuremath{\mathrm{FNO}} &   $\text{\bfseries\num[mode=text,detect-weight=true]{0.70}}  \pm\num{0.03}$ & \num{71.78 +- 0.48} &   $\text{\bfseries\num[mode=text,detect-weight=true]{0.22}}  \pm\num{0.01}$ &   $\text{\bfseries\num[mode=text,detect-weight=true]{25.47}}  \pm\num{0.35}$ &   $\text{\bfseries\num[mode=text,detect-weight=true]{0.07}}  \pm\num{0.00}$ &   $\text{\bfseries\num[mode=text,detect-weight=true]{5.17}}  \pm\num{0.28}$\\[1.2em]
\ensuremath{\mathrm{MACE}_{2}} & \num{0.40} & \num{35.75} & \num{0.10} & \num{12.18} & \num{0.15} & \num{3.35}\\
\ensuremath{\mathrm{MACE}_{2}}--\ensuremath{\mathrm{LES}} & \num{0.34 +- 0.02} &   $\text{\bfseries\num[mode=text,detect-weight=true]{32.38}}  \pm\num{0.05}$ &   $\text{\bfseries\num[mode=text,detect-weight=true]{0.09}}  \pm\num{0.01}$ &   $\text{\bfseries\num[mode=text,detect-weight=true]{10.32}}  \pm\num{0.13}$ &   $\num{0.11}  \pm\num{0.03}$ &   $\num{3.61}  \pm\num{0.22}$\\[0.3em]
\mbox{\ensuremath{\mathrm{MACE}_{2}^{(0)}}--\ensuremath{\mathrm{FNO}^{(\mathrm{l})}}} &   $\num{0.33}  \pm\num{0.01}$ &   $\num{35.37}  \pm\num{0.02}$ &   $\text{\bfseries\num[mode=text,detect-weight=true]{0.09}}  \pm\num{0.00}$ &   $\num{11.37}  \pm\num{0.02}$ &   $\text{\bfseries\num[mode=text,detect-weight=true]{0.07}}  \pm\num{0.00}$ &   $\text{\bfseries\num[mode=text,detect-weight=true]{2.99}}  \pm\num{0.09}$\\
\mbox{\ensuremath{\mathrm{MACE}_{2}^{(0)}}--\ensuremath{\mathrm{FNO}}} &   $\text{\bfseries\num[mode=text,detect-weight=true]{0.31}}  \pm\num{0.01}$ & \num{35.19 +- 0.05} &   $\text{\bfseries\num[mode=text,detect-weight=true]{0.09}}  \pm\num{0.00}$ & \num{11.41 +- 0.02} &   $\text{\bfseries\num[mode=text,detect-weight=true]{0.07}}  \pm\num{0.00}$ &   $\text{\bfseries\num[mode=text,detect-weight=true]{2.89}}  \pm\num{0.09}$\\
\ensuremath{\mathrm{MACE}_{2}}--\ensuremath{\mathrm{FNO}^{(\mathrm{l})}} &   $\num{0.34}  \pm\num{0.02}$ &   $\num{35.42}  \pm\num{0.04}$ &   $\text{\bfseries\num[mode=text,detect-weight=true]{0.09}}  \pm\num{0.00}$ &   $\num{10.95}  \pm\num{0.04}$ &   $\text{\bfseries\num[mode=text,detect-weight=true]{0.07}}  \pm\num{0.00}$ &   $\text{\bfseries\num[mode=text,detect-weight=true]{2.93}}  \pm\num{0.17}$\\
\ensuremath{\mathrm{MACE}_{2}}--\ensuremath{\mathrm{FNO}} &   $\text{\bfseries\num[mode=text,detect-weight=true]{0.30}}  \pm\num{0.01}$ & \num{35.12 +- 0.06} &   $\text{\bfseries\num[mode=text,detect-weight=true]{0.09}}  \pm\num{0.00}$ & \num{10.99 +- 0.02} &   $\text{\bfseries\num[mode=text,detect-weight=true]{0.07}}  \pm\num{0.00}$ &   $\text{\bfseries\num[mode=text,detect-weight=true]{2.94}}  \pm\num{0.06}$\\
\end{tabular}
\end{ruledtabular}
\end{table*}

\subsection{Energy and force accuracy}
\label{sec:physical-accuracy}
As summarized in Table~\ref{tab:materials-results},
FNO augmentation substantially improves the accuracy of the one-interaction backbone for
liquid water, LLZO, and Au$_2$/MgO(001).
With the frozen \ensuremath{\mathrm{MACE}_{1}} backbone, nonlinear FNO already reduces the energy RMSE by
\ensuremath{40\text{--}51\,\%}
and the force RMSE by
\ensuremath{23\text{--}43\,\%}
for the two bulk systems.
The gains are particularly sizable for Au$_2$/MgO(001), reaching
\ensuremath{97\,\%} for energy
and \ensuremath{90\,\%} for forces.
Training the backbone and FNO jointly further lowers the force RMSE by \ensuremath{9\text{--}37\,\%}
in all three systems and the energy RMSE by \ensuremath{10\,\%\text{ and }37\,\%}
for water and LLZO, respectively.
Most of the improvement over the local baseline is, however, already
obtained with linear FNO.
With joint training, \ensuremath{\mathrm{MACE}_{1}}--\ensuremath{\mathrm{FNO}^{(\mathrm{l})}} accounts for
\ensuremath{86\text{--}100\,\%} of the RMSE reduction achieved by \ensuremath{\mathrm{MACE}_{1}}--\ensuremath{\mathrm{FNO}}.
Nonlinearity mainly benefits the forces of LLZO and Au$_2$/MgO(001),
whose RMSEs decrease by a further \ensuremath{18\,\%\text{ and }23\,\%}, respectively,
while both RMSEs of water change by only \ensuremath{3\,\%}.

Increasing the depth of the local backbone largely
reduces the errors of \ensuremath{\mathrm{MACE}_{1}}, lowering the energy RMSE by
\ensuremath{75\text{--}93\,\%}
and the force RMSE by
\ensuremath{66\text{--}94\,\%}.
\ensuremath{\mathrm{MACE}_{2}} is also more accurate than all \ensuremath{\mathrm{MACE}_{1}}--\ensuremath{\mathrm{FNO}} models except for
the energy of Au$_2$/MgO(001), for which the FNO models give lower RMSEs.
Adding FNO to this deeper \ensuremath{\mathrm{MACE}_{2}} backbone further improves the accuracy,
although the gains become smaller, particularly for the forces.
The jointly trained \ensuremath{\mathrm{MACE}_{2}}--\ensuremath{\mathrm{FNO}} lowers the energy RMSE by
\ensuremath{24\,\%\text{, }6\,\%\text{, and }56\,\%}
and the force RMSE by
\ensuremath{2\,\%\text{, }10\,\%\text{, and }12\,\%}
for water, LLZO, and Au$_2$/MgO(001), respectively.
It is clear that joint training becomes less important as the receptive field of the backbone extends.
Whereas joint training reduces the energy and force RMSEs of \ensuremath{\mathrm{MACE}_{1}}--\ensuremath{\mathrm{FNO}} by up to
\ensuremath{37\,\%}, it changes the RMSEs of \ensuremath{\mathrm{MACE}_{2}}--\ensuremath{\mathrm{FNO}}
by less than \ensuremath{4\,\%}.
Nonlinearity likewise loses its advantage with the deeper backbone,
as the force RMSEs of the linear and nonlinear FNO differ by less than
\ensuremath{4\,\%} for all three systems.
The gains from nonlinearity with \ensuremath{\mathrm{MACE}_{1}} in LLZO and Au$_2$/MgO(001)
thus point to the limited one-interaction backbone rather than
to configuration-dependent screening.
The pretrained \ensuremath{\mathrm{MACE}_{2}} representation, in turn, already
supplies most of what the FNO correction requires.

We now compare FNO with LES on the same local backbones.
The accuracies of the two approaches are comparable for the two bulk systems.
With \ensuremath{\mathrm{MACE}_{1}}, FNO gives a lower energy RMSE for water
and agrees with LES within the seed spread for LLZO, while LES
gives a slightly lower force RMSE for water.
With \ensuremath{\mathrm{MACE}_{2}}, FNO retains the lower energy RMSE for water,
and the force RMSEs are lower with LES for both systems.
The learned propagation thus reaches an accuracy comparable to
that of the prescribed Coulomb kernel in LES for the two bulk systems,
where learned charges can partly represent the polarization.

\begin{table*}[!t]
\centering
\caption{Relative total energies and wetting preferences for Au$_2$/MgO(001).
Energies are in \unit{\milli\electronvolt} per cell.
The non-wetting ($E_{\mathrm{nw}}$) and wetting ($E_{\mathrm{w}}$) energies
are evaluated at relaxed geometries obtained with the corresponding model.
The MgO substrate is kept fixed as in the DFT reference throughout.
The force tolerance is \ensuremath{0.01\,\mathrm{eV}\,\text{\AA}^{-1}}.
The DFT non-wetting energy is taken as the common zero for all methods on the same substrate.
$\Delta E_{\mathrm{wet}}=E_{\mathrm{w}}-E_{\mathrm{nw}}$ and
$\delta\Delta E_{\mathrm{wet}}=\Delta E_{\mathrm{wet}}-\Delta E_{\mathrm{wet}}^{\mathrm{DFT}}$.
Spreads are sample standard deviations over five seeds, whereas \ensuremath{\mathrm{MACE}_{1}}
and \ensuremath{\mathrm{MACE}_{2}} are single-seed baselines.
All FNO entries use dense full $z$-mixing.
}
\label{tab:aumgo-switch}
\small

\setlength{\tabcolsep}{2pt}
\begin{ruledtabular}
\begin{tabular}{l@{\hspace{1em}}
S[table-format=-3.1,mode=text,detect-weight=true]@{\extracolsep{0pt}}l@{\extracolsep{\fill}\hspace{0.5em}}
S[table-format=3.1,mode=text,detect-weight=true]@{\extracolsep{0pt}}l@{\extracolsep{\fill}\hspace{0.5em}}
S[table-format=3.1,mode=text,detect-weight=true]@{\extracolsep{0pt}}l@{\extracolsep{\fill}\hspace{0.5em}}
S[table-format=+3.1,mode=text,detect-weight=true]@{\hspace{1.5em}}
S[table-format=-3.1,mode=text,detect-weight=true]@{\extracolsep{0pt}}l@{\extracolsep{\fill}\hspace{0.5em}}
S[table-format=-3.1,mode=text,detect-weight=true]@{\extracolsep{0pt}}l@{\extracolsep{\fill}\hspace{0.5em}}
S[table-format=-3.1,mode=text,detect-weight=true]@{\extracolsep{0pt}}l@{\extracolsep{\fill}\hspace{0.5em}}
S[table-format=+3.1,mode=text,detect-weight=true]}
& \multicolumn{7}{c}{Undoped} & \multicolumn{7}{c}{Al-doped}\\[0.25em]
    \cline{2-8}
    \cline{9-15}
Model & \multicolumn{2}{c}{$E_{\mathrm{nw}}$}
& \multicolumn{2}{c}{$E_{\mathrm{w}}$}
& \multicolumn{2}{c}{$\Delta E_{\mathrm{wet}}$}
& \multicolumn{1}{c}{$\delta\Delta E_{\mathrm{wet}}$}
& \multicolumn{2}{c}{$E_{\mathrm{nw}}$}
& \multicolumn{2}{c}{$E_{\mathrm{w}}$}
& \multicolumn{2}{c}{$\Delta E_{\mathrm{wet}}$}
& \multicolumn{1}{c}{$\delta\Delta E_{\mathrm{wet}}$}\\
\colrule
DFT-PBE\footnotemark[1]
& 0.0 & & 934.8 & & 934.8 &
&
& 0.0 & & -66.9 & & -66.9 &
& \\[0.6em]
\ensuremath{\mathrm{MACE}_{1}}
& 218.9 & & 637.9 & & 419.0 &
& -515.8
& -248.2 & & 170.8 & & 419.0 &
& +486.0 \\
\ensuremath{\mathrm{MACE}_{1}}--\ensuremath{\mathrm{LES}}
& -51.7 & ${}\pm\num{2.1}$
& 890.0 & ${}\pm\num{2.1}$
& 941.7 & ${}\pm\num{4.2}$
& +6.9
& -25.9 & ${}\pm\num{1.7}$
& -103.1 & ${}\pm\num{2.6}$
& -77.2 & ${}\pm\num{4.1}$
& -10.3\\
\ensuremath{\mathrm{MACE}_{1}}--\ensuremath{\mathrm{LES}}\footnotemark[2]
& \multicolumn{2}{c}{} & \multicolumn{2}{c}{} & 919.3 &
& -15.5
& \multicolumn{2}{c}{} & \multicolumn{2}{c}{} & -56.5 &
& +10.4 \\[0.3em]
\mbox{\ensuremath{\mathrm{MACE}_{1}^{(0)}}--\ensuremath{\mathrm{FNO}^{(\mathrm{l})}}}
& -44.2 & ${}\pm\num{1.2}$
& 881.6 & ${}\pm\num{0.4}$
& 925.8 & ${}\pm\num{1.2}$
& -9.0
& -28.3 & ${}\pm\num{0.3}$
& -95.3 & ${}\pm\num{1.2}$
& \bfseries -67.0 & ${}\pm\num{1.0}$
& \bfseries 0.0\\
\mbox{\ensuremath{\mathrm{MACE}_{1}^{(0)}}--\ensuremath{\mathrm{FNO}}}
& -46.5 & ${}\pm\num{0.3}$
& 883.8 & ${}\pm\num{0.6}$
& \bfseries 930.2 & ${}\pm\num{0.5}$
& \bfseries -4.6
& -27.0 & ${}\pm\num{0.2}$
& -96.6 & ${}\pm\num{0.7}$
& -69.6 & ${}\pm\num{0.8}$
& -2.6\\
\ensuremath{\mathrm{MACE}_{1}}--\ensuremath{\mathrm{FNO}^{(\mathrm{l})}}
& -45.3 & ${}\pm\num{0.6}$
& 883.2 & ${}\pm\num{0.7}$
& 928.5 & ${}\pm\num{0.5}$
& -6.4
& -28.0 & ${}\pm\num{0.5}$
& -95.9 & ${}\pm\num{0.6}$
& \bfseries -67.8 & ${}\pm\num{0.9}$
& \bfseries -0.9\\
\ensuremath{\mathrm{MACE}_{1}}--\ensuremath{\mathrm{FNO}}
& -46.7 & ${}\pm\num{0.3}$
& 883.9 & ${}\pm\num{0.4}$
& \bfseries 930.6 & ${}\pm\num{0.5}$
& \bfseries -4.2
& -27.3 & ${}\pm\num{0.2}$
& -97.0 & ${}\pm\num{0.9}$
& -69.6 & ${}\pm\num{0.9}$
& -2.7\\[1.2em]
\ensuremath{\mathrm{MACE}_{2}}
& -33.0 & & 869.7 & & 902.7 &
& -32.1
& -43.4 & & -80.5 & & -37.0 &
& +29.9 \\
\ensuremath{\mathrm{MACE}_{2}}--\ensuremath{\mathrm{LES}}
& -38.6 & ${}\pm\num{5.4}$
& 876.3 & ${}\pm\num{5.9}$
& 914.9 & ${}\pm\num{11.2}$
& -19.9
& -38.4 & ${}\pm\num{5.5}$
& -88.7 & ${}\pm\num{5.3}$
& -50.4 & ${}\pm\num{10.7}$
& +16.6\\[0.3em]
\mbox{\ensuremath{\mathrm{MACE}_{2}^{(0)}}--\ensuremath{\mathrm{FNO}^{(\mathrm{l})}}}
& -47.1 & ${}\pm\num{1.2}$
& 883.8 & ${}\pm\num{0.3}$
& \bfseries 930.9 & ${}\pm\num{1.3}$
& \bfseries -3.9
& -29.0 & ${}\pm\num{0.9}$
& -96.2 & ${}\pm\num{1.2}$
& \bfseries -67.2 & ${}\pm\num{1.5}$
& \bfseries -0.3\\
\mbox{\ensuremath{\mathrm{MACE}_{2}^{(0)}}--\ensuremath{\mathrm{FNO}}}
& -47.0 & ${}\pm\num{0.6}$
& 884.2 & ${}\pm\num{0.6}$
& \bfseries 931.2 & ${}\pm\num{0.4}$
& \bfseries -3.6
& -27.6 & ${}\pm\num{0.6}$
& -96.9 & ${}\pm\num{1.2}$
& \bfseries -69.3 & ${}\pm\num{1.4}$
& \bfseries -2.4\\
\ensuremath{\mathrm{MACE}_{2}}--\ensuremath{\mathrm{FNO}^{(\mathrm{l})}}
& -47.1 & ${}\pm\num{1.1}$
& 884.1 & ${}\pm\num{0.4}$
& \bfseries 931.1 & ${}\pm\num{0.9}$
& \bfseries -3.7
& -28.4 & ${}\pm\num{0.5}$
& -97.7 & ${}\pm\num{3.1}$
& \bfseries -69.3 & ${}\pm\num{3.0}$
& \bfseries -2.4\\
\ensuremath{\mathrm{MACE}_{2}}--\ensuremath{\mathrm{FNO}}
& -46.0 & ${}\pm\num{0.9}$
& 885.0 & ${}\pm\num{0.7}$
& \bfseries 931.0 & ${}\pm\num{0.2}$
& \bfseries -3.9
& -27.9 & ${}\pm\num{1.1}$
& -96.1 & ${}\pm\num{0.9}$
& \bfseries -68.3 & ${}\pm\num{0.5}$
& \bfseries -1.3\\
\end{tabular}
\end{ruledtabular}
\footnotetext[1]{DFT-PBE calculations of Ko et al.~\cite{ko2021fourth}.}
\footnotetext[2]{Wetting preferences from Ref.~\onlinecite{kim2025universal}.
These values are reproduced with our relaxation protocols using the released checkpoint.
}
\end{table*}

The advantage of FNO is more prominent for Au$_2$/MgO(001),
where the energy RMSEs are \ensuremath{36\text{--}41\,\%} lower than
those of LES in all settings on both backbones,
and the force RMSEs are \ensuremath{17\text{--}20\,\%} lower with the
\ensuremath{\mathrm{MACE}_{2}} backbone.
With the shallow \ensuremath{\mathrm{MACE}_{1}} backbone,
only the nonlinear FNO with dense full $z$-mixing
lowers the force RMSE relative to LES (by \ensuremath{19\,\%} with joint training),
whereas the linear FNO and the nonlinear FNO with channel-wise $z$-mixing
give higher force errors
(cf.\ Tables~\ref{tab:materials-results} and~\ref{tab:slab-mixing-comparison}).
Note that dense full $z$-mixing involves about \num{20} million FNO parameters,
compared with \num{0.08} million FNO parameters in the channel-wise $z$-mixing and
\num{0.46} million in the \ensuremath{\mathrm{MACE}_{1}} backbone
(cf.\ Sec.~\ref{sec:inference-cost} and Appendix~\ref{app:slab-mixing-comparison}).
The higher accuracy attained with FNO may reflect the charge transfer
at interfaces as discussed in Sec.~\ref{sec:intro},
for which the effective interaction between learned sources need not
take the Coulomb form.

\subsection{Remote-dopant wetting preference}
\label{sec:wetting-preference}

Remote Al dopants in the MgO(001) substrate present a stringent test for nonlocal MLIPs.
We define the wetting preference for the adsorbed Au$_2$ dimer as the relative energy of
the wetting (parallel) and non-wetting (upright) configurations
\begin{equation}
 \Delta E_{\mathrm{wet}}=
 E(\mathrm{wetting})-E(\mathrm{non\mbox{-}wetting}).
 \label{eq:wetting-preference}
\end{equation}
On undoped MgO, a non-wetting dimer configuration is favored according to
the PBE calculations~\cite{ko2021fourth}.
Al dopants in the substrate lead to a sign reversal in $\Delta E_{\mathrm{wet}}$,
thereby indicating a stabilized wetting configuration.
This reversal cannot be described by local MACE models with a
receptive field too short to couple the dimer to the dopants~\cite{king2025charges,kim2025universal}.
With a cutoff of \ensuremath{5.5\,\text{\AA}}, the one-interaction \ensuremath{\mathrm{MACE}_{1}} couples
only atoms within \ensuremath{11\,\text{\AA}} of each other, whereas the nearest dimer--dopant separation
is \ensuremath{11.1\text{--}11.4\,\text{\AA}} in the non-wetting and wetting geometries.
The second interaction layer of \ensuremath{\mathrm{MACE}_{2}} extends this coupling range to
\ensuremath{22\,\text{\AA}}.
The reach of the local backbones is directly manifested in the wetting
preferences shown in Table~\ref{tab:aumgo-switch}.
While \ensuremath{\mathrm{MACE}_{1}} predicts essentially the same $\Delta E_{\mathrm{wet}}$
on both substrates, \ensuremath{\mathrm{MACE}_{2}} reproduces the sign reversal upon Al doping.

The sign reversal is also reproduced when augmenting \ensuremath{\mathrm{MACE}_{1}} with FNO or LES.
The nonlinear \ensuremath{\mathrm{MACE}_{1}}--\ensuremath{\mathrm{FNO}} predicts wetting preferences within
\ensuremath{5\,\mathrm{meV}} per cell of the DFT values on both substrates,
with seed spreads below \ensuremath{1\,\mathrm{meV}}.
For comparison, \ensuremath{\mathrm{MACE}_{1}}--\ensuremath{\mathrm{LES}} and \ensuremath{\mathrm{MACE}_{2}} deviate by up to \ensuremath{10\,\mathrm{meV}\text{ and }32\,\mathrm{meV}},
respectively.
Linear FNO is at least as accurate on doped MgO but deviates by up to
\ensuremath{9\,\mathrm{meV}} on undoped MgO.
Finally, with the deeper \ensuremath{\mathrm{MACE}_{2}} backbone, FNO augmentation reduces the deviations further
to below \ensuremath{4\,\mathrm{meV}},
while \ensuremath{\mathrm{MACE}_{2}}--\ensuremath{\mathrm{LES}} still deviates by up to \ensuremath{20\,\mathrm{meV}}.

Table~\ref{tab:aumgo-switch} shows that all MLIPs except \ensuremath{\mathrm{MACE}_{1}}
place both endpoints below the reference DFT-PBE energies.
For the FNO models, these offsets amount to \ensuremath{44\text{--}53\,\mathrm{meV}}
on undoped and \ensuremath{27\text{--}31\,\mathrm{meV}} on Al-doped MgO,
equivalent to \ensuremath{0.25\text{--}0.48\,\mathrm{meV}\,\mathrm{atom}^{-1}}, about three to seven times the test energy RMSE.
The offsets are already present at the DFT geometries, as relaxation changes
the endpoint energies of these models by less than \ensuremath{0.2\,\mathrm{meV}}.
We note that the dataset consists of \num{1250} perturbed configurations around each
of the four endpoints but not the relaxed endpoints themselves.
The offsets therefore indicate a common error in extrapolating to the basin minima.
With the FNO models, the offsets are found to be more consistent between the
wetting and non-wetting configurations, thereby largely canceling in $\Delta E_{\mathrm{wet}}$.

\section{Inference time and memory consumption}
\label{sec:inference-cost}

To put the computational cost of MACE--\ensuremath{\mathrm{FNO}} into context, we compare its linear and nonlinear variants
with \ensuremath{\mathrm{MACE}_{1}}--\ensuremath{\mathrm{LES}} and the two local baselines on a single NVIDIA A100-SXM4-40GB GPU.
Table~\ref{tab:inference-cost} compares double-precision energy--force
inference for the three first-principles benchmarks, with one structure
evaluated at a time.
Runtime and memory usage are measured with \textsc{PyTorch} \texttt{2.7.1} and MACE \texttt{0.3.16},
without compilation or \texttt{cuEquivariance} acceleration.

\begin{table}[!t]
\centering
\caption{Energy--force inference runtime and peak tensor memory.
Only joint FNO is shown, using seed \num{17} and $G=1$.
For Au$_2$/MgO(001), linear FNO uses dense full channel--height mixing,
and the nonlinear FNO entries compare channel-wise and dense full
$z$-mixing. All slab FNO entries use four mesh origins and the same benchmarking protocol.}
\label{tab:inference-cost}
\small

\begin{ruledtabular}
\begin{tabular}{lrr}
Model & Runtime & \shortstack{Peak tensor\\memory}\\
& (\unit{\milli\second}) & (\unit{\mebibyte})\\
\colrule
\multicolumn{3}{l}{Liquid water (\num{384} atoms)}\\
\ensuremath{\mathrm{MACE}_{1}} & \num{27} & \num{1399}\\
\ensuremath{\mathrm{MACE}_{2}} & \num{38} & \num{1870}\\
\ensuremath{\mathrm{MACE}_{1}}--\ensuremath{\mathrm{LES}} & \num{31} & \num{1399}\\
\ensuremath{\mathrm{MACE}_{1}}--\ensuremath{\mathrm{FNO}^{(\mathrm{l})}} & \num{36} & \num{1399}\\
\ensuremath{\mathrm{MACE}_{1}}--\ensuremath{\mathrm{FNO}} & \num{40} & \num{1400}\\[0.4em]
\multicolumn{3}{l}{LLZO (\num{192} atoms)}\\
\ensuremath{\mathrm{MACE}_{1}} & \num{14} & \num{298}\\
\ensuremath{\mathrm{MACE}_{2}} & \num{19} & \num{457}\\
\ensuremath{\mathrm{MACE}_{1}}--\ensuremath{\mathrm{LES}} & \num{17} & \num{298}\\
\ensuremath{\mathrm{MACE}_{1}}--\ensuremath{\mathrm{FNO}^{(\mathrm{l})}} & \num{22} & \num{299}\\
\ensuremath{\mathrm{MACE}_{1}}--\ensuremath{\mathrm{FNO}} & \num{31} & \num{305}\\[0.4em]
\multicolumn{3}{l}{Au$_2$/MgO(001) (\num{110} atoms)}\\
\ensuremath{\mathrm{MACE}_{1}} & \num{15} & \num{461}\\
\ensuremath{\mathrm{MACE}_{2}} & \num{29} & \num{1189}\\
\ensuremath{\mathrm{MACE}_{1}}--\ensuremath{\mathrm{LES}} & \num{18} & \num{461}\\
\ensuremath{\mathrm{MACE}_{1}}--\ensuremath{\mathrm{FNO}^{(\mathrm{l})}} (full $z$) & \num{57} & \num{4473}\\
\ensuremath{\mathrm{MACE}_{1}}--\ensuremath{\mathrm{FNO}} (full $z$) & \num{175} & \num{7683}\\
\ensuremath{\mathrm{MACE}_{1}}--\ensuremath{\mathrm{FNO}} (channel-wise) & \num{32} & \num{463}\\
\end{tabular}
\end{ruledtabular}
\end{table}

After five warm-up calls per structure, we report the median of \num{90}
GPU-synchronized calls across three rounds with randomized model order.
The timings include graph construction, energy--force evaluation, and result
transfers, excluding model loading, file input, and stress evaluation.
Predictions are checked against the \textsc{ase} calculators.
We measure memory in a fresh process per model and report the maximum
live-tensor allocation over nine energy--force evaluations after warm-up.
This includes model tensors and force intermediates, but excludes unused
allocator blocks and external CUDA allocations.
All memory values are expressed in MiB ($2^{20}$ bytes).
The FNO models use the production hyperparameters listed in
Table~\ref{tab:computational-settings}, which prioritize predictive accuracy
with relatively generous mesh resolutions and network capacities and have not
been optimized for inference speed.
Frozen and joint nonlinear FNO models differ by less than \ensuremath{1\,\%} in
runtime and have identical tensor-memory peaks.

Overall, LES remains faster than the FNO methods for all three systems.
For the two bulk systems, the cost of FNO augmentation is comparable to that
of LES, particularly with linear FNO.
The tensor-memory peaks of both FNO variants lie within \ensuremath{3\,\%} of
those of \ensuremath{\mathrm{MACE}_{1}} and \ensuremath{\mathrm{MACE}_{1}}--\ensuremath{\mathrm{LES}}, and are
\ensuremath{25\,\%\text{ and }33\,\%}
lower than those of \ensuremath{\mathrm{MACE}_{2}} for liquid water and LLZO, respectively.
The linear FNO takes \ensuremath{17\,\%\text{ and }31\,\%} longer than LES for the two
systems, and the nonlinear FNO \ensuremath{31\,\%\text{ and }82\,\%} longer.
Relative to \ensuremath{\mathrm{MACE}_{2}}, the nonlinear FNO takes
\ensuremath{6\,\%\text{ and }63\,\%}
longer.
FNO augmentation of a shallow MACE backbone thus offers a memory advantage
over a deeper local backbone at the expense of increased evaluation time.
Since the mesh grows with the cell volume at fixed resolution, whether this
advantage persists in larger cells remains to be tested.

For Au$_2$/MgO(001), the nonlinear FNO with dense full $z$-mixing gives the
highest accuracy among the \ensuremath{\mathrm{MACE}_{1}}-based models but is also by far the most
expensive.
Its runtime is
\ensuremath{503\,\%}
longer than that of \ensuremath{\mathrm{MACE}_{2}}, and its tensor-memory peak is
\num{6.5} times larger.
Switching from four mesh origins to one lowers the runtime by about
\ensuremath{66\,\%}.
The displaced meshes are evaluated as a batch, sharing the local MACE
and source-head calculations. This single-mesh comparison changes only the
inference setting and does not imply unchanged prediction accuracy.
The effects of interlacing on translation sensitivity and prediction
accuracy are quantified in Appendix~\ref{app:interlacing-comparison}.

We note that the computational cost associated with Au$_2$/MgO
can be largely mitigated at a modest loss of accuracy.
The linear FNO, which maintains full $z$-mixing, lowers the runtime and
tensor-memory peak to
\ensuremath{57\,\mathrm{ms}} and
\ensuremath{4\,473\,\mathrm{MiB}},
while its force RMSE increases from
\ensuremath{5.17\,\mathrm{meV}\,\text{\AA}^{-1}} to
\ensuremath{6.72\,\mathrm{meV}\,\text{\AA}^{-1}}.
Ultimately, channel-wise $z$-mixing reduces the runtime to
\ensuremath{32\,\mathrm{ms}},
only \ensuremath{11\,\%} longer than that of \ensuremath{\mathrm{MACE}_{2}}, with a tensor-memory peak
essentially equal to that of \ensuremath{\mathrm{MACE}_{1}}.
Its force RMSE of \ensuremath{7.4\,\mathrm{meV}\,\text{\AA}^{-1}} for seed~\num{17} is higher than the
\ensuremath{5.0\,\mathrm{meV}\,\text{\AA}^{-1}} obtained with full $z$-mixing, but it reproduces the wetting
preferences equally well, within \ensuremath{5\,\mathrm{meV}} of DFT
(Appendix~\ref{app:slab-mixing-comparison}).
A low-rank factorization of the full channel--height mixing matrices could
also reduce the cost of full $z$-mixing, which we leave for future work.

\section{Conclusions and outlook}

We have introduced FNO augmentation as a distinct route to nonlocal MLIPs
by learning the global coupling between latent source fields.
Added as a nonlocal correction term to a complete local backbone,
the learned operator captures long-range interactions accurately beyond the local
receptive field, and, where a reference kernel is known, the learned response
can be compared directly with it through spectral analysis in reciprocal space.

A central strength of our framework is its flexibility in
describing long-range interactions without adhering to a specific kernel.
A prescribed Coulomb kernel is a useful prior when the residual
interaction is largely electrostatic, but its predictive power becomes
limited once the effective interaction between the sources departs from
the Coulomb form, e.g., through screening set by the environment.
The learned operator overcomes this limitation by adapting its wavevector dependence to the
interaction at hand.
On a synthetic ionic system with tunable screened Coulomb interactions,
the FNO recovers
\ensuremath{95\text{--}99\,\%}
of the variance of the residual left by the local MACE backbone and,
with joint training,
consistently outperforms a prescribed-kernel long-range model and a
deeper local model at all three screening lengths tested.
It further reproduces the long-wavelength
charge--charge structure factor of the ionic fluid in both
the nearly Coulombic and the strongly screened regimes.
Note that low energy and force errors do not by themselves guarantee
correct long-wavelength charge fluctuations, which underlie dielectric
response and ion solvation.
Reproducing these correlations in its own sampled ensemble without explicitly
including the structure factor in the training indicates that the learned
operator reproduces the collective consequences of the residual interaction,
not only the training labels.

With inhomogeneous screening, the interaction can no longer be described by a
configuration-independent propagation kernel.
Among the models considered here, only the nonlinear FNO captures the
modulation of the coupling between distant ions by the remote screening medium,
which LES, CACE--SOG, and the linear FNO leave unchanged.
Its force RMSE is also
\ensuremath{27\text{--}41\,\%}
lower than those of these models.

We expect learned operators of this kind to be particularly useful where the
effective interaction between learned sources departs from the Coulomb form or
varies with the surrounding configuration.
Among the first-principles benchmarks, Au$_2$/MgO(001) provides an example,
in which remote dopants beyond the reach of the local backbone reverse the
wetting preference of the adsorbed dimer through charge
transfer~\cite{mammen2011morphology,mammen2018wetting}.
The FNO captures this reversal more accurately than both LES and a deeper local
backbone.
For liquid water and LLZO, FNO augmentation also substantially lowers the
errors of the one-interaction backbone, but reaches an accuracy comparable to
that of LES, consistent with environment-dependent charges partly representing
the screening from electronic polarization~\cite{grasselli2026electrostatics}.
Note that these first-principles benchmarks are restricted to static
energies and forces.

Beyond the present benchmarks, learned operators could thus be relevant to
electrolytes and solvated interfaces near electrodes, doped and defective
oxides and semiconductors, and metal--insulator interfaces where charge
transfer and image charges reshape the long-range response.
Finally, connecting the learned response to a physically defined polarization
could enable predictions of Born effective charges through its derivatives
with respect to atomic positions~\cite{gonze1997bec}.
The present MACE--\ensuremath{\mathrm{FNO}} implementation does not identify physical charge densities
in the latent fields.
An extension would require a consistent polarization definition and additional physical
constraints or response supervision.
These observables could both constrain the learned operator and test its
description of nonlocal polarization response.

\begin{acknowledgments}
Computational resources have been provided by the supercomputing facilities
of UCLouvain (CISM/UCL) and the Consortium des \'{E}quipements de Calcul Intensif en F\'ed\'eration Wallonie-Bruxelles (CECI),
funded by the Fonds de la Recherche Scientifique de Belgique (F.R.S.-FNRS)
under Convention 2.5020.11 and by the Walloon Region.
We also acknowledge computational resources made available on Lucia,
the Tier-1 supercomputer funded by the Walloon Region under grant agreement no.\ 1910247.
GPT-6 Astra (OpenAI) was used to assist with code prototyping and data curation under the direction of the authors,
who reviewed and verified the resulting changes.
\end{acknowledgments}

\section*{Data and code availability}

We provide benchmark datasets and trained interatomic potentials
on Zenodo at \url{https://doi.org/10.5281/zenodo.23459502}.
For the synthetic benchmark, the record includes the homogeneous Yukawa and heterogeneous-screening datasets,
scripts for data generation and reference evaluation, remote-medium probes, and MD trajectories.
The same record contains liquid-water, LLZO, and Au$_2$/MgO datasets along with the splits used in this work.
Model checkpoints and training configurations are included for all benchmarks.

The MACE--\ensuremath{\mathrm{FNO}} source code is available at \url{https://github.com/wch3n/mace_fno}.

\appendix
\section{FNO formulation and implementation}
\label{app:technical-formulation}

\subsection{Fourier representation of a nonlocal operator}

A neural operator maps one field to another, with parameters shared over
the spatial domain~\cite{kovachki2023neuraloperator}. Its Fourier
construction starts from a linear integral operator,
\begin{equation}
 (\mathcal K v)(\mathbf r)
 =\int_\Omega\kappa(\mathbf r,\mathbf r')v(\mathbf r')\,\mathrm d\mathbf r',
 \label{eq:integral-operator}
\end{equation}
where $v$ is the input field and $\kappa$ is the real-space coupling
between source and observation points. A general discretization requires
an $M\times M$ matrix for $M$ mesh points.
On a periodic domain, a translation-invariant kernel depends only on
$\mathbf r-\mathbf r'$ and defines a convolution.
Fourier waves then diagonalize the operator
\begin{equation}
 \begin{aligned}
 \mathcal K e^{i\mathbf k\cdot\mathbf r}
 &=\widehat\kappa(\mathbf k)e^{i\mathbf k\cdot\mathbf r},\\
 \widehat{\mathcal K v}(\mathbf k)
 &=R(\mathbf k)\widehat v(\mathbf k),\qquad
 R(\mathbf k)=\widehat\kappa(\mathbf k),
 \end{aligned}
 \label{eq:fourier-multiplier}
\end{equation}
where $\widehat v=|\Omega|^{-1}\int_\Omega v e^{-i\mathbf k\cdot\mathbf r}
\,\mathrm d\mathbf r$,
and $\widehat\kappa$ is the unnormalized kernel integral.
The volume factor is absorbed into $R$ in this convention.

An FNO learns the multiplier within a selected spectral band~\cite{li2021fno}.
For multiple channels, $R$ becomes a matrix that mixes channel coefficients
at the same wavevector. The operation remains nonlocal because each
coefficient and Fourier basis function spans the complete domain.
For example, the homogeneous three-dimensional Coulomb kernel gives
$R(k)=4\pi/k^2$ for $k\ne0$, with boundary conditions and the zero mode
treated separately. Our learned multipliers are not constrained to this form.

On a mesh with dimensions $N_1\times\cdots\times N_d$, we define
\begin{equation}
 \widehat v_a(\mathbf n)=
 \sum_{\mathbf m}v_a(\mathbf m)
 \exp\!\left[-2\pi i\sum_{\alpha=1}^d
 \frac{n_\alpha m_\alpha}{N_\alpha}\right],
 \label{eq:discrete-fourier}
\end{equation}
where $a$ labels channels, $\mathbf m$ labels mesh points, and
$\mathbf n$ labels integer reciprocal modes.
The inverse transform contains $1/M$, with $M=\prod_\alpha N_\alpha$.
Our operator maps $C$ source fields to $H_f$ hidden fields, applies
$L_f$ Fourier layers, and projects back to $C$ response fields
\begin{equation}
 \begin{aligned}
 \mathbf v_0&=P[\boldsymbol\rho;\mathbf q_{\mathbf H}],\\
 \mathbf v_{\ell+1}
 &=\mathrm{GELU}\!\left[
 W_\ell\mathbf v_\ell+
 \mathcal F^{-1}\{R_\ell\widehat{\mathbf v}_\ell\}\right],\\
 \boldsymbol\phi&=Q_2\,\mathrm{GELU}(Q_1\mathbf v_{L_f}),
 \end{aligned}
 \label{eq:fno-stack}
\end{equation}
where $P$ forms the initial hidden fields from the source channels and,
when included, the cell features $\mathbf q_{\mathbf H}$.
These features have the same values at every mesh point and are defined in
Appendix~\ref{app:metric-eqgino}.
The bracket groups the inputs to $P$.
A Fourier layer combines channel-wise mixing through $R_\ell$ with local
channel mixing through $W_\ell$, followed by the GELU activation.
The learned maps $Q_1$ and $Q_2$ convert the final hidden fields into the
$C$ response fields.

The selected Fourier modes determine which wavelengths enter the spectral
branch, while the local branch acts without this spectral truncation.
The source and hidden channels provide multiple learned fields, without
assigned chemical or electrostatic identities.
Different Fourier modes are coupled through GELU,
and the full operator is not generally described by a single linear convolution.
For slabs, Eq.~\eqref{eq:fno-stack} is supplemented by explicit height
mixing, as described in Appendix~\ref{sec:geometries}.

\subsection{Invariant sources and particle--mesh assignment}

For atom $i$, the descriptor $\mathbf h_i$ collects the scalar ($0e$) product
outputs of the MACE interaction layers independent of orientation.
The source head converts these features into source amplitudes
\begin{equation}
 \begin{aligned}
 \widetilde{\mathbf s}_i
 &=A_2\,\mathrm{SiLU}\!\left[A_1\,\mathrm{LN}(\mathbf h_i)+\mathbf b_1\right],\\
 \mathbf s_i&=\widetilde{\mathbf s}_i
 -\frac{1}{N_g}\sum_{j\in g}\widetilde{\mathbf s}_j,
 \end{aligned}
 \label{eq:neutral-source}
\end{equation}
where $g$ labels a structure of $N_g$ atoms,
and layer normalization $\mathrm{LN}$ rescales the descriptors with learned
scale and shift parameters.
The matrix $A_1$ and bias $\mathbf b_1$, followed by the sigmoid linear unit
(SiLU) activation, produce \num{64} hidden features, which the bias-free matrix
$A_2$ maps to the $C$ source amplitudes.
Subtracting the mean makes $\sum_{i\in g}s_{ic}=0$ in each channel,
removing its uniform, zero-wavevector component.

We spread each amplitude smoothly over nearby mesh points using cubic
B-spline weights $w_i(\mathbf m)$, distributing it over \num{64} points
in bulk and slab models.
The finite-height treatment is given in Appendix~\ref{sec:geometries}.
Summing the atomic contributions gives
\begin{equation}
 \begin{aligned}
 \rho_{gc}(\mathbf m)
 &=\frac{1}{\Delta\Omega_g}\sum_{i\in g}s_{ic}w_i(\mathbf m),\\
 \Delta\Omega_g\sum_{\mathbf m}\rho_{gc}(\mathbf m)
 &=\sum_{i\in g}s_{ic}=0,
 \end{aligned}
 \label{eq:latent-source-field}
\end{equation}
where $\Delta\Omega_g$ is the voxel volume.
The weights sum to one and are continuous through their second derivatives.
This assignment follows particle--mesh electrostatics~\cite{darden1993pme,essmann1995pme},
but the amplitudes and fields remain latent variables.

\subsection{Energy derivatives, forces, and backpropagation}

We sum the products of sources and responses
$\boldsymbol\phi=\mathcal G_\omega[\boldsymbol\rho;\mathbf H]$ over channels
and mesh points to obtain the residual energy
\begin{equation}
 \Delta E_{g,\mathrm{FNO}}
 =\frac{\Delta\Omega_g}{2}
 \sum_{c,\mathbf m}\rho_{gc}(\mathbf m)\phi_{gc}(\mathbf m).
 \label{eq:mesh-energy}
\end{equation}
This energy vanishes for zero sources but is not generally quadratic in
$\boldsymbol\rho$, as the response is nonlinear.
Exact energy scaling under cell replication also depends on the mode band,
cell features, and source mean subtraction, not just the volume factor.

Changing the source fields changes both factors in the energy expression.
At fixed cell, differentiation therefore gives
\begin{equation}
 \frac{\delta\Delta E_{\mathrm{FNO}}}{\delta\boldsymbol\rho}
 =\frac12\left[
 \boldsymbol\phi+J_{\mathcal G}(\boldsymbol\rho)^\dagger\boldsymbol\rho
 \right],
 \label{eq:energy-functional-derivative}
\end{equation}
where $J_{\mathcal G}$ describes how the response changes with the sources,
and $\dagger$ denotes its adjoint under the spatial inner product.
For a linear self-adjoint map, this expression reduces to $\boldsymbol\phi$.
We impose neither this property nor Poisson's equation.

Forces follow by differentiating $E_{\mathrm{MACE}}+\Delta E_{\mathrm{FNO}}$
with respect to atomic positions.
Moving an atom changes both where its source is deposited and the source
amplitudes obtained from the MACE descriptors
\begin{equation}
 \frac{\partial\boldsymbol\rho}{\partial\mathbf R_i}
 =\left.\frac{\partial\boldsymbol\rho}{\partial\mathbf R_i}\right|_{\mathbf s}
 +\sum_q
 \frac{\partial\boldsymbol\rho}{\partial\mathbf h_q}
 \frac{\partial\mathbf h_q}{\partial\mathbf R_i}.
 \label{eq:force-paths}
\end{equation}
The first term changes the deposition weights and slab height window at
fixed amplitudes.
The second follows changes in the descriptors, source
head outputs, and channel means, with $q$ indexing atoms.
Both terms are needed with frozen MACE, since its descriptors still
depend on atomic positions.
In joint training, changing the MACE parameters $\theta_{\mathrm M}$
affects both the local energy and the descriptors supplied to the FNO
\begin{equation}
 \frac{\partial E}{\partial\theta_{\mathrm M}}
 =\frac{\partial E_{\mathrm{MACE}}}{\partial\theta_{\mathrm M}}
 +\frac{\partial\Delta E_{\mathrm{FNO}}}{\partial\boldsymbol\rho}
  \frac{\partial\boldsymbol\rho}{\partial\mathbf h}
  \frac{\partial\mathbf h}{\partial\theta_{\mathrm M}}.
 \label{eq:joint-mace-gradient}
\end{equation}
Backpropagation combines these contributions, including derivatives through
the force calculation, to update jointly the local model and FNO.

\subsection{Geometry-specific operators}
\label{sec:geometries}

\subsubsection{Metric-aware EqGINO for bulk and slabs}
\label{app:metric-eqgino}

An unconstrained FNO learns independent channel-mixing matrices for different
wavevectors, so that rotating a field need not rotate the predicted response.
Following EqGINO~\cite{kim2026eqgino}, we impose isotropic weight sharing, in
which modes of the same wavelength share a matrix,
$R_\ell(\mathbf k)=\mathsf R_\ell(|\mathbf k|^2)$, with $|\mathbf k|$ the
in-plane wavevector magnitude for slabs.
Here we adopt this spectral principle in place of the full EqGINO
encoder--decoder architecture.
The response transforms with the input field under rotations and
reflections that preserve the cell, mesh, and selected mode band.
For slabs, this statement applies to in-plane transformations.
This construction does not guarantee equivariance under arbitrary
rotations on a fixed mesh or invariance under arbitrary changes
of lattice basis.
Isotropy constrains the spectral matrices.
The linearized response of the nonlinear operator about a given atomic
configuration can nevertheless be anisotropic, as the spatial
arrangement of the sources can break rotational symmetry.

Since the same integer Fourier index represents different wavelengths in cells
of different sizes or shapes, we evaluate the learned matrices at physical
wavevector magnitudes.
Their values are learned at the reciprocal-space shells of a fixed cubic or
square reference cell and interpolated by a natural cubic spline in
$|\mathbf k|^2$, which reduces to independent shell weights at the reference
cell.
Outside the knot interval, we extrapolate linearly along the endpoint tangents.

All systems in this work use one spectral channel group ($G=1$).
In the nonlinear FNO, the local matrix $W_\ell$ also mixes all channels.
For liquid water, the cell features $\mathbf q_{\mathbf H}$ supply $\ln L$ with
$L=|\det\mathbf H|^{1/3}$ in \unit{\angstrom}, and LLZO additionally receives the six
independent components of $\mathbf H\mathbf H^{\mathsf T}/L^2$ to describe the
cell shape.
With the lattice vectors stored as rows of $\mathbf H$, these features
are constant across the mesh and invariant under rigid rotation
of the whole cell.
The fixed-cell slab has no cell features.

\subsubsection{Slab geometry and full \texorpdfstring{$z$}{z}-mixing}
\label{app:slab-full-height}

The full $z$-mixing in Eq.~\eqref{eq:main-slab-spectral-mixing}
treats the hidden channel and height as a compound index $(a,j)$.
With $H_f=16$ and $N_z=24$, each Fourier layer learns a
$384\times384$ matrix at each radial shell.
We interpolate its entries in $|\mathbf k_\parallel|^2$ using the same
shell-spline construction as for the bulk spectral matrices.
The pointwise matrix $W_\ell$ remains $H_f\times H_f$ and is shared across
all mesh points.

Channel-wise $z$-mixing [Eq.~\eqref{eq:main-slab-global-mixing}]
instead combines an $H_f\times H_f$ matrix at each radial shell with
one wavevector-independent $N_z\times N_z$ height matrix per hidden channel.
For one spectral group and $N_{\mathrm{sh}}$ radial shells, including
the zero mode, the numbers of mixing parameters per Fourier layer are
\begin{equation}
 \begin{aligned}
 P_{\mathrm{full}} &= N_{\mathrm{sh}}(H_fN_z)^2,\\
 P_{\mathrm{cw}} &= N_{\mathrm{sh}}H_f^2+H_fN_z^2.
 \end{aligned}
 \label{eq:slab-mixing-parameters}
\end{equation}
With $H_f=16$, $N_z=24$, and $N_{\mathrm{sh}}=34$ for $m=8$,
these counts are \num{5013504} and \num{17920}, respectively,
a factor of about \num{280}.
Both formulations also have $H_f^2$ pointwise weights per layer and
identical source, lifting, and projection maps.
Appendix~\ref{app:slab-mixing-comparison} compares their accuracy and
the measured GPU cost of the complete models.

For Au$_2$/MgO(001), the \num{24} height points span a fixed
\ensuremath{22\,\text{\AA}} window centered on the instantaneous mean atomic height.
Deposition weights extending beyond the height window are assigned to
the nearest boundary grid point, preserving the total deposited source.
The height window is chosen to ensure that all atoms are enclosed.

\subsection{Training objective and model selection}
\label{app:training-selection}

For a batch of $B$ structures containing $N_{\mathrm{tot}}$ atoms, we minimize
\begin{equation}
 \begin{aligned}
 \mathcal L_{\mathrm{train}}={}&
 \frac{w_E}{B}\sum_{g=1}^{B}
 \left[
 \frac{E_g^{\mathrm{pred}}-E_g^{\mathrm{ref}}}{N_g\sigma_E}
 \right]^2\\
 &+\frac{w_F}{3N_{\mathrm{tot}}}\sum_{i,\alpha}
 \left[
 \frac{F_{i\alpha}^{\mathrm{pred}}-F_{i\alpha}^{\mathrm{ref}}}{\sigma_F}
 \right]^2,
 \end{aligned}
 \label{eq:training-loss}
\end{equation}
where $E_g$ is the total energy, $N_g$ the atom count, and $F_{i\alpha}$
the force component $\alpha$ on atom $i$.
The scales $\sigma_E$ and $\sigma_F$ balance the two errors, with
$w_E=w_F=1$ in every fit.
For the synthetic ionic system, $\sigma_F=\ensuremath{1.0\,\mathrm{eV}\,\text{\AA}^{-1}}$
and $\sigma_E=\ensuremath{0.65\,\mathrm{eV}\,\mathrm{atom}^{-1}\text{, }0.50\,\mathrm{eV}\,\mathrm{atom}^{-1}\text{, and }0.40\,\mathrm{eV}\,\mathrm{atom}^{-1}}$ for
$\kappa=\ensuremath{0.05\,\text{\AA}^{-1}\text{, }0.2\,\text{\AA}^{-1}\text{, and }0.5\,\text{\AA}^{-1}}$, respectively.
Liquid water and LLZO have
$\sigma_E=\ensuremath{0.01\,\mathrm{eV}\,\mathrm{atom}^{-1}}$ and
$\sigma_F=\ensuremath{0.10\,\mathrm{eV}\,\text{\AA}^{-1}}$.
For Au$_2$/MgO, the one-interaction baseline training errors set
$\sigma_E=\ensuremath{0.002\,260\,\mathrm{eV}\,\mathrm{atom}^{-1}}$ and
$\sigma_F=\ensuremath{0.056\,280\,\mathrm{eV}\,\text{\AA}^{-1}}$.
The same scales apply to both backbone depths.

We select FNO checkpoints by minimizing the validation score
\begin{equation}
 \mathcal L_{\mathrm{val}}
 =w_E(\ensuremath{\mathrm{RMSE}}_E/\sigma_E)^2+w_F(\ensuremath{\mathrm{RMSE}}_F/\sigma_F)^2,
 \label{eq:validation-objective}
\end{equation}
where the validation RMSEs use energies per atom and Cartesian force components.
We validate every \num{100} steps for the synthetic system and every
\num{500} steps for the materials benchmarks.
Joint training selects the backbone and FNO parameters from the same step.

The source head and FNO start with an \texttt{Adam} learning rate of \num{1e-3}
for the synthetic ionic system and \num{3e-4} for the materials benchmarks.
The learning rate is halved after four validation evaluations without improvement, down to \num{1e-6}.
The synthetic FNO runs use eight structures per training batch,
and step budgets of \num{100000} and \num{200000} for frozen and joint training, respectively.
For the first-principles benchmarks, both FNO variants use the step budgets in
Table~\ref{tab:hyperparameter-convergence}, except for joint training on the
one-interaction water backbone, which uses \num{200000} steps.
The effective batch sizes are two structures for liquid water and
four for LLZO and Au$_2$/MgO(001).
For nonlinear full $z$-mixing, we accumulate gradients over four single-structure
batches to limit peak memory, preserving this effective batch size.
The weights are fixed during the MACE warm-up.
Shared implementation and training choices are given in
Sec.~\ref{sec:implementation} of the main text.
System-specific settings accompany the benchmarks in
Secs.~\ref{sec:synthetic-results} and~\ref{sec:physical-results}.

The local MACE baselines use validation-selected stage-two checkpoints
throughout.
The exception is \ensuremath{\mathrm{MACE}_{2}} at $\kappa=\ensuremath{0.05\,\text{\AA}^{-1}}$, for which
the stage-one checkpoint gives lower validation energy and force RMSEs.
For the first-principles benchmarks, the local models trained here use
energy--force weights of \num{1}:\num{100} and \num{1000}:\num{100},
and initial learning rates of \num{0.01} and \num{0.001}, in stages one
and two, respectively.
We use \texttt{AMSGrad} and an exponential moving average with decay \num{0.99}.
Element-specific energy offsets are obtained by least-squares fitting
to the training energies and held fixed during model training.
The batch sizes are four structures for water and LLZO and ten for
Au$_2$/MgO(001), with epoch budgets of \num{400}, \num{1000}, and
\num{1000}, respectively. Stage two begins halfway through each budget.
For Au$_2$/MgO(001), we train only \ensuremath{\mathrm{MACE}_{2}} and use the released
\ensuremath{\mathrm{MACE}_{1}} checkpoint~\cite{kim2025universal}.
Table~\ref{tab:materials-results} reports raw test
energy RMSEs without an offset correction.

For LES, we select checkpoints by the stage-two validation loss.
The synthetic models use a \num{1200}-epoch budget.
For the first-principles benchmarks, the loss weights, learning rates, and optimizer
settings match the local MACE fits.
Atomic energy offsets are fitted as for the local MACE models.
The batch sizes are four structures for water and LLZO and ten for Au$_2$/MgO(001).
All three systems use a \num{1000}-epoch budget,
with stage two starting halfway through training.
These LES models use a Gaussian width of \ensuremath{1\,\text{\AA}},
mesh-spacing parameter \texttt{dl}=\ensuremath{2\,\text{\AA}}, and self-interaction
removal.

\subsection{Spectral-response diagnostic}
\label{app:spectral-response-diagnostic}

For the diagnostic in Sec.~\ref{sec:spectral-diagnostic-method}, we keep
atoms, descriptors, and model parameters fixed and perturb the deposited fields.
For a nonzero integer mode $\mathbf n$, the normalized cosine probe is
\begin{equation}
 \begin{aligned}
 c_{\mathbf n\mathbf m}
 &=\cos(\mathbf k_{\mathbf n}\cdot\mathbf r_{\mathbf m}),\\
 f_{\mathbf n}(\mathbf r_{\mathbf m})
 &=\frac{c_{\mathbf n\mathbf m}-\langle c_{\mathbf n\mathbf m}\rangle}
 {\langle(c_{\mathbf n\mathbf m}-\langle c_{\mathbf n\mathbf m}\rangle)^2\rangle^{1/2}}.
 \end{aligned}
 \label{eq:diagnostic-perturbed-field}
\end{equation}
Angle brackets average over mesh points $\mathbf r_{\mathbf m}$, giving a
probe with zero mean and RMS equal to one.
For cell vectors stored as rows of $\mathbf H$, the physical wavevector is
$\mathbf k_{\mathbf n}=2\pi\mathbf H^{-1}\mathbf n$.
The equivalent cosine modes $\pm\mathbf n$ are counted once.

Writing $E_{\mathbf n}(\mathbf a)=
\Delta E_{\mathrm{FNO}}[\boldsymbol\rho+f_{\mathbf n}\mathbf a]$,
we evaluate the Hessian from positive and negative perturbations.
Let $\mathbf e_c$ select channel $c$, with $E_0=E_{\mathbf n}(\mathbf0)$,
$E_c^\pm=E_{\mathbf n}(\pm\epsilon\mathbf e_c)$, and
$E_{cc'}^{st}=E_{\mathbf n}[\epsilon(s\mathbf e_c+t\mathbf e_{c'})]$
for $s,t\in\{-1,1\}$. Central differences give
\begin{equation}
 \begin{aligned}
 H_{cc}&\simeq
 \frac{E_c^++E_c^--2E_0}{\mathcal V_{\mathrm{eff}}\epsilon^2},\\
 H_{cc'}&\simeq
 \frac{E_{cc'}^{++}+E_{cc'}^{--}-E_{cc'}^{+-}-E_{cc'}^{-+}}
 {4\mathcal V_{\mathrm{eff}}\epsilon^2}\quad(c\ne c'),
 \end{aligned}
 \label{eq:diagnostic-hessian-differences}
\end{equation}
where $\mathcal V_{\mathrm{eff}}$ is defined in
Sec.~\ref{sec:spectral-diagnostic-method}.
We set the perturbation amplitude $\epsilon$ to a fraction $\eta$ of the
typical source-field magnitude, measured by its RMS over all channels and
mesh points.
For the synthetic benchmark, we use $\eta=\num{0.05}$,
i.e., \ensuremath{5\,\%} of this magnitude.

We diagonalize the Hessians independently at the sampled wavevectors before averaging their positive
$\lambda_+$ over shells.
Modes with no positive eigenvalue are recorded but excluded from the
logarithmic fit, without discarding any model or structure.
Equal-magnitude shells have relative tolerance \num{2e-6}
and absolute tolerance $\ensuremath{10^{-10}\,\text{\AA}^{-1}}$.

For a given structure, we fit the logarithms of the finite, positive shell
averages by unweighted least squares with a free intercept.
The negative slope gives $p$, with equal weights across shells.
We first average the structure-level exponents separately for the trained
models, then report the mean and sample standard deviation across training seeds.
Eigenvalues depend on the learned channel basis, so their scaling alone
does not identify physical charges.

\section{Numerical validation and computational cost}
\label{app:numerical-validation-scaling}

\subsection{Mesh interlacing and translation sensitivity}
\label{app:interlacing-comparison}
\begin{table*}[t]
\centering
\caption{Effect of mesh interlacing on \mbox{\ensuremath{\mathrm{MACE}_{1}^{(0)}}--\ensuremath{\mathrm{FNO}}} and \ensuremath{\mathrm{MACE}_{1}}--\ensuremath{\mathrm{FNO}} with fixed inference weights.
$T_E$ and $T_F$ measure translation sensitivity. RMSEs are relative to DFT.
Costs refer to energy--force inference.}
\label{tab:interlacing-comparison}
\small
\begin{ruledtabular}
\begin{tabular}{lrrrrrrr}
Model & Mesh origins & $T_E$ & $T_F$ & Energy RMSE & Force RMSE & Time & Tensor peak\\
 & & (\unit{\milli\electronvolt\per\atom}) & (\unit{\milli\electronvolt\per\angstrom}) & (\unit{\milli\electronvolt\per\atom}) & (\unit{\milli\electronvolt\per\angstrom}) & (\unit{\milli\second}) & (\unit{\mebibyte})\\
\colrule
\multicolumn{8}{l}{Liquid water}\\
\mbox{\ensuremath{\mathrm{MACE}_{1}^{(0)}}--\ensuremath{\mathrm{FNO}}} & 1 & \num{0.058} & \num{9.573} & \num{0.76} & \num{82.12} & \num{40} & \num{1400}\\
\mbox{\ensuremath{\mathrm{MACE}_{1}^{(0)}}--\ensuremath{\mathrm{FNO}}} & 8 & \num{0.001} & \num{0.463} & \num{0.77} & \num{81.85} & \num{44} & \num{1400}\\
\ensuremath{\mathrm{MACE}_{1}}--\ensuremath{\mathrm{FNO}} & 1 & \num{0.059} & \num{9.496} & \num{0.66} & \num{71.87} & \num{40} & \num{1400}\\
\ensuremath{\mathrm{MACE}_{1}}--\ensuremath{\mathrm{FNO}} & 8 & \num{0.001} & \num{0.453} & \num{0.67} & \num{71.58} & \num{44} & \num{1400}\\[0.4em]
\multicolumn{8}{l}{LLZO}\\
\mbox{\ensuremath{\mathrm{MACE}_{1}^{(0)}}--\ensuremath{\mathrm{FNO}}} & 1 & \num{0.045} & \num{3.815} & \num{0.35} & \num{40.36} & \num{32} & \num{305}\\
\mbox{\ensuremath{\mathrm{MACE}_{1}^{(0)}}--\ensuremath{\mathrm{FNO}}} & 8 & \num{0.001} & \num{0.388} & \num{0.35} & \num{40.29} & \num{41} & \num{985}\\
\ensuremath{\mathrm{MACE}_{1}}--\ensuremath{\mathrm{FNO}} & 1 & \num{0.021} & \num{2.571} & \num{0.22} & \num{25.01} & \num{32} & \num{305}\\
\ensuremath{\mathrm{MACE}_{1}}--\ensuremath{\mathrm{FNO}} & 8 & \num{0.001} & \num{0.232} & \num{0.22} & \num{24.95} & \num{41} & \num{985}\\[0.4em]
\multicolumn{8}{l}{Au$_2$/MgO(001)}\\
\mbox{\ensuremath{\mathrm{MACE}_{1}^{(0)}}--\ensuremath{\mathrm{FNO}}} & 1 & \num{0.281} & \num{11.579} & \num{0.24} & \num{12.99} & \num{32} & \num{462}\\
\mbox{\ensuremath{\mathrm{MACE}_{1}^{(0)}}--\ensuremath{\mathrm{FNO}}} & 4 & \num{0.011} & \num{0.461} & \num{0.07} & \num{8.88} & \num{33} & \num{463}\\
\ensuremath{\mathrm{MACE}_{1}}--\ensuremath{\mathrm{FNO}} & 1 & \num{0.311} & \num{10.496} & \num{0.27} & \num{11.52} & \num{32} & \num{462}\\
\ensuremath{\mathrm{MACE}_{1}}--\ensuremath{\mathrm{FNO}} & 4 & \num{0.009} & \num{0.422} & \num{0.07} & \num{7.37} & \num{33} & \num{463}\\[0.4em]
\end{tabular}
\end{ruledtabular}
\end{table*}

Mesh discretization can make the predicted energy depend on the placement
of the grid relative to the atoms.
To reduce this dependence, we average FNO energies over shifted mesh origins
for a given set of atomic positions.
For $n_\alpha$ mesh points along a periodic cell vector $\mathbf a_\alpha$,
the mesh spacing is $\mathbf a_\alpha/n_\alpha$.
The allowed offsets are zero or half this spacing,
$\mathbf a_\alpha/(2n_\alpha)$, chosen independently along the $d$ periodic directions, giving
\begin{equation}
 \Delta E_{\mathrm{int}}(\mathbf R)
 =2^{-d}\sum_{\boldsymbol\delta\in\{0,1/2\}^d}
 \Delta E(\mathbf R;\boldsymbol\delta),
 \label{eq:slab-interlacing}
\end{equation}
where $\mathbf R=\{\mathbf R_i\}$ denotes the atomic positions and
$\boldsymbol\delta$ specifies offsets in units of the mesh spacings.
The quantity $\Delta E(\mathbf R;\boldsymbol\delta)$ is the residual energy
evaluated on that mesh.
We redeposit the sources and evaluate the same FNO separately for the different mesh origins.
There are four origins for a slab ($d=2$) and eight for periodic bulk ($d=3$).

We compare the effect of interlacing using the same validation-selected, seed-\num{17}
frozen and joint \ensuremath{\mathrm{MACE}_{1}}--\ensuremath{\mathrm{FNO}} checkpoints with the hyperparameters in
Table~\ref{tab:computational-settings}.
The Au$_2$/MgO checkpoints use the channel-wise height mixing
[Eq.~\eqref{eq:main-slab-global-mixing}].
The trained weights remain unchanged.
Energy and force RMSEs are evaluated using the full test sets covering \num{134} water, \num{200} LLZO, and \num{500}
Au$_2$/MgO structures.

For each benchmark system, we additionally sample \num{32} structures without replacement
and apply \num{16} rigid translations per structure in order to study the
translation sensitivity.
For a given structure $s$ associated with atomic positions $\mathbf R_s$,
we apply the same displacement $\mathbf t_{sj}$ to every atom for translation $j$, relative to
the fixed mesh origins.
The shift is in-plane for a slab.
We then evaluate
\begin{equation}
 \begin{aligned}
 \Delta_t E_s(j)&=E(\mathbf R_s+\mathbf t_{sj})-E(\mathbf R_s),\\
 \Delta_t F_{si\alpha}(j)
 &=F_{i\alpha}(\mathbf R_s+\mathbf t_{sj})-F_{i\alpha}(\mathbf R_s),\\
 T_E&=\operatorname{RMS}_{s,j}[\Delta_t E_s(j)/N_s],\\
 T_F&=\operatorname{RMS}_{s,j,i,\alpha}[\Delta_t F_{si\alpha}(j)],
 \end{aligned}
 \label{eq:mesh-translation-statistics}
\end{equation}
where $i$ and $\alpha$ label atoms and Cartesian components, respectively,
and $N_s$ is the atom count.

Table~\ref{tab:interlacing-comparison} shows reduced translation sensitivity
for every checkpoint, by factors of
\numrange{24}{45} for $T_E$
and \numrange{10}{25} for $T_F$.
Only the slab shows appreciable DFT-RMSE improvement.
Bulk changes stay below \ensuremath{0.005\,\mathrm{meV}\,\mathrm{atom}^{-1}} and
\ensuremath{0.30\,\mathrm{meV}\,\text{\AA}^{-1}}.

The interlacing cost measurements follow the protocol in Sec.~\ref{sec:inference-cost}.
We process the shifted meshes as a single FNO batch, while the MACE
backbone and source head run only once per structure.
Graph construction and result transfers are also shared across mesh origins.
Latency increases by about \ensuremath{9\,\%} for water, \ensuremath{28\,\%} for
LLZO, and \ensuremath{2\,\%} for Au$_2$/MgO.
Peak tensor memory changes by less than \ensuremath{1\,\%} for water and
Au$_2$/MgO and increases by \ensuremath{223\,\%} for LLZO.

\begin{table*}[t]
\centering
\caption{Validation RMSEs of \mbox{\ensuremath{\mathrm{MACE}_{1}^{(0)}}--\ensuremath{\mathrm{FNO}}} for the synthetic ionic system.
$N$ is the number of mesh points per direction ($N^3$ points in total), $m$ the mode parameter,
$C$ the number of source channels, and $L$ the number of Fourier layers.
The first row gives the production settings of
Table~\ref{tab:computational-settings}, followed by the tested variants.
All fits share the frozen seed-\num{17} backbone at a given $\kappa$ and a \num{100000}-step budget.
RMSEs are evaluated at validation-selected checkpoints. Multi-seed entries give means and sample standard deviations.
Energy and force RMSEs are in \unit{\milli\electronvolt\per\atom} and \unit{\milli\electronvolt\per\angstrom}, respectively.
Seed counts follow the order of the three $\kappa$ columns. Dashes mark untested combinations.}
\label{tab:synthetic-hyperparameters}
\small
\setlength{\tabcolsep}{3pt}
\begin{ruledtabular}
\begin{tabular}{crrrrrrrrrl}
 & & & & \multicolumn{2}{c}{$\kappa=\ensuremath{0.05\,\text{\AA}^{-1}}$} & \multicolumn{2}{c}{$\kappa=\ensuremath{0.2\,\text{\AA}^{-1}}$} & \multicolumn{2}{c}{$\kappa=\ensuremath{0.5\,\text{\AA}^{-1}}$} & \\[0.3em]
\cline{5-6}\cline{7-8}\cline{9-10}
$N$ & $m$ & $C$ & $L$ & Energy & Forces & Energy & Forces & Energy & Forces & Seeds\\[0.3em]
\colrule
\num{44} & \num{11} & \num{4} & \num{2} & \num{0.12 +- 0.01} & \num{16.13 +- 0.15} & \num{0.14 +- 0.00} & \num{15.77 +- 0.22} & \num{0.14 +- 0.01} & \num{13.39 +- 0.16} & 5/5/5\\[0.3em]
\num{44} & \num{11} & \num{8} & \num{2} & \num{0.13 +- 0.02} & \num{16.07 +- 0.31} & \num{0.15} & \num{15.79} & \num{0.14} & \num{13.27} & 3/1/1\\
\num{44} & \num{11} & \num{16} & \num{2} & \num{0.12 +- 0.01} & \num{15.71 +- 0.10} & \num{0.13} & \num{15.53} & \num{0.14} & \num{13.09} & 3/1/1\\
\num{44} & \num{11} & \num{4} & \num{4} & \num{0.13 +- 0.01} & \num{16.40 +- 0.18} & \num{0.14} & \num{16.10} & \num{0.14} & \num{13.54} & 5/1/1\\
\num{44} & \num{8} & \num{4} & \num{4} & \num{0.13} & \num{16.13} & \multicolumn{1}{c}{--} & \multicolumn{1}{c}{--} & \multicolumn{1}{c}{--} & \multicolumn{1}{c}{--} & 1/--/--\\
\num{32} & \num{11} & \num{4} & \num{4} & \num{0.13} & \num{16.80} & \multicolumn{1}{c}{--} & \multicolumn{1}{c}{--} & \multicolumn{1}{c}{--} & \multicolumn{1}{c}{--} & 1/--/--\\
\end{tabular}
\end{ruledtabular}
\end{table*}

\subsection{Hyperparameter sensitivity and convergence}
\label{app:hyperparameter-convergence}

We assess the mesh resolution $N$, the mode parameter $m$, the source-channel count $C$,
and the number of Fourier layers $L$ with a frozen \ensuremath{\mathrm{MACE}_{1}} backbone.
Tables~\ref{tab:synthetic-hyperparameters} and~\ref{tab:hyperparameter-convergence}
summarize the synthetic and physical benchmarks, respectively.
$N$ denotes the number of mesh points per periodic direction.
The slab height-grid size $N_z$ is specified separately.
The mode parameter $m$ includes integer indices
$n_\alpha=-(m-1),\ldots,m-1$ along the Fourier-transformed directions.
The complete parameter screens and model-selection records remain available
with the source data.

For the synthetic system, we compare the FNO settings of
Table~\ref{tab:computational-settings} ($m=11$, two Fourier layers, a
$44^3$ mesh, and $C=4$) with variants in which one or two of these settings
change.
All variants are trained with the frozen seed-\num{17} \ensuremath{\mathrm{MACE}_{1}} backbone at
the corresponding $\kappa$ and the \num{100000}-step protocol of the synthetic benchmark,
and Table~\ref{tab:synthetic-hyperparameters} reports the validation RMSEs at
the selected checkpoints.
Changes in the force RMSE are quoted as differences paired by FNO seed, with
the standard error of the mean difference.

None of the variants lowers the force RMSE by more than about
\ensuremath{3\,\%}.
Increasing the number of source channels from four to eight changes it by
\ensuremath{(0.0 \pm 1.1)\,\%} for
$\kappa=\ensuremath{0.05\,\text{\AA}^{-1}}$, whereas sixteen channels lower it by
\ensuremath{(2.2 \pm 0.5)\,\%}, and by
\ensuremath{2.0\text{--}2.6\,\%}
for single seeds at the other two $\kappa$, at about three times the training
cost per step.
Four Fourier layers increase the force RMSE by
\ensuremath{(1.7 \pm 0.6)\,\%}, and neither the lower mode
parameter nor the coarser mesh, both tested with four layers, improves on the
production setting.
The energy RMSEs show no systematic change across the variants, and the test
RMSEs give the same ranking.
The production setting thus represents a compromise between accuracy and cost,
with sixteen source channels as the only variant that lowers the force error
appreciably.

For liquid water, LLZO, and Au$_2$/MgO(001), Table~\ref{tab:hyperparameter-convergence}
compares one-parameter variations around the production settings.
We average over three paired FNO seeds at a fixed training budget within
each panel, using the same data splits and backbones as the first-principles benchmarks.
We select checkpoints with Eq.~\eqref{eq:validation-objective} and report
raw energy RMSEs without offset corrections.
All fits have hidden width $H_f=16$ and a source head with \num{64} hidden units.
The Au$_2$/MgO(001) convergence data are obtained with the channel-wise
height mixing in Eq.~\eqref{eq:main-slab-global-mixing}, using a fixed
\ensuremath{22\,\text{\AA}} window.
All fits use one spectral channel group ($G=1$).
Loss normalization and interlacing remain fixed within a system.

\begin{table*}[t]
\centering
\caption{Validation RMSEs of \mbox{\ensuremath{\mathrm{MACE}_{1}^{(0)}}--\ensuremath{\mathrm{FNO}}} for liquid water, LLZO, and Au$_2$/MgO(001).
$N$, $m$, $C$, and $L$ follow Table~\ref{tab:synthetic-hyperparameters}.
Bulk meshes have $N^3$ points, and slab meshes have $N\times N\times N_z$ points.
The first row in a panel gives the production values of the listed hyperparameters,
followed by one-parameter variations.
Entries give means and sample standard deviations over FNO seeds~\numlist{29;41;53}.
The backbone, data split, and training budget are fixed within each panel.
The Au$_2$/MgO(001) data use the channel-wise height mixing.
Energy and force RMSEs are in \unit{\milli\electronvolt\per\atom} and \unit{\milli\electronvolt\per\angstrom}, respectively,
    at the lowest-loss validation checkpoints.}
\label{tab:hyperparameter-convergence}
\footnotesize
\setlength{\tabcolsep}{1.5pt}
\begin{ruledtabular}
\begin{tabular}{rrrrrr@{\hspace{1em}}rrrrrr@{\hspace{1em}}rrrrrrr}
\multicolumn{6}{c}{(a) Liquid water} & \multicolumn{6}{c}{(b) LLZO} & \multicolumn{7}{c}{(c) Au$_2$/MgO(001)}\\
\multicolumn{6}{c}{\num{120000} steps} & \multicolumn{6}{c}{\num{160000} steps} & \multicolumn{7}{c}{\num{240000} steps}\\[0.3em]
$N$ & $m$ & $C$ & $L$ & Energy & Forces & $N$ & $m$ & $C$ & $L$ & Energy & Forces & $N$ & $N_z$ & $m$ & $C$ & $L$ & Energy & Forces\\[0.3em]
\colrule
\num{24} & \num{8} & \num{16} & \num{2} & \num{0.53 +- 0.03} & \num{74.55 +- 0.56} & \num{32} & \num{8} & \num{16} & \num{4} & \num{0.34 +- 0.00} & \num{39.13 +- 0.26} & \num{24} & \num{24} & \num{8} & \num{16} & \num{4} & \num{0.05 +- 0.00} & \num{8.76 +- 0.08}\\[0.3em]
\num{32} & \num{8} & \num{16} & \num{2} & \num{0.53 +- 0.01} & \num{75.13 +- 0.77} & \num{40} & \num{8} & \num{16} & \num{4} & \num{0.34 +- 0.01} & \num{39.45 +- 0.90} & \num{32} & \num{24} & \num{8} & \num{16} & \num{4} & \num{0.06 +- 0.00} & \num{9.11 +- 0.10}\\
\num{24} & \num{10} & \num{16} & \num{2} & \num{0.56 +- 0.02} & \num{76.24 +- 0.48} & \num{32} & \num{10} & \num{16} & \num{4} & \num{0.34 +- 0.01} & \num{40.09 +- 0.86} & \num{24} & \num{24} & \num{10} & \num{16} & \num{4} & \num{0.06 +- 0.00} & \num{8.95 +- 0.27}\\
\num{24} & \num{8} & \num{8} & \num{2} & \num{0.55 +- 0.02} & \num{74.36 +- 0.88} & \num{32} & \num{8} & \num{8} & \num{4} & \num{0.35 +- 0.02} & \num{39.93 +- 1.35} & \num{24} & \num{24} & \num{8} & \num{8} & \num{4} & \num{0.06 +- 0.00} & \num{9.32 +- 0.50}\\
\num{24} & \num{8} & \num{32} & \num{2} & \num{0.55 +- 0.06} & \num{74.11 +- 1.08} & \num{32} & \num{8} & \num{32} & \num{4} & \num{0.35 +- 0.02} & \num{39.26 +- 0.55} & \num{24} & \num{24} & \num{8} & \num{32} & \num{4} & \num{0.06 +- 0.00} & \num{8.80 +- 0.10}\\
\num{24} & \num{8} & \num{16} & \num{4} & \num{0.68 +- 0.05} & \num{77.06 +- 0.39} & \num{32} & \num{8} & \num{16} & \num{6} & \num{0.34 +- 0.01} & \num{38.75 +- 0.09} & \num{24} & \num{24} & \num{8} & \num{16} & \num{6} & \num{0.05 +- 0.00} & \num{8.54 +- 0.21}\\
 &  &  &  &  &  &  &  &  &  &  &  & \num{24} & \num{16} & \num{8} & \num{16} & \num{4} & \num{0.06 +- 0.00} & \num{9.53 +- 0.29}\\
 &  &  &  &  &  &  &  &  &  &  &  & \num{24} & \num{32} & \num{8} & \num{16} & \num{4} & \num{0.06 +- 0.00} & \num{8.57 +- 0.15}\\
\end{tabular}
\end{ruledtabular}
\end{table*}

Finer periodic meshes and a larger mode parameter give no systematic improvement.
Changing the source-channel count from \num{16} to \num{8} or \num{32}
also provides no consistent benefit.
Adding two Fourier layers lowers the mean force RMSE by about
\ensuremath{1\,\%} for LLZO and \ensuremath{3\,\%} for Au$_2$/MgO(001).
For water, the same increase in depth raises both mean RMSEs at the fixed training budget.
For the slab, increasing $N_z$ from \num{24} to \num{32} lowers the force
RMSE by about \ensuremath{2\,\%}, while reducing it to \num{16} increases
the error by about \ensuremath{9\,\%}.
These matched-budget comparisons support the production settings as a practical
accuracy--cost compromise, without establishing a unique converged optimum.

\subsection{Full and channel-wise \texorpdfstring{$z$}{z}-mixing}
\label{app:slab-mixing-comparison}

We compare channel-wise and dense full $z$-mixing for jointly trained
\ensuremath{\mathrm{MACE}_{1}}--\ensuremath{\mathrm{FNO}} models with seed $17$ (Table~\ref{tab:slab-mixing-comparison}).
Both use the same starting backbone, training and validation structures,
and four Fourier layers with $H_f=16$.
The mesh contains $24\times24\times24$ points, with $m=8$, $C=16$,
and four in-plane mesh origins.
Both models completed \num{240000} training steps.

\begin{table*}[t]
\centering
\caption{Accuracy and GPU cost of channel-wise and full $z$-mixing
for joint \ensuremath{\mathrm{MACE}_{1}}--\ensuremath{\mathrm{FNO}} on Au$_2$/MgO(001).
FNO parameter counts include the source head, and total counts also
include the backbone.}
\label{tab:slab-mixing-comparison}
\small
\setlength{\tabcolsep}{4pt}
\begin{ruledtabular}
\begin{tabular}{lrrrrrrrr}
 & \multicolumn{2}{c}{Test RMSE} & \multicolumn{2}{c}{Wetting preference (meV)} & \multicolumn{2}{c}{Parameters} & & \\
\cline{2-3}\cline{4-5}\cline{6-7}
Mixing & \shortstack{Energy\\(\unit{\milli\electronvolt\per\atom})} & \shortstack{Forces\\(\unit{\milli\electronvolt\per\angstrom})} & Undoped & Al-doped & FNO & Total & \shortstack{Runtime\\(ms)} & \shortstack{Peak tensor\\memory (MiB)}\\
\colrule
Channel-wise & \num{0.0725} & \num{7.37} & \num{930.47} & \num{-69.23} & \num{83008} & \num{545984} & \num{32.4} & \num{462.9}\\
Full $z$ & \num{0.0691} & \num{4.98} & \num{929.89} & \num{-69.32} & \num{20065344} & \num{20528320} & \num{175.1} & \num{7683.2}\\
\end{tabular}
\end{ruledtabular}
\end{table*}

\subsection{Reference solver for inhomogeneous screening}
\label{app:heterogeneous-solver}

We solve Eq.~\eqref{eq:het-poisson-k} by expanding the potential
$\varphi$ in plane waves up to $|\mathbf k|\le\ensuremath{3.5\,\text{\AA}^{-1}}$,
including $\mathbf k=0$.
In a neutral cell, the zero-mode potential vanishes for a uniform medium.
An inhomogeneous medium instead couples it to the nonzero modes through
\[
\kappa_0^2\widehat\varphi(\mathbf 0)
+\kappa_1^2\langle c_S\varphi\rangle
=4\pi\widehat\rho_\sigma(\mathbf 0),
\]
where $\langle\cdot\rangle$ denotes the cell average.
The zero mode must therefore be included even when the total charge is zero.

The medium field $c_S$ is a sum of Gaussians centered on the $S$ particles
[Eq.~\eqref{eq:het-kappa}],
and its Fourier coefficients contain the smoothing factor $e^{-\ell^2k^2/2}$.
We evaluate these coefficients analytically
from the particle positions and truncate the medium-field expansion
where its Gaussian smoothing factor falls below $10^{-14}$, which leads to
$|\mathbf k|\le B=\sqrt{2\ln10^{14}}/\ell$.
For $\ell=\ensuremath{3\,\text{\AA}}$, $B=\ensuremath{2.68\,\text{\AA}^{-1}}$.

We evaluate the convolution in Eq.~\eqref{eq:het-poisson-k} by forming
the product $c_S\varphi$ in real space, rather than directly in reciprocal space.
Both expansions are transformed to a grid of $N_{\mathrm{grid}}$ points per axis,
multiplied pointwise, and transformed back, keeping only the modes of the
$\varphi$ basis~\cite{boyd2001spectral}.
The product contains integer mode components up to $n_\varphi+n_c$, where
$n_\varphi$ and $n_c$ are the largest absolute integer mode components of the
two expansions (\num{17} and \num{13} for the \ensuremath{32\,\text{\AA}} cell and
$\ell=\ensuremath{3\,\text{\AA}}$).
On a finite grid, modes beyond its resolution are aliased onto lower ones, and
the condition $N_{\mathrm{grid}}>2n_\varphi+n_c$ ensures that none of them falls
into the retained band.
To satisfy this condition, all labels use $N_{\mathrm{grid}}=\num{48}$.
We solve the linear system by conjugate gradients, preconditioned by the
diagonal inverse of the uniform-medium operator,
$(k^2+\kappa_0^2+\kappa_1^2)^{-1}$.
A relative residual of $10^{-12}$ is reached in \numrange{22}{26}
iterations per configuration.

To reduce the sensitivity of the energy to the finite solver tolerance,
we use the stationary expression
\begin{equation}
E_{\mathrm{scr}}=C_{\mathrm{Coul}}\left[\int_\Omega\rho_\sigma\varphi\,\mathrm d\mathbf r
-\frac{1}{8\pi}\int_\Omega\varphi\,A\varphi\,\mathrm d\mathbf r\right],
\label{eq:het-stationary-energy}
\end{equation}
where $A=-\nabla^2+\kappa^2(\mathbf r)$.
This expression equals Eq.~\eqref{eq:het-energy} at the exact solution,
and its error is second order in the solver residual.
At convergence, stationarity removes the terms involving $\delta\varphi$
from the energy variation~\cite{gilson1993forces}
\begin{equation}
\delta E_{\mathrm{scr}}=C_{\mathrm{Coul}}\left[\int_\Omega\varphi\,\delta\rho_\sigma\,\mathrm d\mathbf r
-\frac{1}{8\pi}\int_\Omega\varphi^2\,\delta\kappa^2\,\mathrm d\mathbf r\right].
\label{eq:het-energy-variation}
\end{equation}
The forces can therefore be evaluated without differentiating the solver
iterations. Displacing ion $i$ changes $\rho_\sigma$, whereas displacing
screening particle $m\in S$ changes $\kappa^2=\kappa_0^2+\kappa_1^2c_S$.
Their screened force contributions are
\[
\begin{aligned}
\mathbf F_i^{\mathrm{scr}}&=-C_{\mathrm{Coul}}\int_\Omega\varphi\,
\frac{\partial\rho_\sigma}{\partial\mathbf R_i}\,\mathrm d\mathbf r,\\
\mathbf F_m^{\mathrm{scr}}&=\frac{C_{\mathrm{Coul}}\kappa_1^2}{8\pi}
\int_\Omega\varphi^2\,\frac{\partial c_S}{\partial\mathbf R_m}\,\mathrm d\mathbf r.
\end{aligned}
\]
The coefficients of $\varphi^2$ needed for the $S$ forces are obtained on the
same alias-free grid. We multiply these contributions by $a_{\mathrm h}$
and add the Lennard--Jones forces to obtain the reference forces.
The constant $E_{\mathrm{self}}^{(0)}$ is evaluated in the $S$-free medium
($\kappa=\kappa_0$) using the nonzero modes of the same plane-wave basis,
so that the self-energy subtraction is consistent with the Fourier cutoff.
One energy and force evaluation takes about \ensuremath{0.7\,\mathrm{s}} on a single
CPU core.

Before labeling, we verified the uniform Yukawa limit,
grid independence for $N_{\mathrm{grid}}=\numlist{48;64;96}$,
symmetries under translation and cell-compatible rotations, and the vanishing net force.
Analytic forces agree with finite differences within \ensuremath{10^{-8}\,\mathrm{eV}\,\text{\AA}^{-1}}.

\bibliographystyle{apsrev4-2}
\bibliography{references}
\end{document}